\documentclass{article}
\usepackage[a4paper, total={6.5in,9in}]{geometry}
\usepackage[utf8]{inputenc}
\usepackage{amsmath}
\usepackage{amssymb}
\usepackage{graphicx}
\usepackage{authblk}
\usepackage{url}
\usepackage[authoryear]{natbib}
\usepackage{adjustbox}
\usepackage{rotating}
\usepackage{comment}
\usepackage{longtable}
\usepackage{xcolor}
\usepackage{ulem}
\usepackage[english]{babel}
\usepackage{setspace}
\usepackage{multirow}
\usepackage{amsfonts}
\usepackage{array} 
\usepackage{flushend} 
\usepackage{stfloats} 
\usepackage{color}
\usepackage{chngpage} 
\usepackage{totcount} 
\usepackage{hyperref}
\usepackage{fix-cm} 
\usepackage{algorithm}
\usepackage{algorithmicx} 
\usepackage{algpseudocode} 
\usepackage{listings} 
\usepackage{crop}
\usepackage{subfloat} 
\usepackage{subfig}
\usepackage{tikz} 
\usepackage{hyperref} 
\usepackage{footnote}
\usepackage{mathrsfs} 
\usepackage{wrapfig} 
\usepackage{amsthm} \usepackage[most]{tcolorbox}
\usepackage{listings}  
\tcbuselibrary{listings, breakable}

\lstdefinelanguage{Stata}{
	morekeywords={clear, set, obs, seed, gen, rnormal, twoway, scatter, line, local, range, sqrt, aspect, graphregion, name, legend, xlabel, ylabel, xtitle, ytitle,mcolor,msize},
	sensitive=false,
	morecomment=[l]{*},
	morecomment=[l]{//},
	morestring=[b]",
}

\newtcolorbox[auto counter, number within=section]{mybox}[2][]{%
	enhanced,
	colback=blue!5!white,
	colframe=blue!75!black,
	fonttitle=\bfseries,
	coltitle=white,
	title=Box~\thetcbcounter: #2,
	listing only,
	listing options={language=R},
	unbreakable,
	boxsep=1pt,
	left=2pt,      
	right=2pt,
	top=2pt,
	bottom=2pt,
	sharp corners,
	#1
}

\begin{document}
	
	\vspace{-40pt}
	\title{ballmapper: Applying Topological Data Analysis Ball Mapper in Stata}
	\vspace{-30pt}

	\author[1]{Simon Rudkin \thanks{\textbf{Corresponding Author} Full Address: Department of Social Statistics, School of Social Sciences, University of Manchester, Oxford Road, Manchester, M13 9PL, United Kingdom. Tel: +44 (0)7955 109334 Email:simon.rudkin@manchester.ac.uk}}
	\affil[1]{School of Social Sciences, University of Manchester, United Kingdom}
	
	\vspace{-30pt}
	\author[2]{Wanling Rudkin \thanks{Full Address: University of Exeter Business School, University of Exeter, Streatham Court, Rennes Drive, Exeter, EX4 4PU, United Kingdom.  Email: w.rudkin@exeter.ac.uk}}
	\affil[2]{University of Exeter Business School, University of Exeter, United Kingdom}
	\vspace{-40pt}
	\vspace{-20pt}
	
	\doublespacing
	
	\maketitle
	
	\begin{abstract}		
		Topological Data Analysis Ball Mapper (TDABM) offers a model free visualization of multivariate data which does not necessitate the information loss associated with dimensionality reduction. TDABM \cite{dlotko2019ball} produces a cover of a multidimensional point cloud using equal size balls, the radius of the ball is the only parameter. A TDABM visualization retains the full structure of the data. The graphs produced by TDABM can convey coloration according to further variables, model residuals, or variables within the multivariate data. An expanding literature makes use of the power of TDABM across Finance, Economics, Geography, Medicine and Chemistry amongst others. We provide an introduction to TDABM and the \texttt{ballmapper} package for Stata.
	\end{abstract}


\section{Introduction}

Topological Data Analysis Ball Mapper \cite{dlotko2019ball} is an algorithm for the visualization of multi-dimensional datasets. Unlike methods such as T-SNE or UMAP, there is no dimensionality reduction and hence there is no loss of information from the dataset. For an exposition of the comparison between TDABM, T-SNE, UMAP and the original mapper algorithm of \cite{singh2007topological}, see \cite{dlotko2019ball} and \cite{dlotko2022topological}. Viewing the structures within data gives information on potential correlations and relationships. The merits of visualizing data to explore structure are well understood \citep{anscombe1973graphs,matejka2017same}.  \cite{anscombe1973graphs} further highlights how data visualisation after modelling supports more effective model evaluation\footnote{The \cite{anscombe1973graphs} example concerns 4 datasets each with the same correlation between $X$ and $Y$, and each with the same first and second moments for $X$ and $Y$. Moreover, the 4 datasets have the same ordinary least squares regression line and model R-squared. Only on viewing the data can you see that the relationship is only linear in one of the 4.}. The challenge presented to the empirical researcher is that data visualization is limited at 2 or 3 dimensions unless advanced algorithms are used. Where tools like Principal Components Analysis (PCA), UMAP, TSNE, and panel plots provide visualizations, TDABM offers an intuitive and easy to implement alternative. 

Let us define a dataset $X$ in terms of $K$ variables. Each variable $k \in \left[1,K\right]$ should have sufficiently many values to make the plotting of a scatterplot a sensible option\footnote{We do not prescribe a numeric value to define sufficient, but recommend caution is exercised when handling variables with fewer than 10 possible values.}. Within $X$, each individual data point $x_i$ has a value $x_{ik}$ for variable $k$. Because TDABM is a distance-based approach, the algorithm will not work with missing values. Hence, it is required that $x_{ik}$ be non-missing for $i \in \left[1,N \right]$ and $k \in \left[1,K\right]$. Further, the dataset must be scaled to ensure that all $x_k$ have similar support. Scaling is common for Machine Learning algorithms and other distance based statistical modelling. All discussion in this paper relates to $X_k$ as the appropriately scaled dataset.

Intuition for the use of balls derives from the idea that points within a given radius are ``similar''. By representing $X$ as a $K$-dimensional point cloud, any point within radius $\varepsilon$ of a point can be considered similar to that point. The TDABM algorithm covers the data in balls such that each point is in at least one ball. When discussing balls, we are able to talk about sets of similar points. There are analogies with clustering algorithms, but there is no variation in the group sizes. A ball also handles outliers differently. Where clustering algorithms like k-means \citep{hartigan1979algorithm} place outliers into their nearest cluster, TDABM will leave outliers in their own ball unless the radius is sufficiently large. When optimising the number of clusters, k-means provides fewer clusters than TDABM uses balls. To date there is no optimisation for the ball radius $\varepsilon$\footnote{The intuition for not providing an optimal radius is that we wish to understand the structure of data at many levels. For local structure a small radius is needed. Meanwhile, to see the full structure of the data a large radius is needed. The natural analogy is to mapping the Earth's surface. Different scale maps are needed for different purposes.}. \cite{otway2024shape}, \cite{dlotko2021financial} and \cite{rudkin2023economic} all present comparisons of TDABM results and clustering algorithms.

The objective of TDABM is to produce a topologically faithful representation of $X$ which can be visualised in two-dimensions. TDABM achieves this by covering the space in balls. The first ball, ball 1, is centred on the first observation in the dataset. A ball of radius $\varepsilon$ is drawn. Points within the ball are covered. The algorithm continues by selecting a point from the uncovered set and constructing a new ball around that point. The algorithm stops when all points are covered.  Because of the way in which the centers of the balls are chosen, overlap between balls is possible. A ball, $j$, written as $B_j(X,\epsilon)$, in two-dimensions is shown as a disc. The data is covered in a series of $L$ balls, such that there are $L$ discs on the TDABM graph. Edges between balls in the TDABM graph form where there is overlap of the balls. To convey the density of the data within a part of the joint distribution, balls are sized according to the number of points within the ball. The mechanism through which the abstract representation is achieved is expanded upon in Section \ref{sec:method}.

Because the TDABM algorithm retains information about which points sit within each ball the user is able to color the discs within the plot according to any function on the data contained within. A common coloration is to use an additional variable, $Y$. By applying the average value of $Y$ within the ball as coloration, the TDABM graph shows how $Y$ varies across the distribution of $X$. Using coloration also allows the user to see how each $x_k$ is varying across the space. Further, membership information allows the user to link back to their data. To assess models, the use of residuals or predicted values as coloring variables enable model assessment. Where there are identifiers for the data points, such as the region names or firm names, the user can then talk about which identities are in each ball and what the coloration of that ball informs. The Stata implementation provides options to change coloration and link back to the underlying data set. 

The ability of TDABM to generate interpretable visualizations of multivariate data has seen the methodology employed across economics and finance. Finance papers have focused on the use of TDABM to reappraise existing models. \cite{qiu2020refining} show that firm failures exist in a small subset of the space defined by the \cite{altman1968financial} Z-score default zone. \cite{charmpi2023topological} builds upon \cite{qiu2020refining} to show how TDABM can be used to effectively forecast firm failure based on the proportion of firms within a neighbourhood who did fail historically. Under the efficient markets hypothesis \cite{fama1973risk}, the direction of financial returns should not be predictable. \cite{rudkin2024return} illustrate that the ability to forecast future Bitcoin return directions depends upon where in the space of past trajectories the forecast is made. In both cases, the value derives from seeing where past models are underperforming and being able to relate that underperformance back to the data. \cite{dlotko2021financial} demonstrates how visualising model fit across the explanatory variable space can reveal where Machine Learning models outperform the established Ordinary Least Squares models. By showing that the residuals from Machine Learning models are only significantly lower in absolute terms in the extremities of the space, the value of giving up the interpretability of OLS for improved Machine Learning fit is revisited. 

Economics work has a greater focus on the ability to visualise stories in the data using TDABM. For example, \cite{rudkin2023economic} shows that the United Kingdom vote to leave the European Union in the 2016 is concentrated in a group of highly homogenous (large balls with many connections) constituencies, where the vote to remain in the European Union was much more fragmented. \cite{otway2024shape} then demonstrates how the patterns persist over time and link to general election results. Further applications include the study of the digital divide across European regions \citep{rudkin2024topology} and the study of migration patterns in \cite{tubadji2025cultural}. A recent paper by \cite{benites2025topology} expands upon variability in educational outcomes across the socio-demographic space of small scale local geographies.  \cite{rudkin2023regional} studies the development trajectories and the resilience of regions to the global financial crisis. The demonstration of extreme variability of resilience within each ball underscores that trajectories do not align with resilience. In a complex ecosystem like the economy, the choice of variables is important to the interpretation of TDABM. Often TDABM indicates that more variables would be needed. 

In the natural sciences, TDABM is also gaining traction. For example, \cite{madukpe2025comparative} apply TDABM in environmental monitoring to show how the approach can guide more efficient pollution monitoring; the authors note that the single parameter of TDABM provides an advantage over the traditional mapper of \cite{singh2007topological}. The ability to see multi-dimensional datasets in a single plot is further shown to be valuable in understanding reactions in analytical chemistry \cite{koljancic2025untargeted}. Relatedly, the multidimensional datasets of biology make visualisation complex. Presenting the example of learning from fish monoliths, TDABM is used to demonstrate influence that dimensionality reduction has on understanding decay \citep{valerio2025topological}. \cite{han2025topological} uses TDABM to view trajectories of cranial pressure, leveraging the ability to see the dimensions of the data as lagged values as well as being different variables observed at the same time period. The trajectory approach is similar to that employed by \cite{rudkin2023regional} and \cite{rudkin2024return}.

The remainder of this paper is organised as follows. Section \ref{sec:method} provides more depth on the TDABM methodology. Section \ref{sec:intuit} provides intuition for the TDABM methodology with a bivariate example. Section \ref{sec:art} discusses the artificial data which is used in the example constructions and builds TDABM graphs upon that data. Section \ref{sec:extend} introduces some of the further stages in analysis that can be undertaken using TDABM. Section \ref{sec:census} presents an example analysis of data from Stata's built in auto dataset of 1978 cars and prices. Section \ref{sec:summary} concludes.

\section{Methodology}
\label{sec:method}

Within this guide there is a dataset $X$ with $K$ variables. We will consider the data as a point cloud $P$. Indexing each data point as $i$, $i \in \lbrace 1, ..., N \rbrace$, we can define the location of $x_i$ in $P$ by the values $x_{ik}$, $k \in \lbrace 1, ..., K \rbrace$. Intuitively this is the way that a bivariate dataset is plotted onto the 2-dimensional space of a scatterplot. A point cloud is a generalisation of the scatterplot idea. The TDABM algorithm requires that there be a further variable, $Y$, which is used to color the datapoints. The coloring variable $Y$ may be one of the $X$ variables, or may be a distinct variable within the overall dataset. The final input to the TDABM algorithm is the radius of the balls used in the cover $\varepsilon$. An algorithm to identify the optimal $\varepsilon$ is an ongoing research agenda. However, the strong recommendation is to understand how the data is structured at multiple values of $\varepsilon$. This section describes how TDABM creates an abstract 2-dimensional representation of $K-$dimensional data. 

The algorithm begins by selecting a point at random from $P$. The first selected point becomes the first landmark, $l_1$. A ball of radius $\varepsilon$ is drawn centred on $l_1$. The first ball drawn becomes ball 1, $B_1(X,\varepsilon)$. Note that the labelling of balls is purely to allow the subsequent discussion of the TDABM graph and that the number 1 has no further interpretation. The points of $P$ that are contained within $B_1(X,\varepsilon)$ are considered covered and become the first members of the covered set. A second landmark is selected at random from the uncovered points in $P$. The new landmark is $l_2$ and is the center for ball 2, $B_2(X, \varepsilon)$. Any points within  $B_2(X, \varepsilon)$ that were not covered by ball 1 are added to the covered set. The combination of $B_1(X, \varepsilon)$ and $B_2(X, \varepsilon)$ becomes the start of the overall cover $B(X, \varepsilon)$. If there are still uncovered points then a further landmark, $l_3$, is selected from the uncovered set $B'(X, \varepsilon)$. A ball, $B_3(X, \varepsilon)$ is drawn, adding further points to $B(X, \varepsilon)$. The process of selecting landmarks and drawing balls continues until all points are covered by at least one ball, that is $B'(X, \varepsilon) = \emptyset$.  In total the number of landmarks is $L$.

Each ball, $B_b'(X, \varepsilon)$, $b \in \left[1, L\right]$ retains knowledge of the points contained within. The number of points in ball $b$, $n_b$, informs on the density of the dataset in the area covered by that ball. The average value of $Y$ amongst the points within the ball, $\bar{y}_b$, provides information on $Y$ in the part of data space covered by ball $B_b(X, \varepsilon)$. Knowledge of points also allows the algorithm to identify points which feature in more than one ball. Where the intersection of two balls, $q$ and $s$, is non-empty $B_q(X, \varepsilon) \cap B_s(X, \varepsilon) \neq \emptyset$, an edge is drawn between  $B_q(X, \varepsilon)$ and $B_s(X, \varepsilon)$. When the TDABM graph is plotted, balls are represented as discs sized proportionally to $n_b$ and connected with edges where identified. Because the algorithm is converting a $K$ dimensional dataset into 2-dimensions, the resulting plot has no interpretation for the horizontal and vertical directions on the page. The plot is instead an abstract 2-dimensional representation of the data.

In order to apply TDABM in Stata, it is necessary to install the \texttt{ballmapper} package from the GitHub site of Simon Rudkin. The command line to make the installation is provided in Box \ref{box:install}.

\begin{mybox}[label=box:install]{Installing \texttt{ballmapper}}
	The \texttt{ballmapper} is currently available from the GitHub repository of Simon Rudkin
	\begin{lstlisting}[language=Stata]
		net install ballmapper, from("https://raw.githubusercontent.com/srudkin12/statabm/main") replace
	\end{lstlisting}
	
\end{mybox}

\section{Intuition}
\label{sec:intuit}

To understand the construction of the TDABM plots in practice, this short section considers a bivariate example. A .do file to produce the anlaysis, \texttt{intuition.do} is available on the accompanying GitHub for the \texttt{ballmapper} package. The data contains 1000 points, $N=1000$ and two variables, $K=2$. Both of the two variables are drawn at random from the standard normal distribution and are independent. Hence we have $X_1 \sim N(0,1)$ and $X_2 \sim N(0,1)$. Using a Gaussian cloud like this allows the example to align with many of the assumed data generating processes of economics and finance. The Gaussian cloud also has structural properties that we can illustrate with TDABM. Firstly, the centre of the cloud is dense, corresponding to the points that are within a short distance of the mean on both $X_1$ and $X_2$. Moving away from the mean the cloud becomes sparser. Keeping one variable closer to its mean and letting the other vary wider means fewer points. However, allowing both variables to move away from the mean makes the joint probability of observing a point much lower. In practical terms, the centre of the scatterplot of $X_1$ and $X_2$ is dense, the areas along the $X_1=0$ and $X_2=0$ axes are sparse and there are almost no points in the corners of the plot. Contours of the density of a Gaussian cloud are circular. A TDABM plot of the cloud must therefore display these properties. Box \ref{box:dataset} provides the necessary Stata code to generate and plot the data.

\begin{mybox}[label=box:dataset]{Stata Code for Dataset}
	An initial bivariate dataset is produced, beginning with the generation of the initial values of $X_1$ and $X_2$ from $N \sim \left(0,1\right)$. The first step is to clear the environment to ensure that the new data is the data used by Stata.
	\begin{lstlisting}[language=Stata]
		clear
		set seed 1
		set obs 1000
		
		gen x1 = rnormal(0, 1)
		gen x2 = rnormal(0, 1)
	\end{lstlisting}
	A plot of the data is generated to show the results
	\begin{lstlisting}[language=Stata]
		twoway (scatter x2 x1, mcolor(gs12%50) msize(small)), xtitle("X{sub:1}") ytitle("X{sub:2}") xlabel(-4(1)4) ylabel(-4(1)4) aspect(1) graphregion(color(white)) name(scatter_orig, replace)
	\end{lstlisting}
	
\end{mybox}

For this example, $\varepsilon=1$ is selected, corresponding to the standard deviation of 1 of the variables. Using 1 also ensures that the code is simple to follow. At this $\varepsilon=1$, the 'balls' are large enough to bridge the gap between individual observations to ensure a connected graph. $\varepsilon=1$ is also small enough to capture the decaying density of the distribution as we move away from the origin. While lower values of $\varepsilon$ would provide a more granular view of the point cloud, $\epsilon = 1$ successfully reduces the 1,000-point sample into a parsimonious 'skeleton'. We show that the example actually has $L=21$ landmarks.

\begin{figure}
	\begin{center}
		\caption{Bivariate Normal Example Data}
		\label{fig:statabuild0}
		\includegraphics[width=12cm]{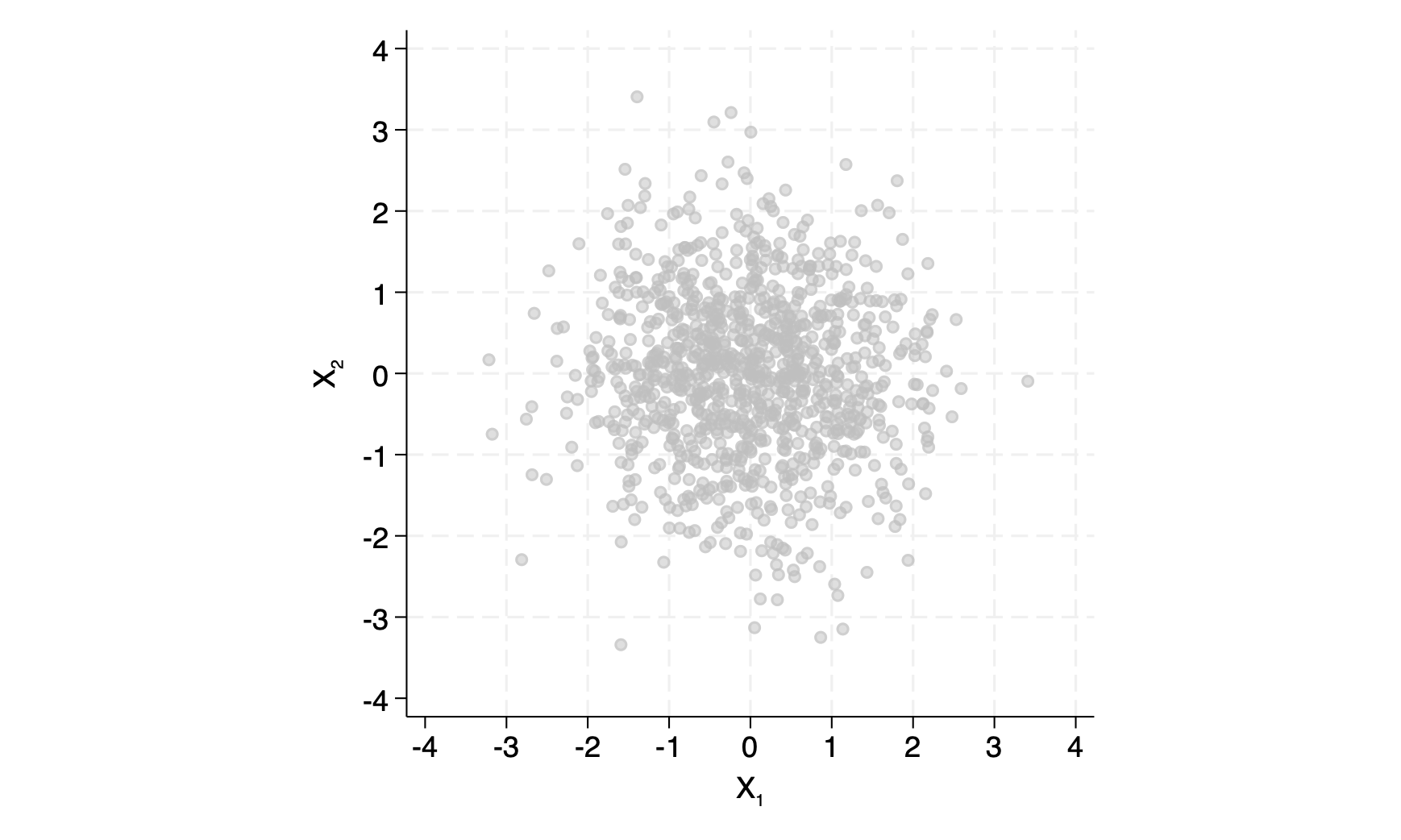}
	\end{center}
	\raggedright
	\footnotesize{Notes: Data points used in the artificial bivariate example. Each variable is drawn independently at random from a standard normal distribution, $X_1 \sim N(0,1)$ and $X_2 \sim N(0,1)$. $N=1000$.}
\end{figure}

Figure \ref{fig:statabuild0} presents the dataset that is used in this demonstration of the TDABM method. The Gaussian cloud is in clear evidence. The centre of the cloud is dense, whilst the peripheries have few points. Some points are far from the centre, isolated in the extremes of the plot. To represent this data accurately would therefore require capturing of the density and the overall proximity of most points. To begin the process of constructing the TDABM representation let us draw a ball around point 1 in the dataset. The code for drawing the first ball is in Box \ref{box:dataset}.

\begin{mybox}[label=box:ball1]{Stata Code for Dataset}
	To demonstrate the drawing of a ball around a single datapoint, take the first datapoint in the dataset and generate a distance to that point for all other points. The single datapoint is landmark 1.
	\begin{lstlisting}[language=Stata]
		local x11 = x1[1]
		local x21 = x2[1]
		gen dist_to_p1 = sqrt((x1 - `x11')^2 + (x2 - `x21')^2)
	\end{lstlisting}
	The example has $\epsilon=1$, so identify all points which are within a distance 1 of the landmark.
	\begin{lstlisting}[language=Stata]
		gen in_ball = (dist_to_p1 <= 1)
	\end{lstlisting}
	For the plotting we need a circle to represent the ball
	\begin{lstlisting}[language=Stata]
		range phi 0 2*_pi 100
		gen circle_x = `x11' + cos(phi)
		gen circle_y = `x21' + sin(phi)
	\end{lstlisting}
	Finally the scatterplot is generated. Note there are a lot more elements in the command now.
	\begin{lstlisting}[language=Stata]
		twoway (scatter x2 x1 if in_ball==0, mcolor(gs14%40) msize(small)) (scatter x2 x1 if in_ball==1 & _n > 1, mcolor(red%30) msize(small)) (scatter x2 x1 if _n==1, mcolor(red) msize(medium) msymbol(D)) (line circle_y circle_x, lcolor(red) lwidth(medium)), xtitle("X{sub:1}") ytitle("X{sub:2}") xlabel(-4(1)4) ylabel(-4(1)4) aspect(1) legend(order(3 "Point 1 (Landmark)" 2 "Points within {&epsilon}=1" 4 "Ball Boundary") pos(6) rows(1)) graphregion(color(white)) name(ball_demo, replace)
	\end{lstlisting}
\end{mybox}

\begin{figure}
	\begin{center}
		\caption{Building the Topological Data Analysis Ball Mapper Plot}
		\label{fig:statabuild}
		\begin{tabular}{c cc }
			\includegraphics[height=4cm]{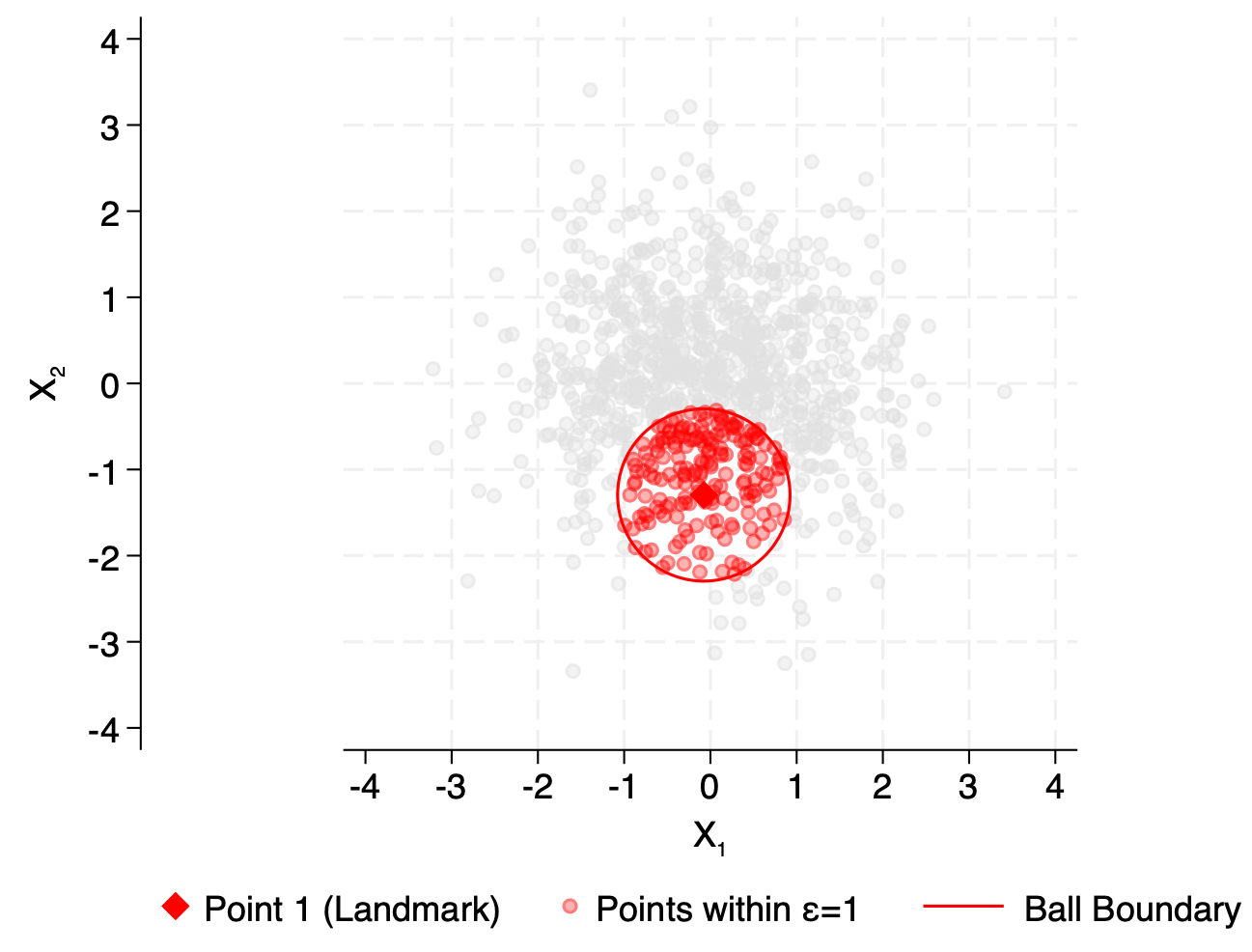}&
			\includegraphics[height=4cm]{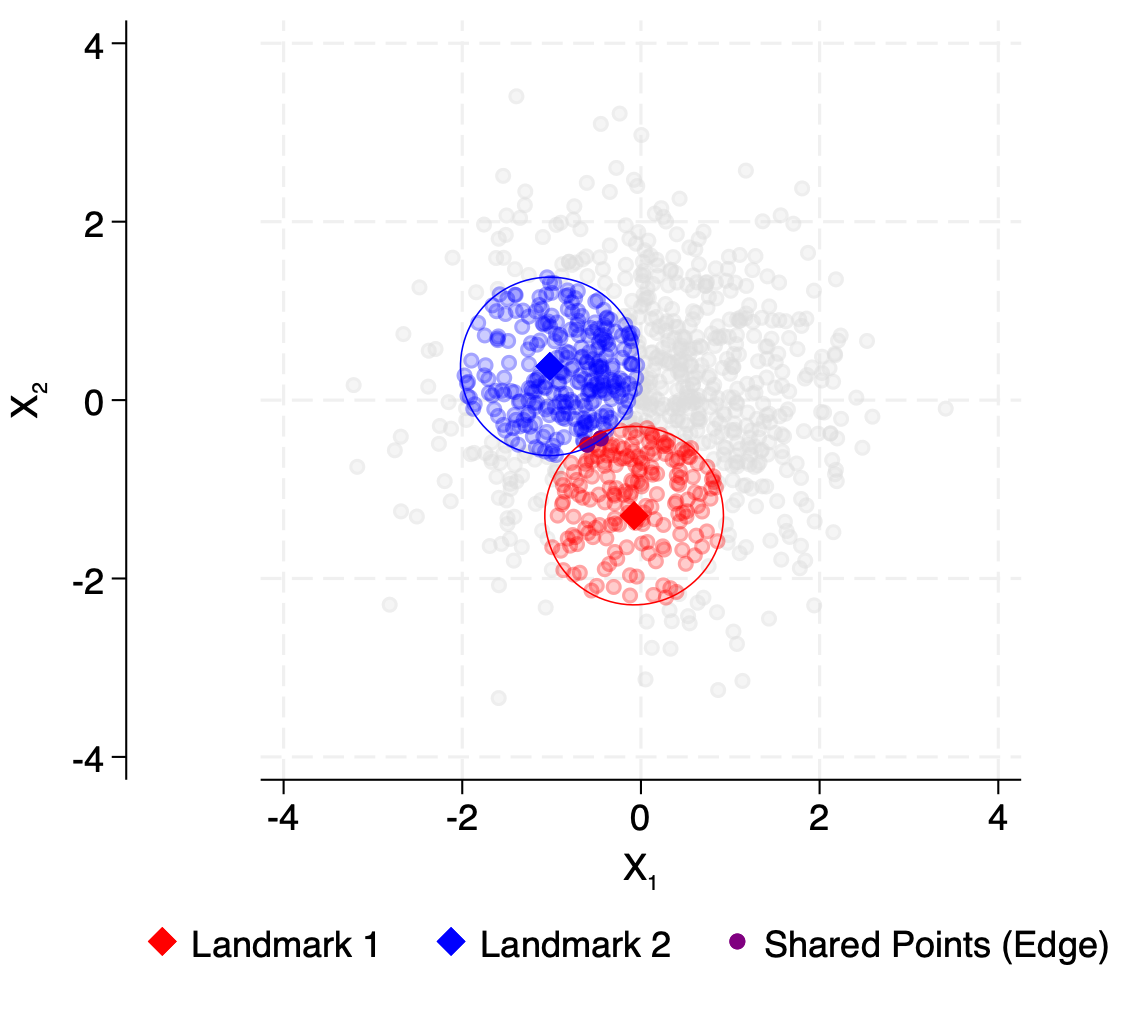}&
			\includegraphics[height=4cm]{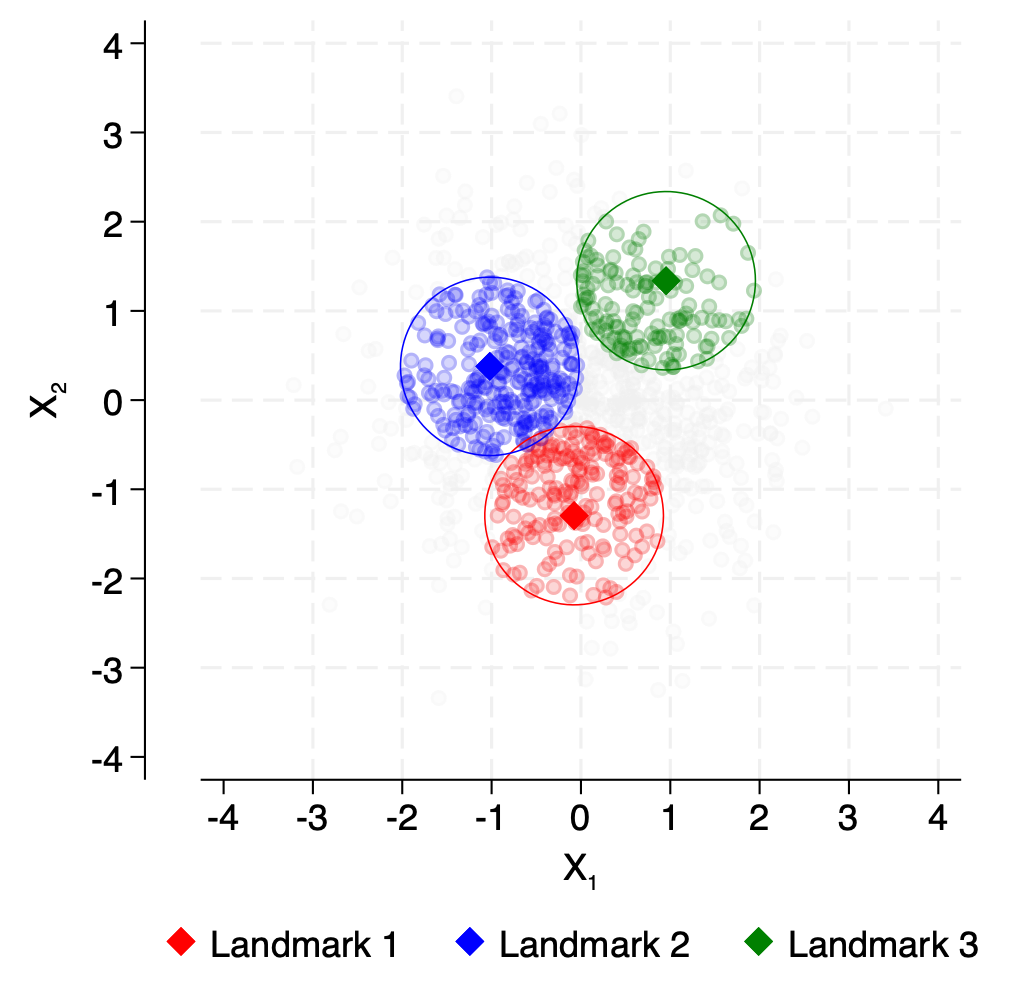}\\
			(a) Ball 1 & (b) Example Ball with Overlap & (c) Adding Ball 2 \\
		\end{tabular}
	\end{center}
	\raggedright
	\footnotesize{Notes: Construction of balls using the Topological Data Analysis Ball Mapper (TDABM) plot as implemented in the Stata package \texttt{ballmapper}. Panel (a) shows the construction of a ball of radius 1 around the first data point, labelled landmark 1. The ball is labelled ball 1. Panel (b) shows a second ball around a data point chosen to have overlap with ball 1. Here the landmark point is labelled landmark 2. Panel (c) shows the ball around the second datapoint in the data frame, here labelled landmark 3.  Data has $X_1 \sim N(0,1)$ and $X_2 \sim N(0,1)$. $N=1000$.}
\end{figure}

After running the code in Box \ref{box:ball1} we obtain panel (a) of Figure \ref{fig:statabuild}. The single ball is to the lower center of the overall cloud. Because this is a dense part of the space, there are a large number of points covered. However, the uncovered set of points remains large. To show an overlap, a second ball is drawn around one of the points that is close to Ball 1. In a random implementation of the TDABM algorithm there is a probability that the point shown would be selected. In most software implementations of TDABM, the algorithm selects the points in the order that they appear within the dataset. For this example, the second point is that shown in panel (c) of Figure \ref{fig:statabuild}. There is no overlap with either of the two previously selected balls. 

The specific points selected as landmarks in Figure \ref{fig:statabuild} depend on the random selection process. However, the resulting TDABM graphs are stable to the order of landmark selection. In the TDABM papers published to date, plots are used to illustrate the consistency of TDABM graphs over 1000s of iterations of the selection \citep[for example]{rudkin2023economic}. To see different perturbations of the landmark selection, you can reorder the data prior to supplying the dataframe to the \texttt{ballmapper()} function. To further demonstrate the construction of the TDABM plot for this bivariate dataset, an Appendix is provided using a dataset which produced with a different seed for the construction of $X_1$ and  $X_2$. Full \texttt{Stata} code is available in the accompanying GitHub to create the Appendix.

\section{Artificial Data}
\label{sec:art}

As an illustration of the need to consider both local and global structure, we consider a bivariate dataset $X$ with 900 observations. The variables $X_1$ and $X_2$ are drawn independently from standard normal distributions. Once drawn, subsets are subject to translation to create a large X shape. 100 observations are moved by -6 on $X_1$ and increased by $6$ on $X_2$. One group of 100 observations is moved by -6 on $X_1$ and -6 on $X_2$. A third group is moved 6 on $X_1$ and 6 on $X_2$, whilst a fourth group is moved 6 on $X_1$ and -6 on $X_2$. These four groups become the end of the X shape. Completing the global X structure are 4 groups shifted by 3, in the same pattern as the 6 shifts, and a final group which remains in the center of the X. Because the dataset has an X shape, it is henceforth referred to as the X dataset.

Following the shifts we have $\bar{X}_1 = -0.032$ and $\bar{X}_2 = 0.024$, the means remain close to 0. The standard deviations of $X_1$ and $X_2$ are increased by the shifts, such that $\sigma_1 = 4.592$ and $\sigma_2 = 4.601$. The Pearson correlation between the two variables is $\rho_{12}=0.003$. The summary statistics are consistent with a Gaussian cloud drawn from distributions with a higher standard deviation. There is no information in the first two moments that suggests the X shape. The X Dataset also serves as a reminder of the importance of visualizing data in the spirit of \cite{matejka2017same}.

\begin{figure}
	\begin{center}
		\caption{Artificial X Shaped Data with Outcome}
		\label{fig:x1}
		\includegraphics[width=10cm]{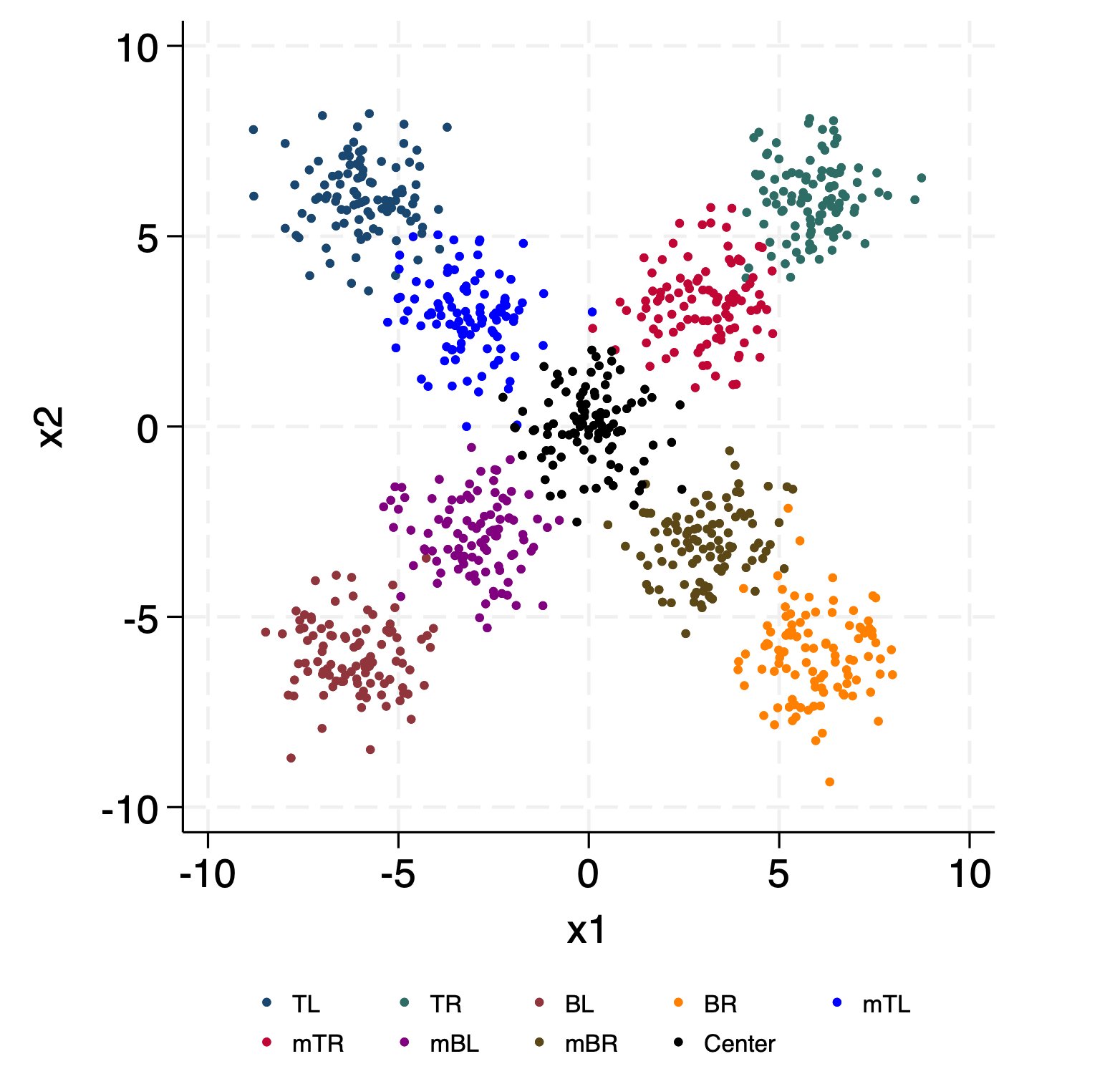}
	\end{center}
	\raggedright
	\footnotesize{Notes: Figure plots the X dataset. The underlying dataset has 900 observations on 2 variables, $X_1$ and $X_2$. The data is translated to have 9 groups of 100 points centred on (-6,6) (-3,3), (3,3), (6,6), (0,0), (-3,-3), (3,-3), (-6,-6), and (6,-6). The coloration variable, $Y_2$ is based on the group in which a data point is found, ranging from 1 to 9. Labels in the legend are for identification only.}
\end{figure}

\begin{figure}
	\begin{center}
		\caption{Artificial X Shaped Data with Alternative Outcomes}
		\label{fig:x2}
		\begin{tabular}{c c}
			\includegraphics[height=6cm]{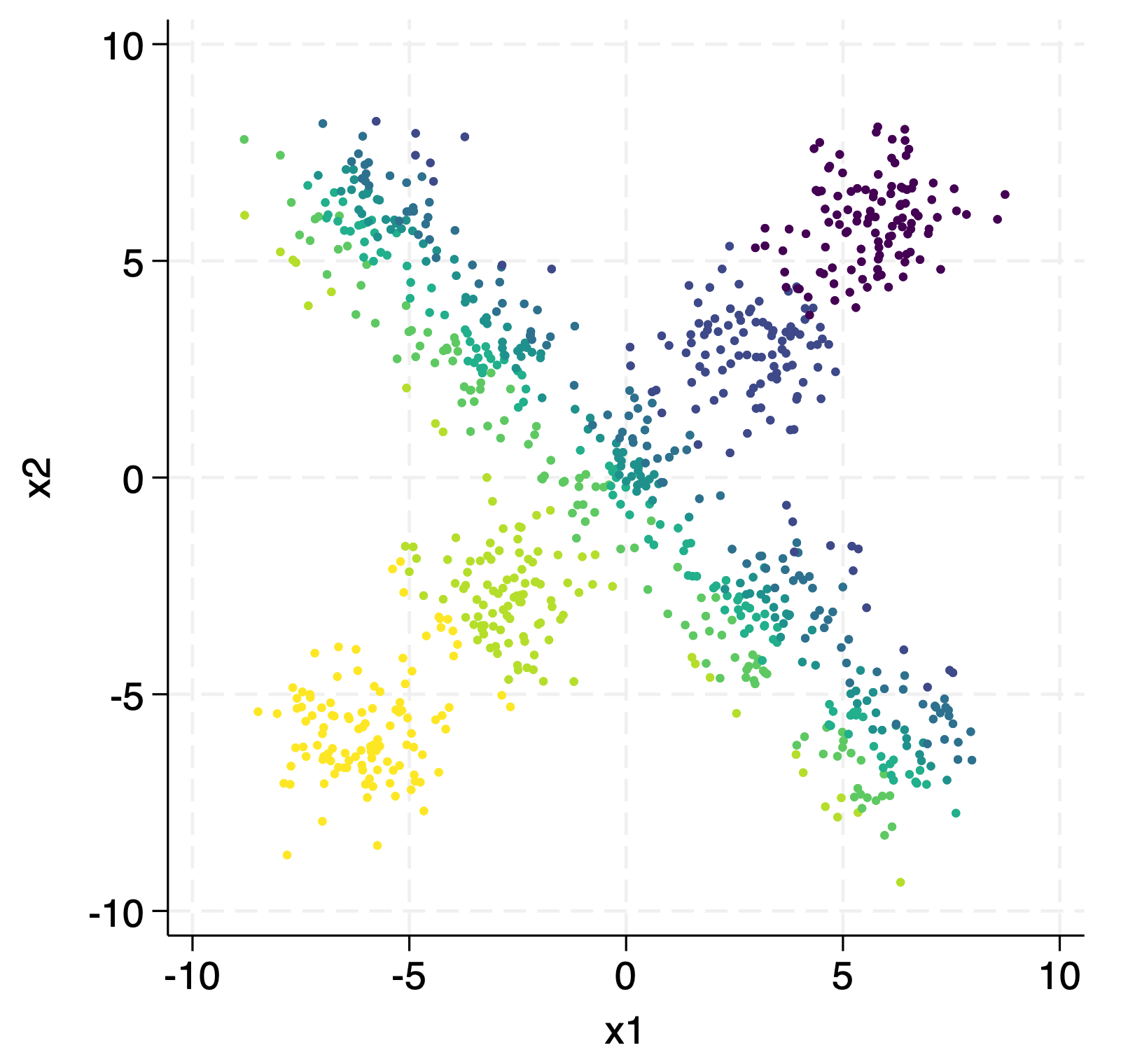}&
			\includegraphics[height=6cm]{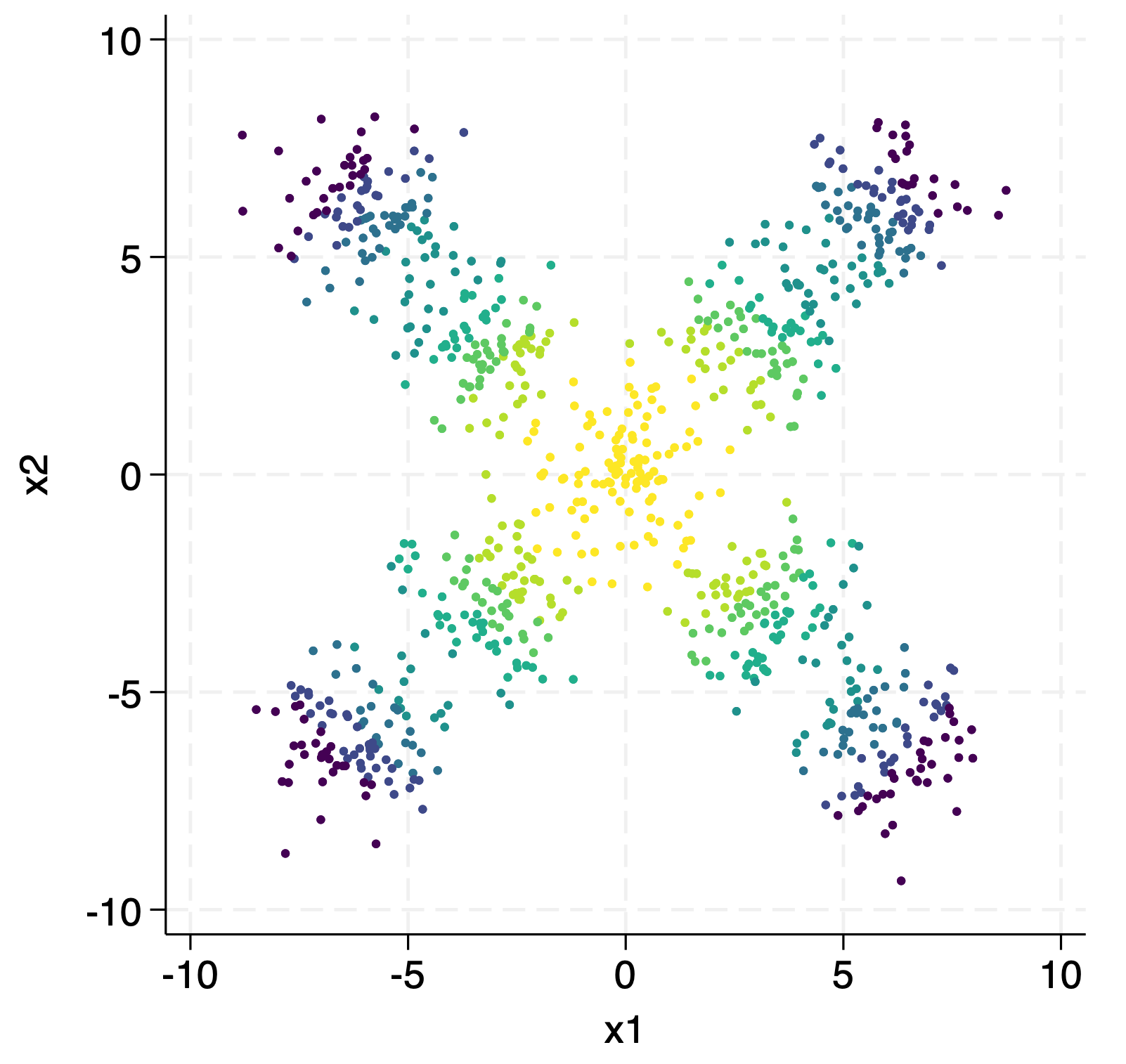}\\
			(a) $Y_1$ - Linear  & (b) $Y_3$ - Quadratic \\
			\includegraphics[height=6cm]{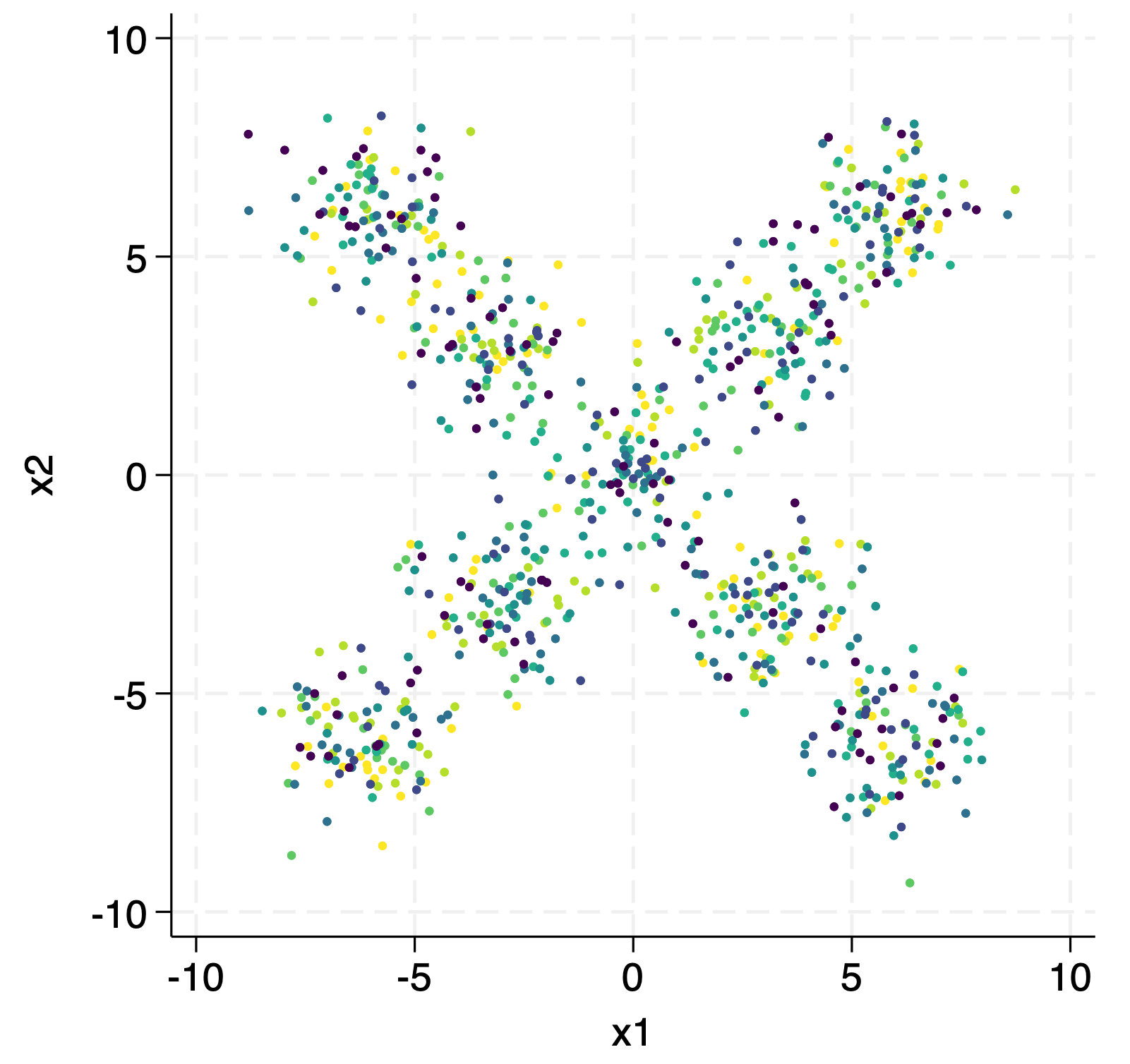}&
			\includegraphics[height=6cm]{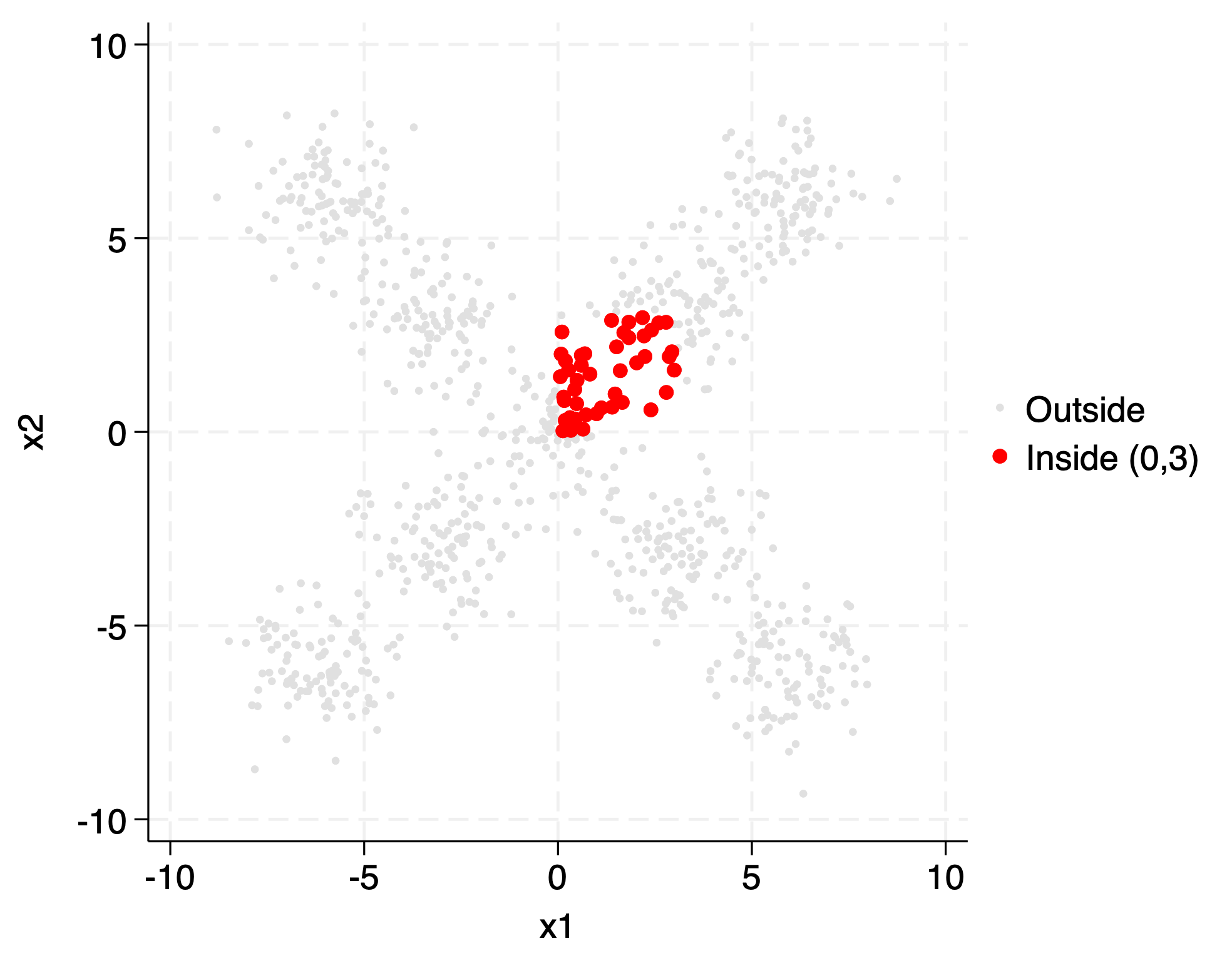}\\
			(c) $Y_4$ - Noise & (d) $Y_5$ - Restricted Range \\
		\end{tabular}
	\end{center}
	\raggedright
	\footnotesize{Notes: Scatterplots of the X dataset with alternative colorations. The underlying dataset has 900 observations on 2 variables, $X_1$ and $X_2$. The data is translated to have 9 groups of 100 points centred on (-6,6) (-3,3), (3,3), (6,6), (0,0), (-3,-3), (3,-3), (-6,-6), and (6,-6). The colorations are given as $Y_1 = X_1 + X_2 + \theta$ where $\theta \sim N(0,0.2)$, $Y_2$ is the group number, $Y_3 = X_1^2 + X_2^2 + \theta$ where again $\theta \sim N(0,0.2)$,  $Y_4 = \phi$ where $\phi \sim N(0,1)$, and $Y_5$ takes the value 1 when $0<X_1<3$ and $0<X_2<3$ are both satisfied.}
\end{figure}

The outcomes $Y_i$ are given according to the expressions in Table \ref{tab:x1}. These represent a linear function of the $X$'s, group membership, a quadratic function of the $X$'s, noise, and a binary indicator for a subset of the space. By having 5 distinct patterns, we can see how TDABM represents these outcomes on the X data points cloud. Figures \ref{fig:x1} and \ref{fig:x2} illustrate. 

\begin{table}
	\begin{center}
		\caption{Outcomes for X Dataset}
		\label{tab:x1}
		\begin{tabular}{l l l l}
			\hline
			Outcome & Short & Equation & Assumptions \\
			\hline
			$Y_1$ & Linear & $Y_1 = X_1 + X_2 + \theta$ & $\theta \sim N(0,0.2)$\\
			$Y_2$ & Group & - & -\\
			$Y_3$ & Quadratic & $Y_3 = X_1^2 + X_2^2 + \theta$& $\theta \sim N(0,0.2)$\\
			$Y_4$ & Noise & $Y_4 = \phi$ & $\phi \sim N(0,1)$\\
			$Y_5$  & Restricted Range & $Y_5 = \begin{cases} 1 & \text{if } 0 < X_1 < 3 \text{ and } 0 < X_2 < 3 \\ 0 & \text{otherwise} \end{cases}$& \\
			\hline
		\end{tabular}
	\end{center}
	\raggedright
	\footnotesize{Notes: Outcomes used in the analysis of the X dataset. $X_1$ and $X_2$ are variables in the dataset. $\theta$ and $\phi$ are noise terms as defined on the relevant lines. $N=900$.}
\end{table}

Figure \ref{fig:x1} shows how the X shape emerges naturally. The sub-clouds are visible in the scatter plot. The 9 dense centers can be picked out immediately. To emphasize the structure, the coloration in Figure \ref{fig:x1} is according to the group number. Colors are arranged such that neighboring groups are sufficiently contrasting. Drawing $X_1$ and $X_2$ from the standard normal distribution means that in the limit 95\% of observations lie within plus or minus 2 standard deviations of the mean, that is between -2 and 2. Within 3 standard deviations we find 99\% of observations in the limit. By having the clouds shifted in multiples of 3, there is a possibility for overlap between the clouds. The distance between two adjacent centres is $\sqrt{18} \approxeq 4.24$ so 2 standard deviations from the centers does not overlap, but 3 does. Looking closely at Figure \ref{fig:x1} reveals that there is indeed some overlap. 

Figure \ref{fig:x2} shows the same X dataset colored according to the other 4 coloration rules. In panel (a), the highest values of $Y_1$ are where $X_1$ and $X_2$ are at their highest in the top right. Meanwhile the lowest values of $Y_1$ are found to the lower left. Along the top left to bottom right diagonal there is a gradient of color, consistent with the fact that for each sub-cloud on the diagonal the lowest values are to the bottom left of that sub-cloud. Panel (b) shows the highest values at the ends of the X as expected. Panel (c) demonstrates no patterns because $Y_4$ is a noise term independent of $X_1$ and $X_2$. Finally, panel (d) shows the points with $0<X_1<3$ and $0<X_2<3$ in red, distinct from all of the other points. Because we are working in 2-dimensions here, the scatterplots are sufficient to show the pattern across the space.

We now wish to demonstrate how TDABM captures the patterns observed within the panels of Figure \ref{fig:x2}. To implement ballmapper on the data, the code in Box \ref{box:xcode2} is followed. The command for \texttt{ballmapper} accepts inputs for the axis variables, $X$, the coloring variable, $Y$, the radius $\varepsilon$. There are further optional arguments, layout repulsion and attraction which control how the algorithm displays the TDABM graph in the plot. If attraction is too strong then edges can pull balls too close together. If repulsion is too strong then everything gets pushed to the edges. Since attraction and repulsion are display parameters, they do not change the inference provided by the TDABM graph. The user may adjust attraction and repulsion to suit their data. Finally, the filename argument specifies the name of the .png file that will be created by the algorithm. The .png file is saved in the working directory that is in use for the Stata session. On completion, details of the edge strengths, landmarks and ball sizes are placed in a new Stata frame \texttt{BM\_RESULTS}. The information about which point is in which ball, together with a merge back to the underlying dataset is placed in \texttt{BM\_MERGED}. At the exploratory stage, only the TDABM graph is essential. 

\begin{mybox}[label=box:xcode2]{Implementing \texttt{ballmapper} on X Dataset}
	To demonstrate the drawing of a ball around a single datapoint, take the first datapoint in the dataset and generate a distance to that point for all other points. The single datapoint is landmark 1.
	\begin{lstlisting}[language=Stata]
		ballmapper x1 x2, epsilon(0.8) color(y1) layout repulsion(0.05) attraction(0.01) filename("xy1bm08")
	\end{lstlisting}
	The example has $\epsilon=0.8$, but we can also fit with any other epsilon. We also work with the other y values. So $y4$ with radius $\varepsilon=1.2$ is entered as
	\begin{lstlisting}[language=Stata]
		ballmapper x1 x2, epsilon(1.2) color(y4) layout repulsion(0.05) attraction(0.01) filename("xy4bm12")
	\end{lstlisting}
	When the code completes, Stata shows:
	\begin{lstlisting}[language=Stata]
		Ball Mapper Successful.
		-> Graph data stored in frame: BM_RESULTS
		-> Original data + Ball IDs in frame: BM_MERGED
	\end{lstlisting}
\end{mybox}

\begin{figure}
	\begin{center}
		\caption{TDABM Plots for X Dataset}
		\label{fig:x3}
		\begin{tabular}{c c }
			\multicolumn{2}{c}{\includegraphics[width=9cm]{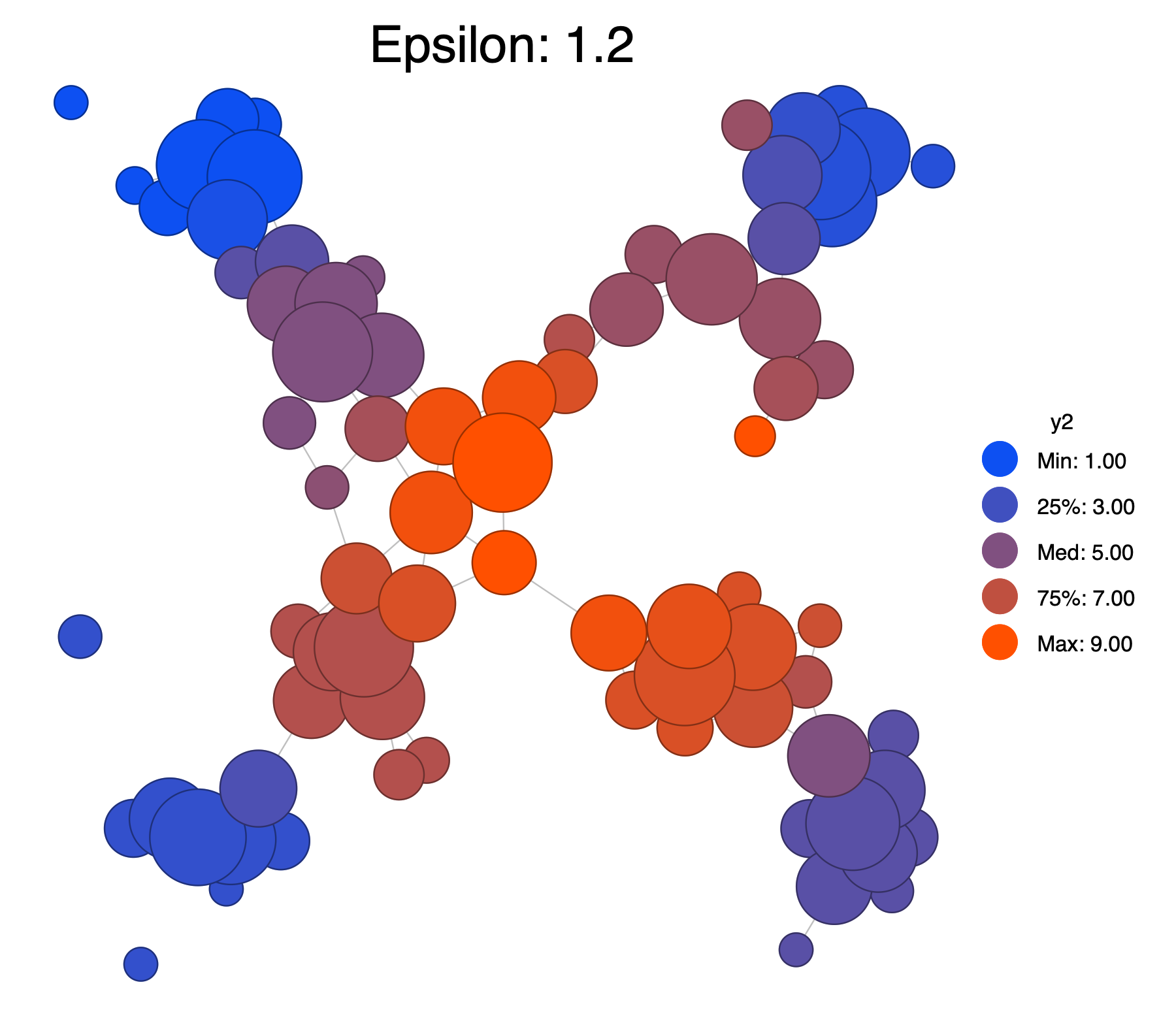}}\\
			\multicolumn{2}{c}{(a) Colored by Group Membership}\\
			\includegraphics[height=5.5cm]{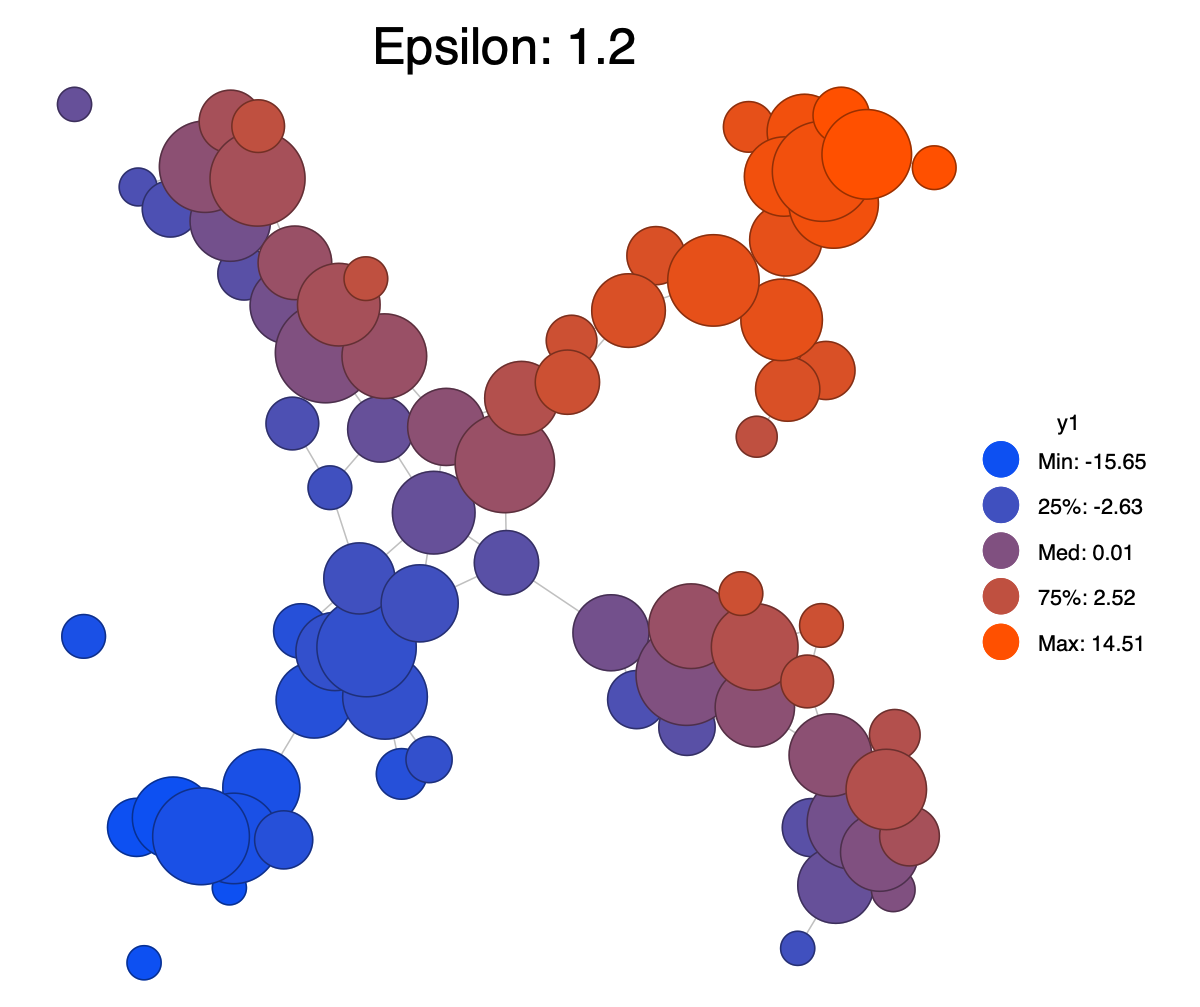}&
			\includegraphics[height=5.5cm]{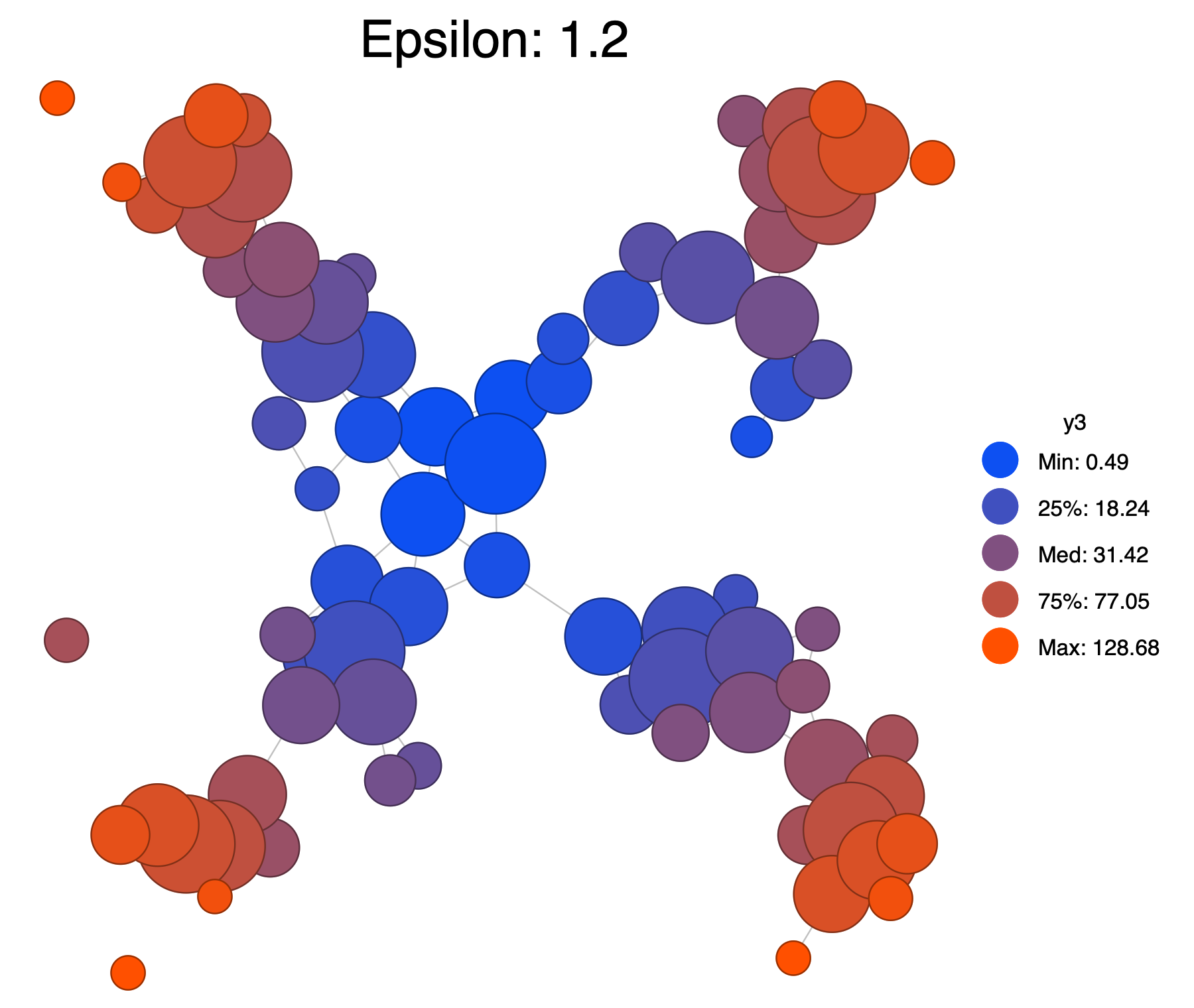}\\
			(b) $Y_1$ - Linear & (c) $Y_3$ - Quadratic\\
			\includegraphics[height=5.5cm]{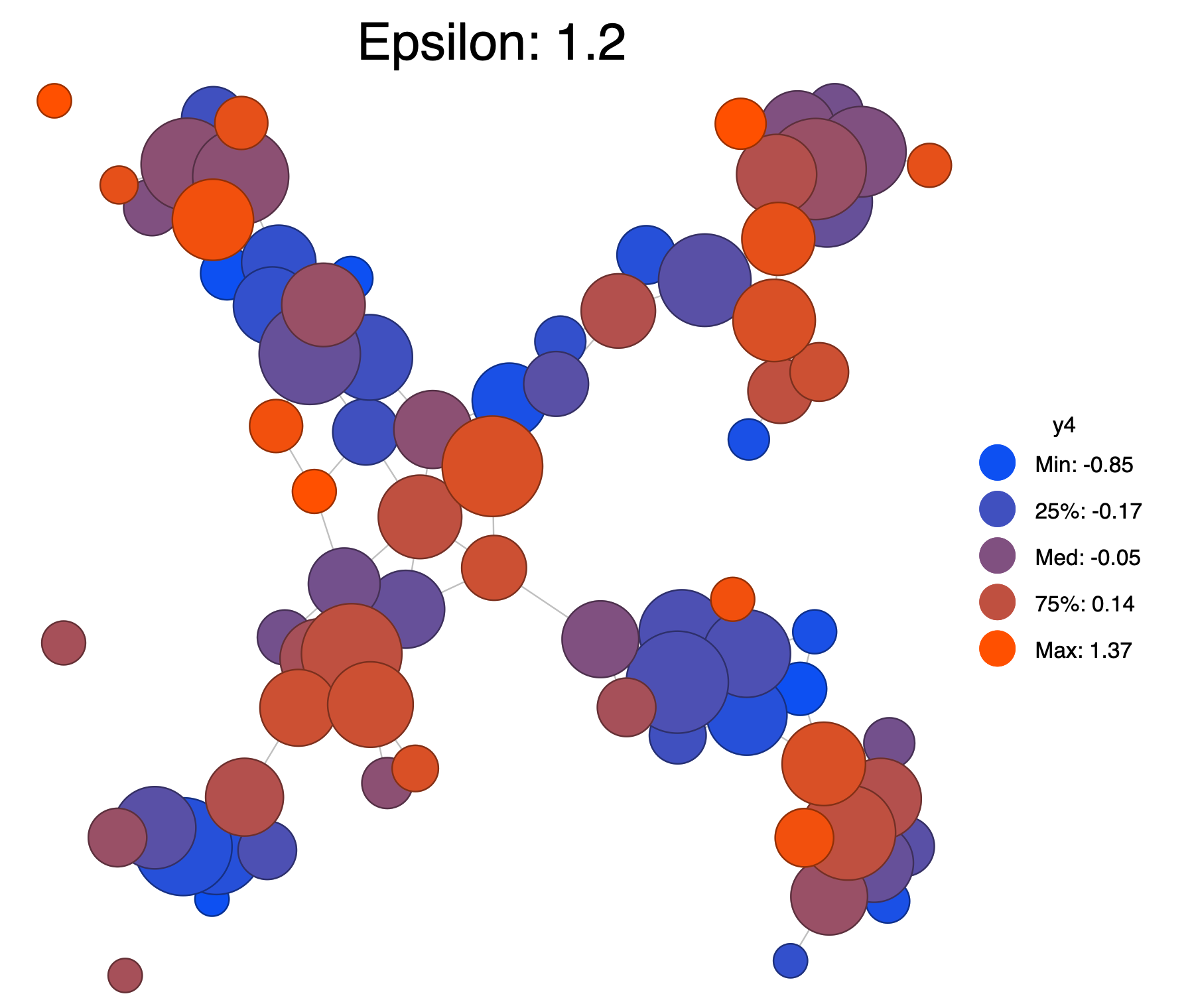}&
			\includegraphics[height=5.5cm]{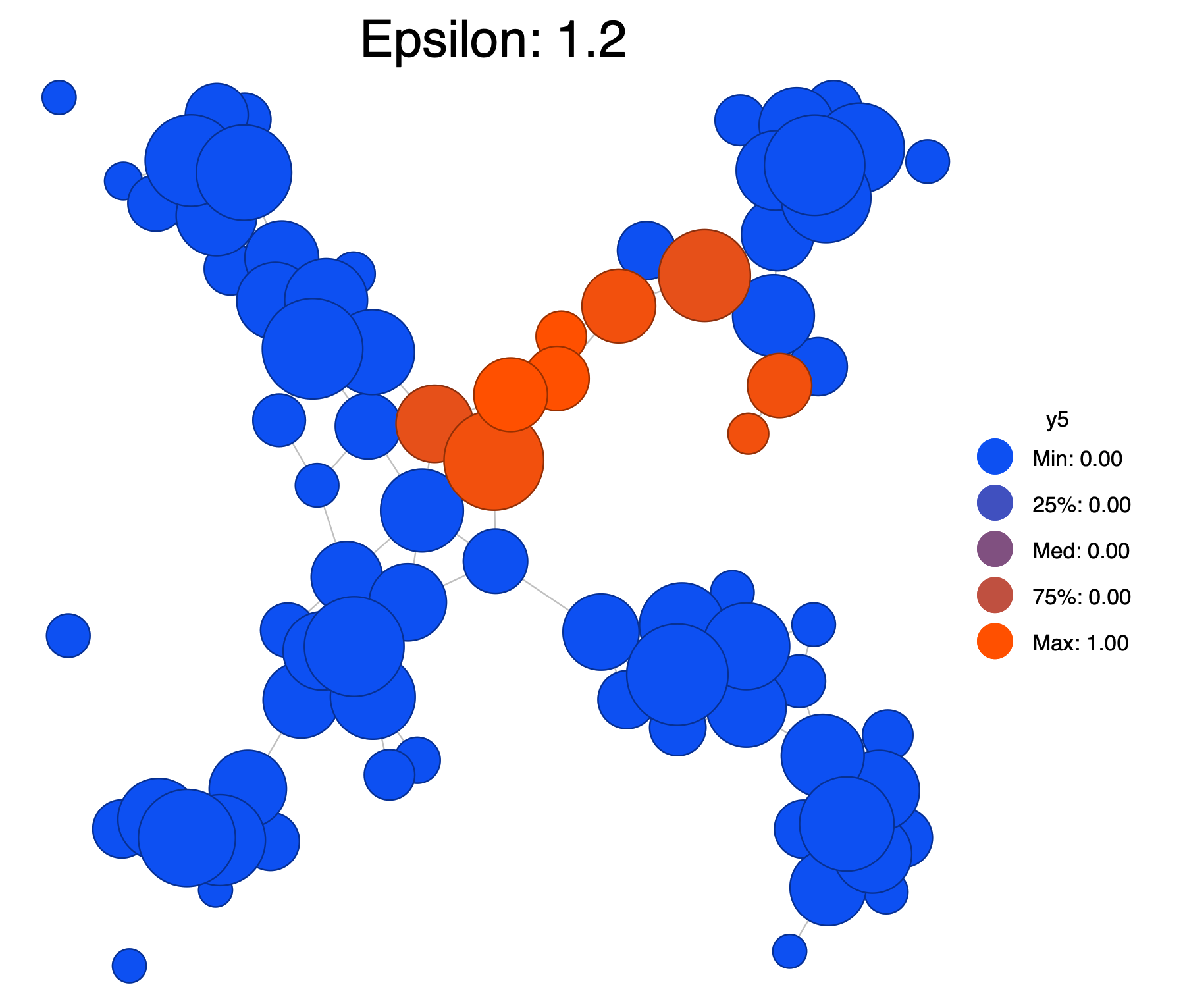}\\
			(d) $Y_4$ - Noise & (e) $Y_5$ - Restricted Range\\
		
		\end{tabular}
	\end{center}
	\raggedright
	\footnotesize{Notes: The underlying dataset has 900 observations on 2 variables, $X_1$ and $X_2$. The data is translated to have 9 groups of 100 points centred on (-6,6) (-3,3), (3,3), (6,6), (0,0), (-3,-3), (3,-3), (-6,-6), and (6,-6). The colorations are given as $Y_1 = X_1 + X_2 + \theta$ where $\theta \sim N(0,0.2)$, $Y_2$ is the group number, $Y_3 = X_1^2 + X_2^2 + \theta$ where again $\theta \sim N(0,0.2)$,  $Y_4 = \phi$ where $\phi \sim N(0,1)$, and $Y_5$ takes the value 1 when $0<X_1<3$ and $0<X_2<3$ are both satisfied. }
\end{figure}

\begin{figure}
	\begin{center}
		\caption{Role of Radius on X Dataset}
		\label{fig:x4}
		\begin{tabular}{c c}
			\includegraphics[height=5.5cm]{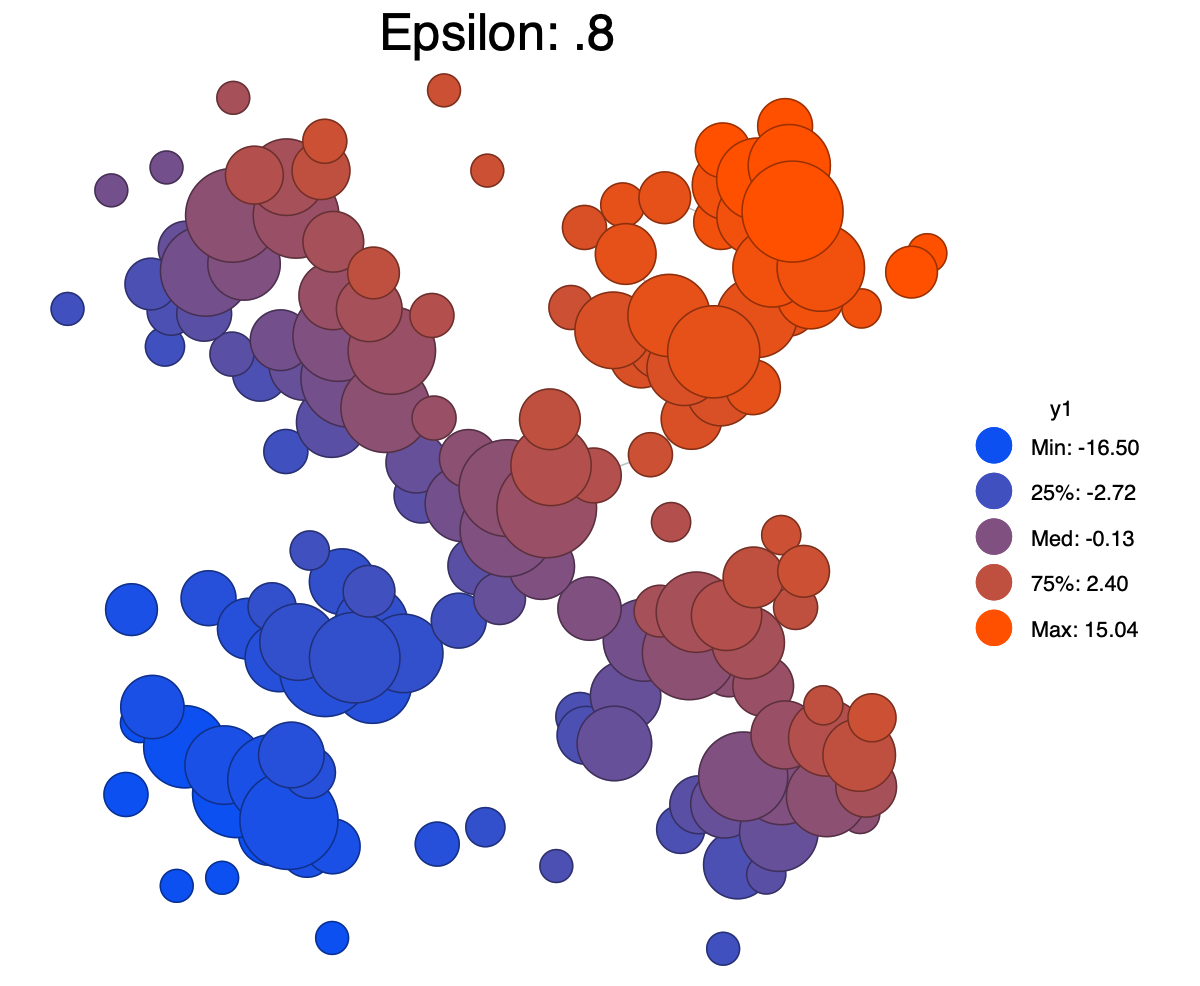}&
			\includegraphics[height=5.5cm]{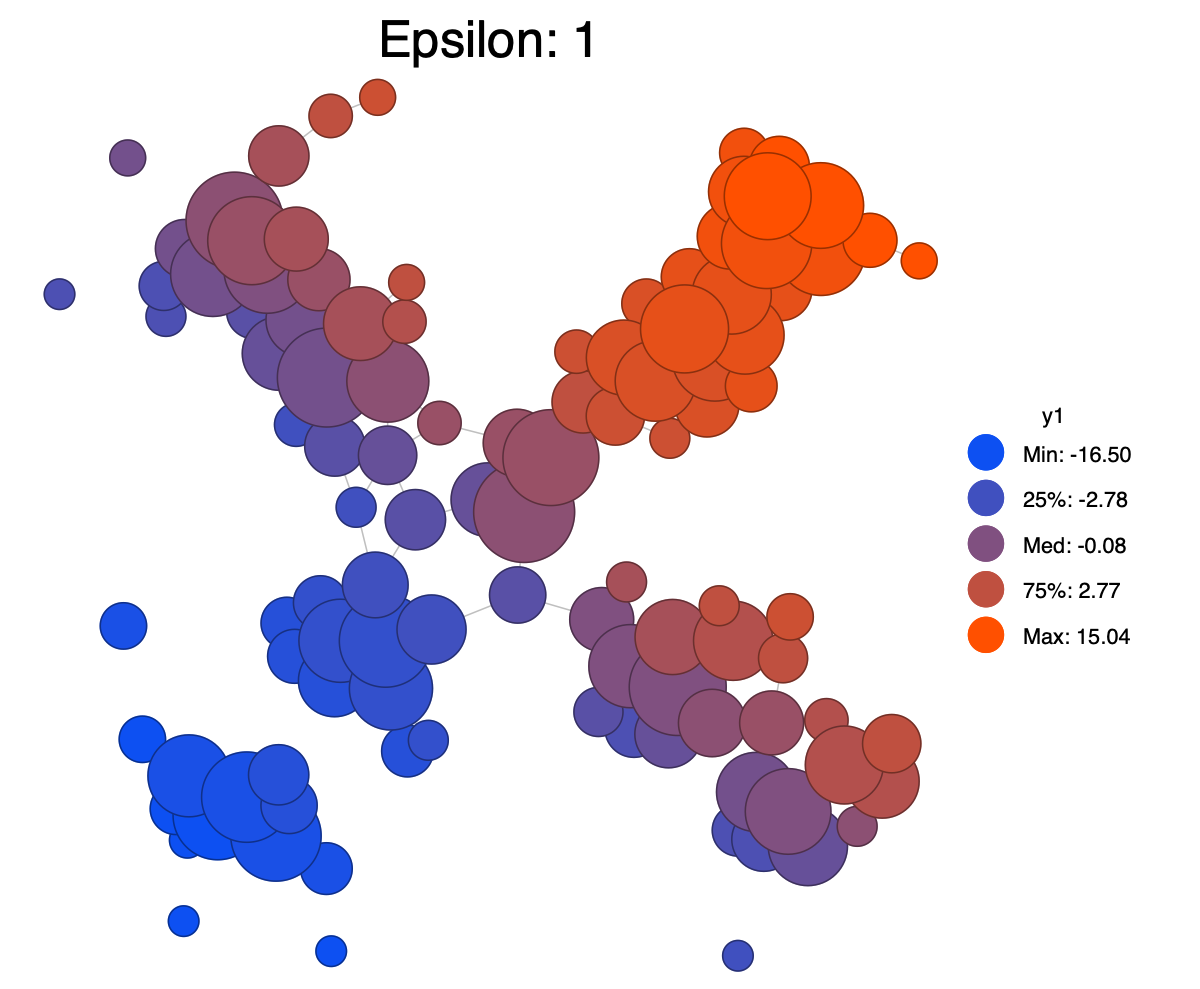}\\
			(a) $\varepsilon=0.80$ & (b) $\varepsilon=1.00$ \\
			\includegraphics[height=5.5cm]{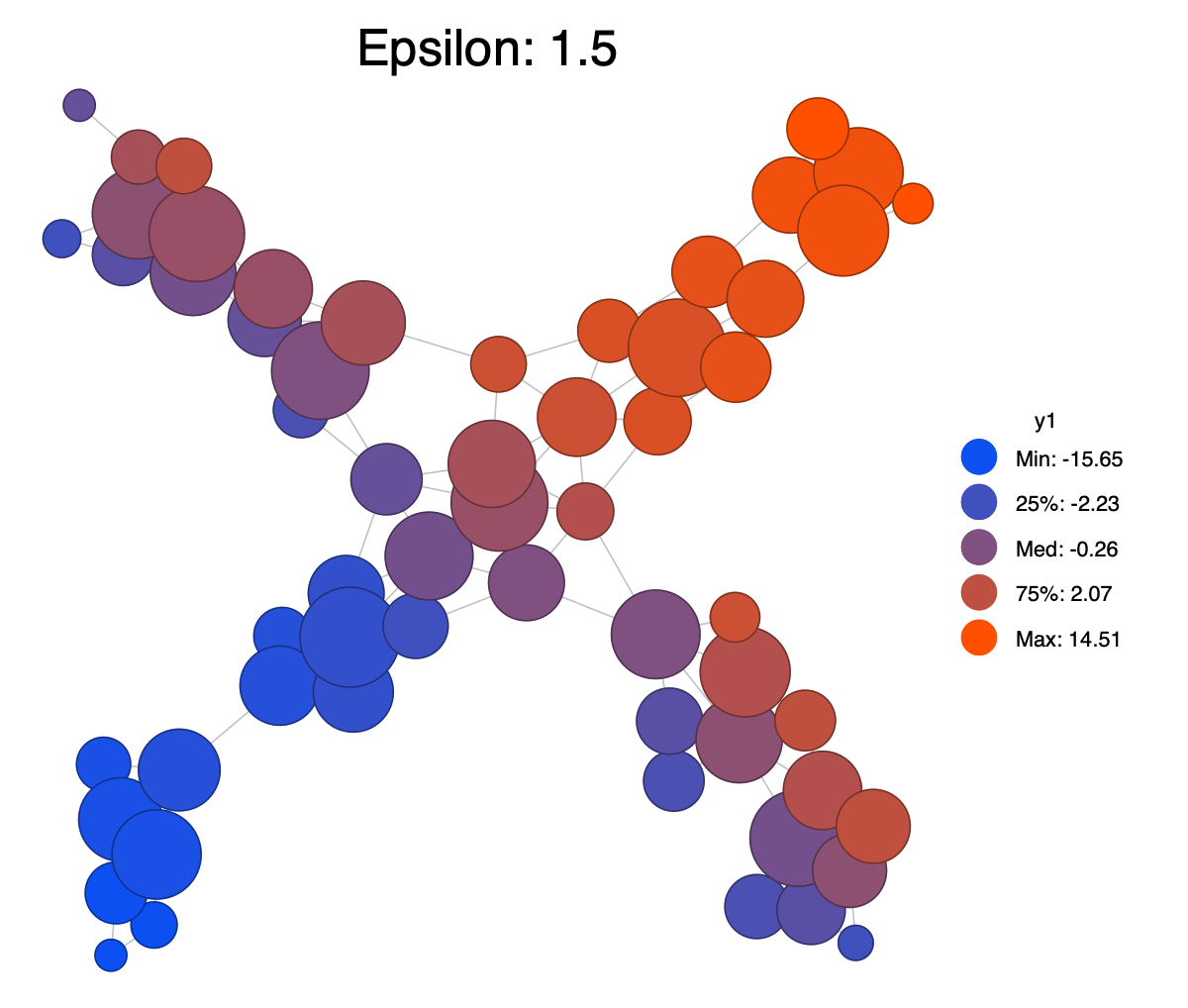}&
			\includegraphics[height=5.5cm]{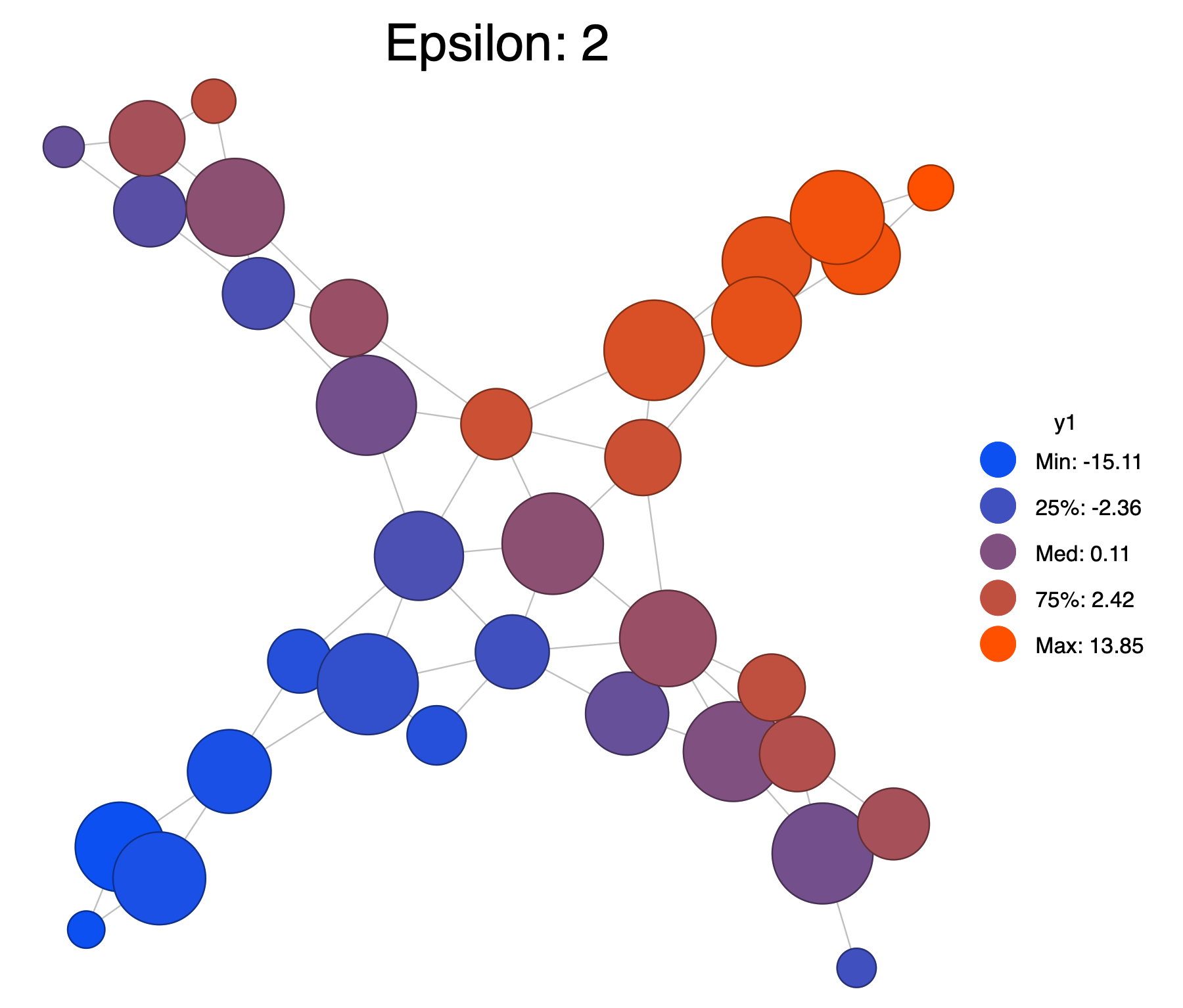}\\
			(c) $\varepsilon=1.50$ & (d) $\varepsilon=2.00$ \\
		\end{tabular}
	\end{center}
	\raggedright
	\footnotesize{Notes: Role of radius assessed through 4 different values of the ball radius $\varepsilon$. The underlying dataset has 900 observations on 2 variables, $X_1$ and $X_2$. The data is translated to have 9 groups of 100 points centred on (-6,6) (-3,3), (3,3), (6,6), (0,0), (-3,-3), (3,-3), (-6,-6), and (6,-6). Ball colorations are given as $Y_1 = X_1 + X_2 + \theta$ where $\theta \sim N(0,0.2)$. }
\end{figure}

Figure \ref{fig:x3} has 5 panels, one for each of the outcome variables. Because these plots are based on the same random variables, the structure of the data is identical. A radius of $\varepsilon=1.20$ is used here to show both the local structure of the sub-clouds and the global structure of the X. The decision on radius is explored further in Figure \ref{fig:x4}. Panel (a) is colored by group membership. Group membership is complicated by having 9 different values, but the structure is clear. Where a ball has a group membership that is not a whole number, this is because there are points from different sub-clouds combined within the ball. We see overlap in the areas between each sub-cloud as would be expected. At $\varepsilon=1.20$, there are still some balls which are not connected to any of the groups. These are in the tails of the distribution. It is common to observe such outliers when plotting Gaussian clouds. 

Working around the other 4 panels of Figure \ref{fig:x3}, we see that the expected patterns do emerge. Panel (b) is colored by $Y_1$ and has the expected gradient from the lowest values at the bottom left of the X to the highest values top right. The other diagonal has more color gradient within, having lower values to the lower left and higher to the upper right all along the bar. Notice that we could rotate the X and still understand which was the bottom left to top right diagonal of the X. Panel (c) shows $Y_3$ complete with the expected higher values at the ends of the arms of the X. Panel (d) does show noise as there are mixed colors across all parts of the X. Panel (e) is colored with just two colors since all the values are 0 or 1. The 1's are all in the top right of the center sub-cloud and in parts of the sub-cloud centred on (3,3). We see a small arm of balls coming down towards the center in the sub-cloud centered on (3,3). This arm of balls is in a sparser part of the space between the two sub-clouds and does not connect to any other points. We can actually see from panel (a) that part of this arm reaches into outliers from the central sub-cloud.  These are artificial datasets, Figure \ref{fig:x3} reassures that TDABM is picking up the expected behaviors. 

The 4 plots in Figure \ref{fig:x4} represent 4 different $\varepsilon$ values to complement $\varepsilon=1.20$. Adding additional radii is an important part of ensuring that the inference drawn from a TDABM plot is robust. When $\varepsilon$ is small there are more balls in the sparse regions of the point cloud that are disconnected from the main shape. Panel (a) shows $\varepsilon=0.80$. The smaller balls also mean that more balls are needed to cover the space relative to a larger radius. As the radius increases to $\varepsilon=1.00$ in panel (b), the number of disconnected balls falls and the main shape of the X becomes clearer. At $\varepsilon=1.00$, the sub-cloud centered on (-6,-6) is not connected to the rest of the sub-clouds. As noted in the set up, the choice of a distance of 3 on each axis means that there is limited overlap. Only higher radii are able to find the overlap. We see overlap at $\varepsilon=1.20$ in the main results. Increasing the radius further reduces the balls and starts to lose some of the shape of the sub-clouds. At $\varepsilon=1.50$, panel (d) shows that there are still dense centers of subclouds, but at $\varepsilon=2.00$, panel (e) has difficulty demonstrating the local structure. 

The dataset shown in this section has a deliberate global and local structure. The exercise of verifying that the structures are captured by TDABM is a proof of concept for the algorithm. In real world settings, the structure would not be known. Therefore we encourage the use of multiple radii. The range of radii to explore are determined by the range of the data. In this case the maximum likely distance between two points is approximately 25\footnote{3 standard deviations from the furthest apart centers gives a distance of 18 on each axis, a total distance of $\sqrt{18^2+18^2}\approxeq25.46$. }. 1.2 gives 20 balls along the diagonal. 

\section{Building on \texttt{ballmapper}}
\label{sec:extend}

\begin{figure}
	\begin{center}
		\caption{X Dataset with Labels}
		\label{fig:x5}
		\includegraphics[width=12cm]{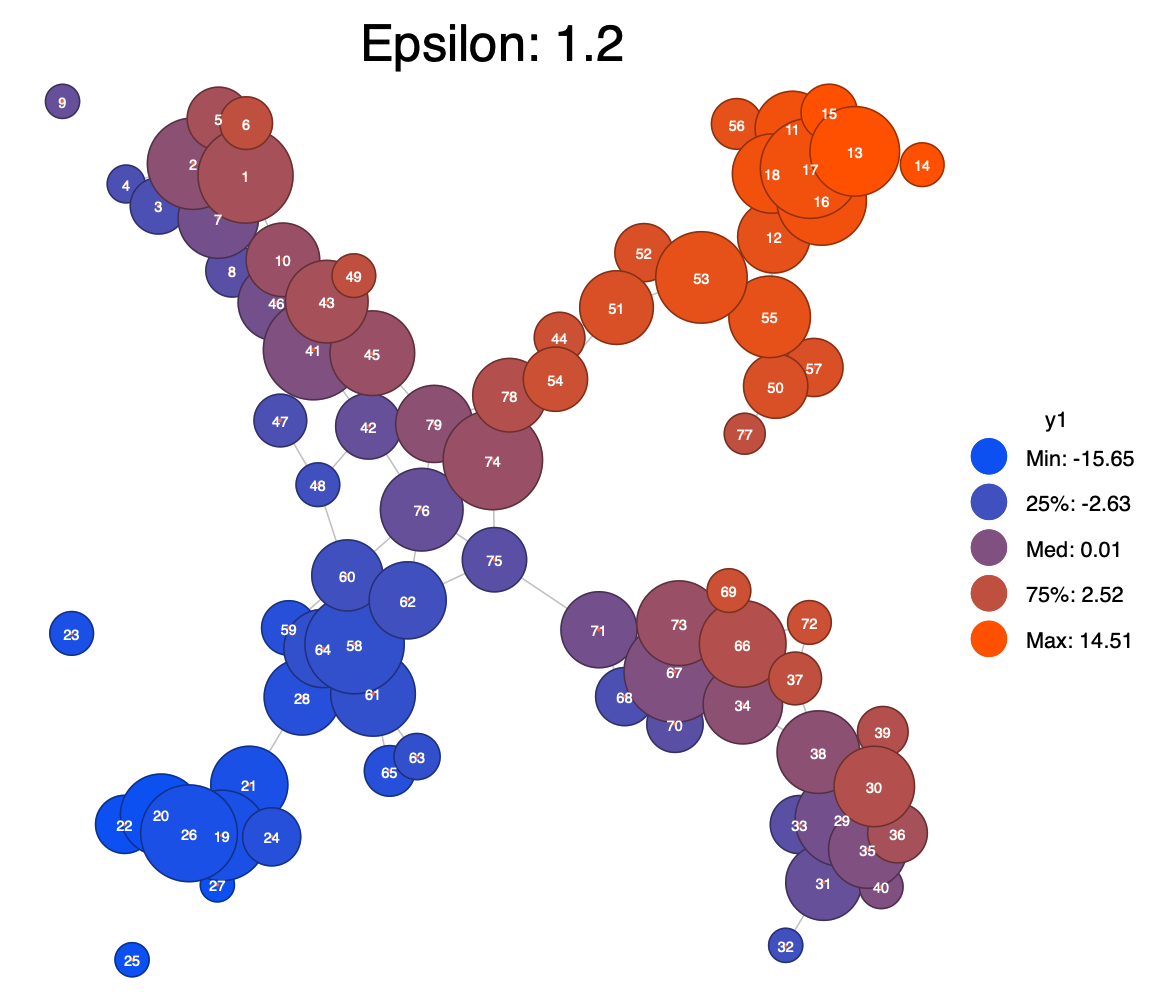}
	\end{center}
	\raggedright
	\footnotesize{Notes: The underlying dataset has 900 observations on 2 variables, $X_1$ and $X_2$. The data is translated to have 9 groups of 100 points centred on (-6,6) (-3,3), (3,3), (6,6), (0,0), (-3,-3), (3,-3), (-6,-6), and (6,-6). }
\end{figure}

For this section we will focus on the $Y1$ coloration of the X dataset. To allow the analysis of the TDABM graphs, the first step is to add the option \texttt{labels}. The labels option tells Stata to include the ball numbers on the TDABM graph. The resultant figure with $\varepsilon=1.20$ is shown in Figure \ref{fig:x5}. The only difference with panel (b) of Figure \ref{fig:x3} is that the ball numbers are now included on each ball. 

Because of the labels, we can now talk about the higher values of $Y1$ appearing in balls 11 to 17. Notice that the consecutive numbering here is due to the fact that we manually adjusted the data in blocks. So the next uncovered point is highly likely to be within the same sub-cloud. The overlap of clouds creates an interesting result where a point from the sub-cloud centred on (3,3) ends up as a landmark for the ball attached to the sub-cloud centered on (6,6). Ball 56 has a number which is in the set otherwise with the sub-cloud centered on (3,3). Similar patterns can be found across all of the sub-clouds. Being able to describe the TDABM graph is the first step, but we want to exploit the fact that the algorithm knows precisely which points are in each ball.

\subsection{Outputs from \texttt{ballmapper()}}

The \texttt{ballmapper()} function creates two new frames, \texttt{BM\_RESULTS} and \texttt{BM\_MERGED}. The \texttt{BM\_RESULTS} frame contains the information about the TDABM graph. Within the frame there is the ball number, the co-ordinates of the point in the TDABM plot, the number of points in the ball and the size value that is used in the plotting. Because Stata uses a binned coloration, the color bin is included as a column in the \texttt{BM\_RESULTS} dataframe. Below the list of balls there is a list of edges. The edges are not numbered, but their start and finish co-ordinates are provided. A type column informs whether a line represents a node or an edge in the TDABM graph. 

The \texttt{BM\_MERGED} frame contains a merged dataset in which the membership of each ball is linked back to the underlying dataset. This merged dataframe is the one which permits the analysis of variables in the underlying dataset that were not included in the TDABM graph. For example, where the points have textual identifiers, the dataframe within \texttt{BM\_MERGED} provides the list of which points are in which ball. To generate summary statistics on each of the balls, the \texttt{BM\_MERGED} dataframe can be combined with grouping by ball. 

Switching between the frames can be managed using the menus in Stata or the \texttt{frame change} command. To see data within either frame, the data editor browse option can be used once the frame has focus. It is important to note that these frames are overwritten every time the \texttt{ballmapper()} function is called. Therefore, ensure that the most recent run is indeed the one that you wish to analyze before commencing any of the analyses in this section.

\subsection{Summarising \texttt{ballmapper()} Outputs}

Because we have the merged dataframe between the underlying dataset and the TDABM graph ball membership, further analysis of the balls can be undertaken using the information in \texttt{BM\_MERGED}. It is is possible for users to create their own functions on the combined data. Two summary functions are provided within the base package. Firstly there is a command to summarize the mean values for multiple variables across the balls, \texttt{ballsummary()}. The second command, \texttt{variablesummary()} allows a detailed summary of a single variable across the balls. Box \ref{box:ballsummary} begins with the former.

\begin{mybox}[label=box:ballsummary]{Constructing Summary of Multiple Variables with \texttt{ballsummary}}
	The \texttt{ballsummary} function works with a list of specified variables to produce means for each of the variables across the balls. The output is saved in the specified .csv file:
	\begin{lstlisting}[language=Stata]
		ballsummary y1 y2 y3 y4 y5, csvfile("ymeans12")
	\end{lstlisting}
	Example shows a summary of all of the $Y$ variables in the X dataset.
\end{mybox}

The code in box \ref{box:ballsummary} produces a .csv file with 1 row for each of the 79 balls. Table \ref{tab:x2} provides the first five lines as an example. It can be seen that all of the first 5 balls are in the sub-cloud centered on (-6,6), which Stata has numbered 1. Hence $Y_2$ is 1 for all of the balls and no points have $Y_5=1$. We see ball 1 has 44 points and ball 4 has just 3 points. The resulting difference in size is apparent in Figure \ref{fig:x5}.

\begin{table}
	\begin{center}
		\caption{Summary of Y Variables X Dataset}
		\label{tab:x2}
		\begin{tabular}{l c c c c c c}
			\hline
			Ball & $Y_1$ & $Y_2$ & $Y_3$ & $Y_4$ & $Y_5$ & Size \\
			\hline
			1	&0.656	&1&	65.43&	-0.016 &	0&	44\\
			2	&0.136	&1&	82.75&	-0.039 &	0&	40\\
			3	&-2.102&1&	79.77&	-0.060&	0&	11\\
			4	&-2.329	&1&	101.5&	0.384&	0&	3\\
			5	&1.110	&1&	90.63&	-0.195&	0&	15\\
			\hline
		\end{tabular}
	\end{center}
	\raggedright
	\footnotesize{Notes: Table provides the mean values of each of the 5 $Y$ variables used in coloration of the X dataset. The underlying dataset has 900 observations on 2 variables, $X_1$ and $X_2$. The data is translated to have 9 groups of 100 points centred on (-6,6) (-3,3), (3,3), (6,6), (0,0), (-3,-3), (3,-3), (-6,-6), and (6,-6). The colorations are given as $Y_1 = X_1 + X_2 + \theta$ where $\theta \sim N(0,0.2)$, $Y_2$ is the group number, $Y_3 = X_1^2 + X_2^2 + \theta$ where again $\theta \sim N(0,0.2)$,  $Y_4 = \phi$ where $\phi \sim N(0,1)$, and $Y_5$ takes the value 1 when $0<X_1<3$ and $0<X_2<3$ are both satisfied.}
\end{table}

The second task that is regularly undertaken for a TDABM graph is the consideration of variation within a ball. If a TDABM graph captures the variation in the coloration variable across the space, we would expect to see limited ranges of $Y$ within each of the balls. A limited overlap of the $Y$ values across the set of balls is also evidence of there being systematic variation in $Y$ across the space. Box \ref{box:varsummary} provides the code to produce a summary of $Y_5$ in the dataset.

\begin{mybox}[label=box:varsummary]{Detailed Summary of Variable using  \texttt{variablesummary}}
	To construct a more detailed summary of a single variable across the balls, the \texttt{variablesummary()} command is used. An optional boxplot can be produced to summarise the variable in each ball. The output is saved in the specified csv file. 
	\begin{lstlisting}[language=Stata]
		variablesummary y5, boxplot boxfile("y5_12_box") csvfile("y5_12_stats")
	\end{lstlisting}
	Example shows a detailed summary of the $Y_5$ variable in the X dataset.
\end{mybox}

\begin{figure}
	\begin{center}
		\caption{Boxplots of $Y$ across Balls X Dataset}
		\label{fig:x6}
		\begin{tabular}{c c}
			\includegraphics[width=7cm]{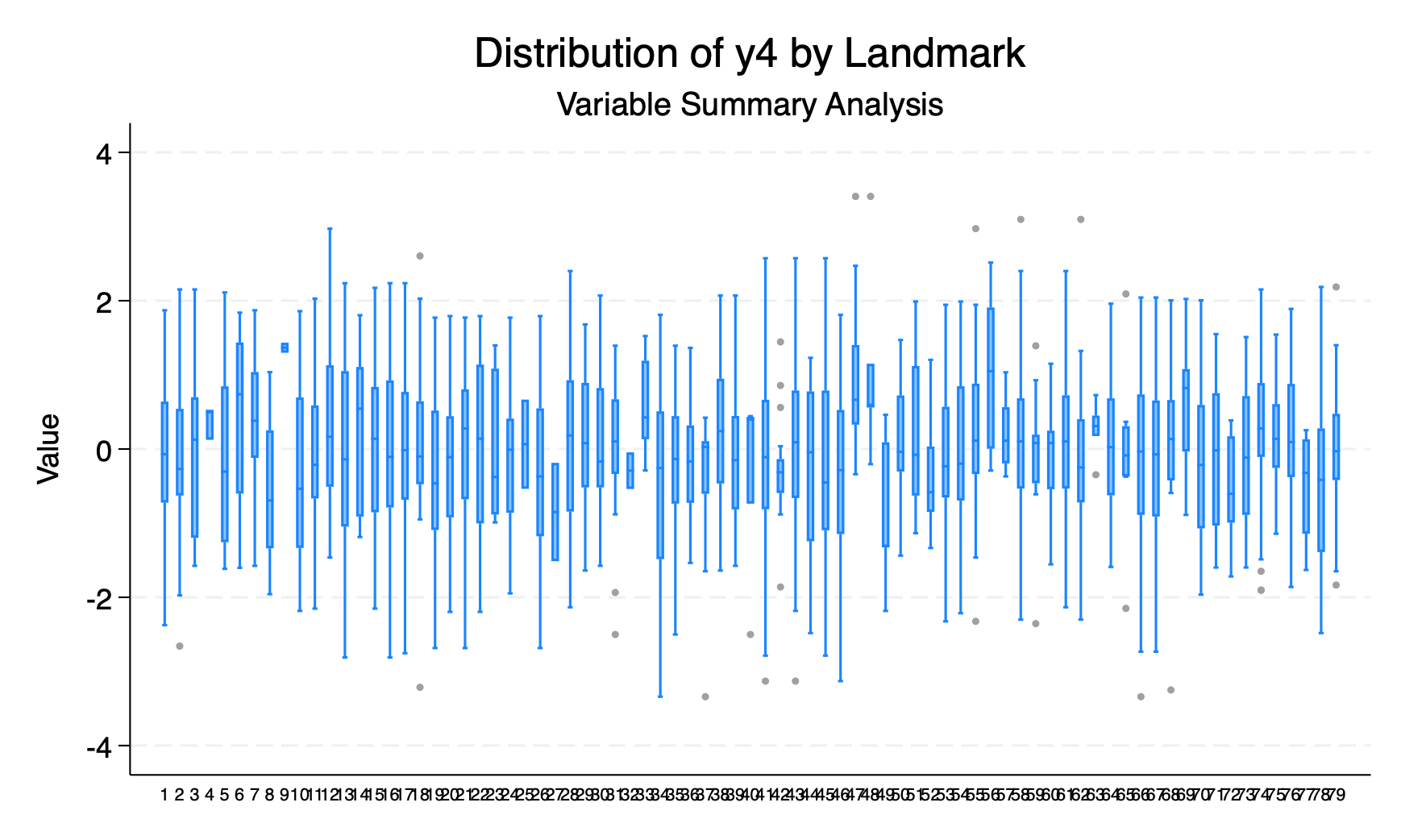}&
			\includegraphics[width=7cm]{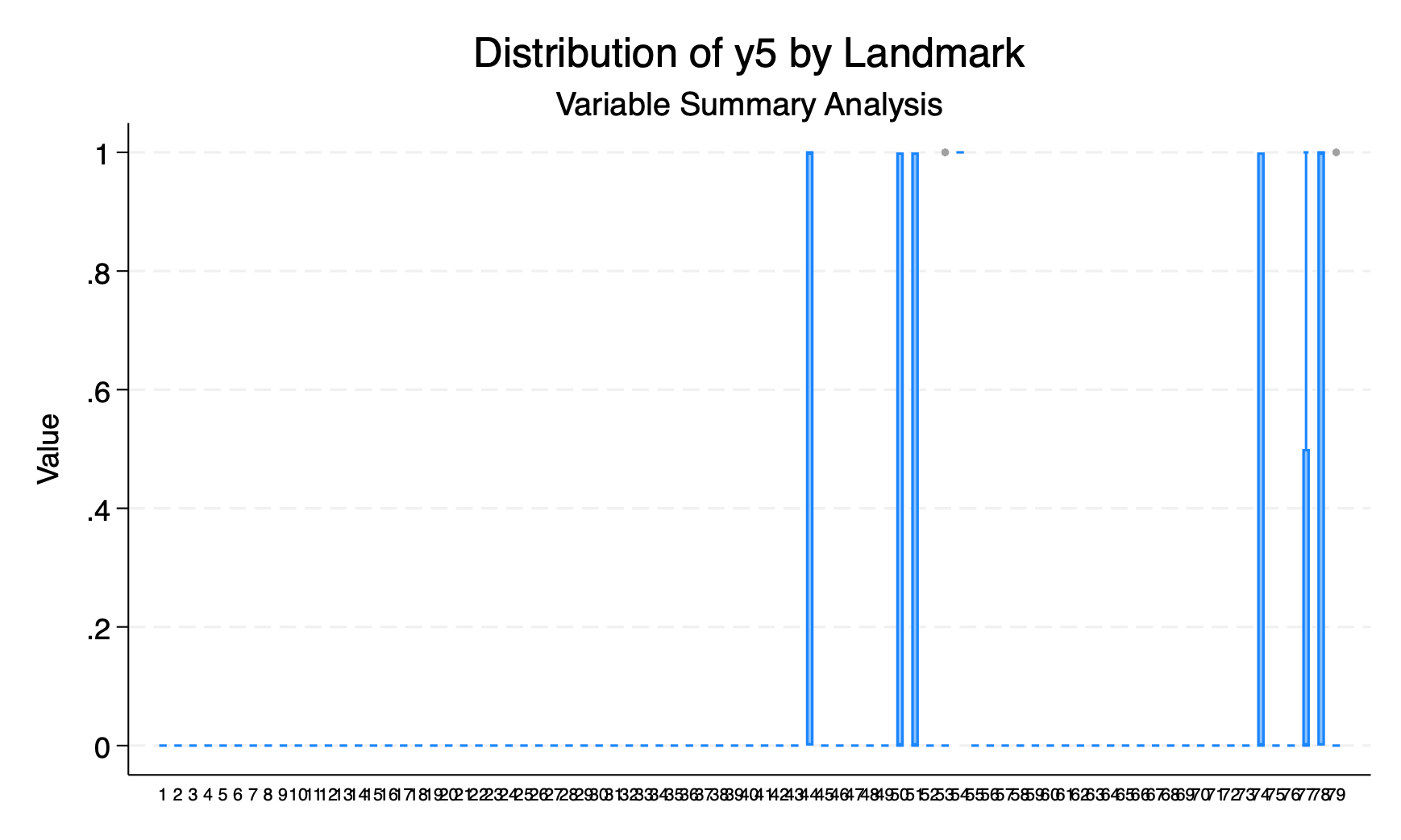}\\
			(a) $Y_4$ - Noise & (b) $Y_5$ - Restricted Range \\
		\end{tabular}
	\end{center}
	\raggedright
	\footnotesize{Notes: Figures plot the minimum, maximum, quartiles and median of the stated variable for each ball in the TDABM plot. The underlying dataset has 900 observations on 2 variables, $X_1$ and $X_2$. The data is translated to have 9 groups of 100 points centred on (-6,6) (-3,3), (3,3), (6,6), (0,0), (-3,-3), (3,-3), (-6,-6), and (6,-6). $Y_4 \sim N(0,1)$ is fully independent of the ball number. $Y_5$ is 1 for any point which has $X_1$ and $X_2$ between 0 and 3.}
\end{figure}

Using the code in Box \ref{box:varsummary}, we can obtain a detailed summary of a single variable across the balls. The summary produces the mean, standard deviation, minimum, maximum, quartiles and median of the variable within each ball. Balls may have low numbers of points, so caution is urged in interpreting the values. To help with interpretation, the number of points in each ball is written in the final column of the resulting table. As with the \texttt{ballsummary()} function, the code is applied on the most recently produced TDABM graph. For the $Y_1$ variable, the summary for balls 1 to 5 are given in Table \ref{tab:x3}. The boxplots generated for $Y_4$ and $Y_5$ from applying the \texttt{variablesummary()} command are provided in Figure \ref{fig:x6}.

\begin{table}
	\begin{center}
		\caption{Example Table Lines From $Y_1$ Summary X Dataset}
		\label{tab:x3}
		\begin{tabular}{l c c c c c c c c}
			\hline
		    Ball & Mean & Std. Dev. & Min & q25 & q50 & q75 & Max & Size\\
		    \hline
			1	&0.656	&0.770	&-0.696 &	0.030 &	0.542&	1.228 &	2.699&	44\\
			2	&0.136	&0.775 &-1.652 &	-0.578 &	0.158 &	0.785&	1.732 &	40\\
			3	&-2.101&	0.752&	-3.379&	-2.608 &	-2.279 &	-1.431 &	-0.924&	11\\
			4	&-2.329&	0.593&	-2.756&	-2.756&	-2.579 &	-1.652 &	-1.652 &	3\\
			5	&1.110	&0.520&	0.344&	0.918&	1.053 &	1.376&	2.456 &	15\\
			\hline
		\end{tabular}
	\end{center}
	\raggedright
	\footnotesize{Notes: Values provide the mean, standard deviation, the minimum, maximum, quartiles and median of $Y_1$ for each ball in the TDABM plot. $Y_1 = X_1 + X_2 + \theta$, where $\theta \sim N(0,0.2)$ is a random noise component. The underlying dataset has 900 observations on 2 variables, $X_1$ and $X_2$. The data is translated to have 9 groups of 100 points centred on (-6,6) (-3,3), (3,3), (6,6), (0,0), (-3,-3), (3,-3), (-6,-6), and (6,-6). }
\end{table}

Balls 1 to 5 are all within the sub-cloud centred on (-6,6). Balls close to the center of this sub-cloud would have a value of $Y_1$ close to 0. At the edges of the sub-cloud the total value of $X_1$ and $X_2$ is influenced by the difference in the two, hence we see smaller balls having $Y_1$ further from 0. Ball 3 can be seen to be an example of a ball which covers a part of the cloud where $Y_1$ is negative, whilst balls 1 and 2 have negative and positive values of $Y_1$ included. Ball 5 represents an example of a ball which is entirely in a part of the space where $Y_1$ is positive, but close enough to the center of the sub-cloud that there are still 15 points. The full table, included in the appendix, 

Figure \ref{fig:x6} is produced...

\section{Auto Data Example}
\label{sec:census}

To illustrate TDABM on a built-in dataset from Stata, we will consider the dataset of cars. The auto dataset contains observations on models of cars which were available for sale in 1978. There are 12 variables in the dataset, with a total of 74 observations. However, the make variable rep78 is missing for 5 observations. We drop rep78. The make variable is text based so is missing from the summary statistics table and not suitable for TDABM. The foreign variable is a dummy and is suitable for coloring only. Headroom only has a limited number of values and so is dropped from the analysis. Summary statistics for the variables used as $X$ in the TDABM analysis is provided in Table \ref{tab:auto1}. Code for this section is available in the file \texttt{autodata.do} on the GitHub site.

\begin{table}
	\begin{center}
		\caption{Auto Dataset Summary Statistics}
		\label{tab:auto1}
		\begin{tabular}{l l c c c c }
			\hline
			Variable & Short &  Mean   &  Std. dev. &       Min &       Max\\
			\hline
			
			Recommended Retail Price & price &       6165 &    2950 &       3291  &    15906\\
			Fuel Economy & mpg &  21.30&    5.786 &     12    &     41\\
			Headroom & headroom &   2.993 &    .846 &       1.5       &   5\\
			Trunk Volume & trunk &    13.76&    4.277   &       5     &    23\\
			Axle Weight (Kg) & weight&3020 &   777.2    &   1760  &     4840\\
			Vehicle Length (Inches) & length   & 187.9 &   22.27&        142&        233\\
			Turning Circle (Yards) & turn & 39.65&    4.399    &     31       &  51\\
			Displacement & displacement & 197.3 &   91.84&         79      &  425\\
			Gear Ratio & gear\_ratio&    3.015 &   0.456  &      2.19  &     3.89\\
			Non-American Manufacturer & foreign&    .297 &   .461 &        0        &  1\\
			\hline
		\end{tabular}
	\end{center}
	\raggedright
	\footnotesize{Notes: Summary statistics of the variables used in our analysis of the built in auto dataset of 1978 cars. For further descriptions of the variables see the Stata documentation. $N=74$.}
\end{table}

Table \ref{tab:auto1} informs that 29.7\% of the cars within the dataset are produced by foreign manufacturers. The average price for the cars is \textdollar6165. The lowest price is \textdollar 3291 and the highest is \textdollar 15906. Low prices are representative of the fact that the data is from 1978. The axis variables for the TDABM plot are all on different scales. For example the highest gear\_ratio is 3.89, but the lowest weight is 1760. Hence it is necessary to standardize the variables prior to running the \texttt{ballmapper()} function. Code for standardizing is included in the .do file for this section. 

\begin{table}
	\begin{center}
		\caption{Auto Dataset Correlation Matrix}
		\label{tab:auto2}
		\begin{tabular}{l c c c c c c c c c}
			\hline
			           &   price&      mpg  &  trunk  & weight &  length  &   turn &displacement & gear\_ratio &  foreign\\
			\hline
			price &   1.0000&&&&&&&&\\
			mpg &  -0.4686 &  1.0000&&&&&&&\\
			trunk &   0.3143  &-0.5816   &1.0000&&&&&&\\
			weight &   0.5386 & -0.8072   &0.6722 &  1.0000&&&&&\\
			length &   0.4318 & -0.7958  & 0.7266&   0.9460  & 1.0000&&&&\\
			turn &  0.3096  &-0.7192   &0.6011&   0.8574&   0.8643  & 1.0000&&&\\
			displacement &   0.4949  &-0.7056&   0.6086 &  0.8949 &  0.8351&   0.7768&   1.0000&&\\
			gear\_ratio &  -0.3137 &  0.6162&  -0.5087 & -0.7593 & -0.6964 & -0.6763 & -0.8289   &1.0000&\\
			foreign &   0.0487  & 0.3934  &-0.3594  &-0.5928 & -0.5702  &-0.6311&  -0.6138  & 0.7067  & 1.0000\\
			\hline
			
		\end{tabular}
	\end{center}
	\raggedright
	\footnotesize{Notes: Correlation matrix for the variables used in our analysis of the built in auto dataset of 1978 cars. For further descriptions of the variables see the Stata documentation and Table \ref{tab:auto1}. $N=74$.}
\end{table}

The correlation matrix in Table \ref{tab:auto2} shows that there are strong negative correlations between fuel economy, length, weight, the turning circle and displacement. Length is positively correlated with the trunk size, weight, turning circle, displacement and gear\_ratio. Weight has a positive correlation with turning circle and displacement. Finally, there is a positive strong correlation between turning circle and displacement, and negative strong correlations are observed between gear\_ratio, displacement and turning circle. Price is positively correlated with the trunk size, weight, length, turning circle and displacement. Meanwhile there is a negative correlation between price and fuel economy and gear\_ratio. Foreign brands are associated with higher mpg, lower trunk size, lower weight, shorter length, smaller turning circles and smaller displacement. There is a positive association between foreign and the gear\_ratio.

\begin{figure}
	\caption{Selected Pairwise Scatter Plots}
	\label{fig:auto1}
	\begin{tabular}{c c c}
		\includegraphics[width=5cm]{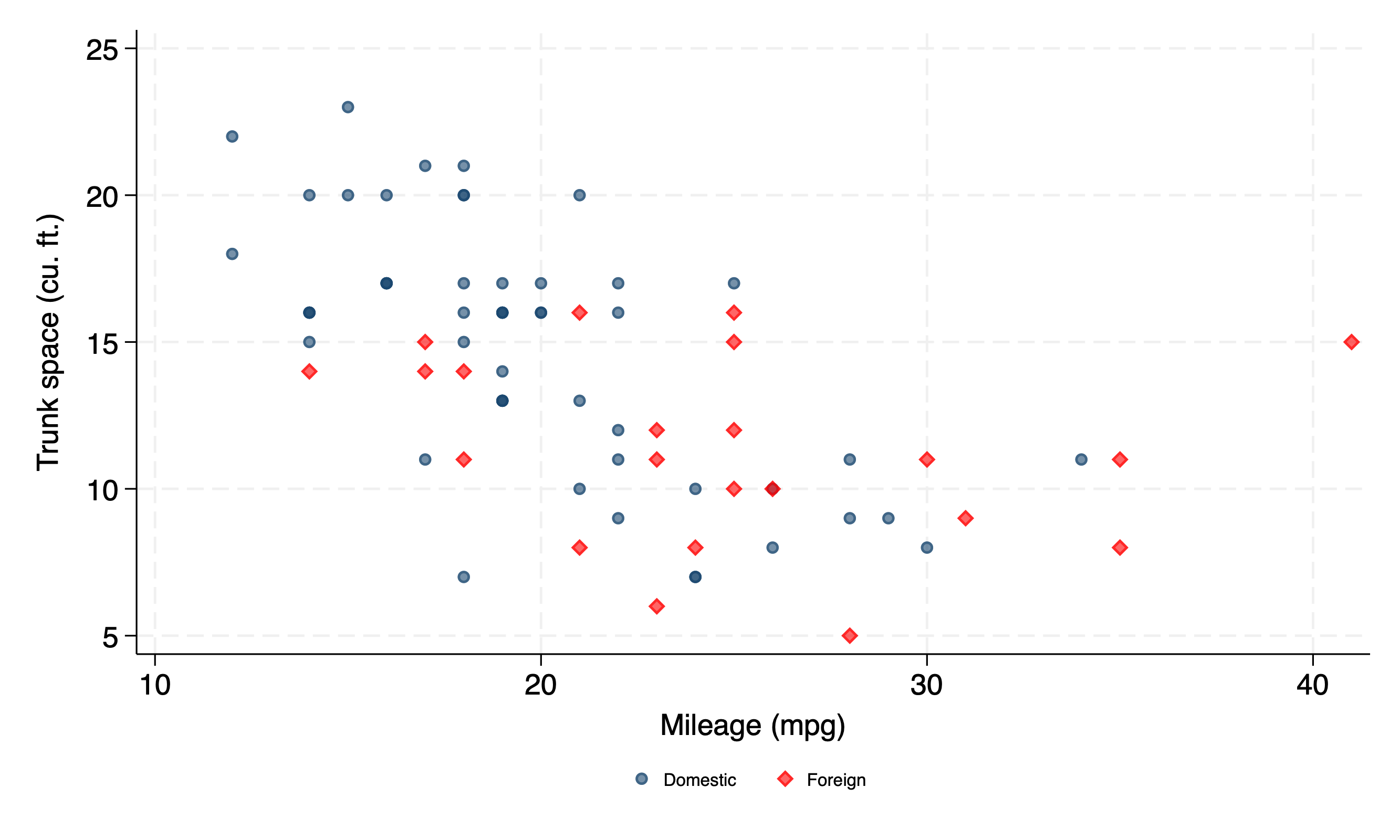}&
		\includegraphics[width=5cm]{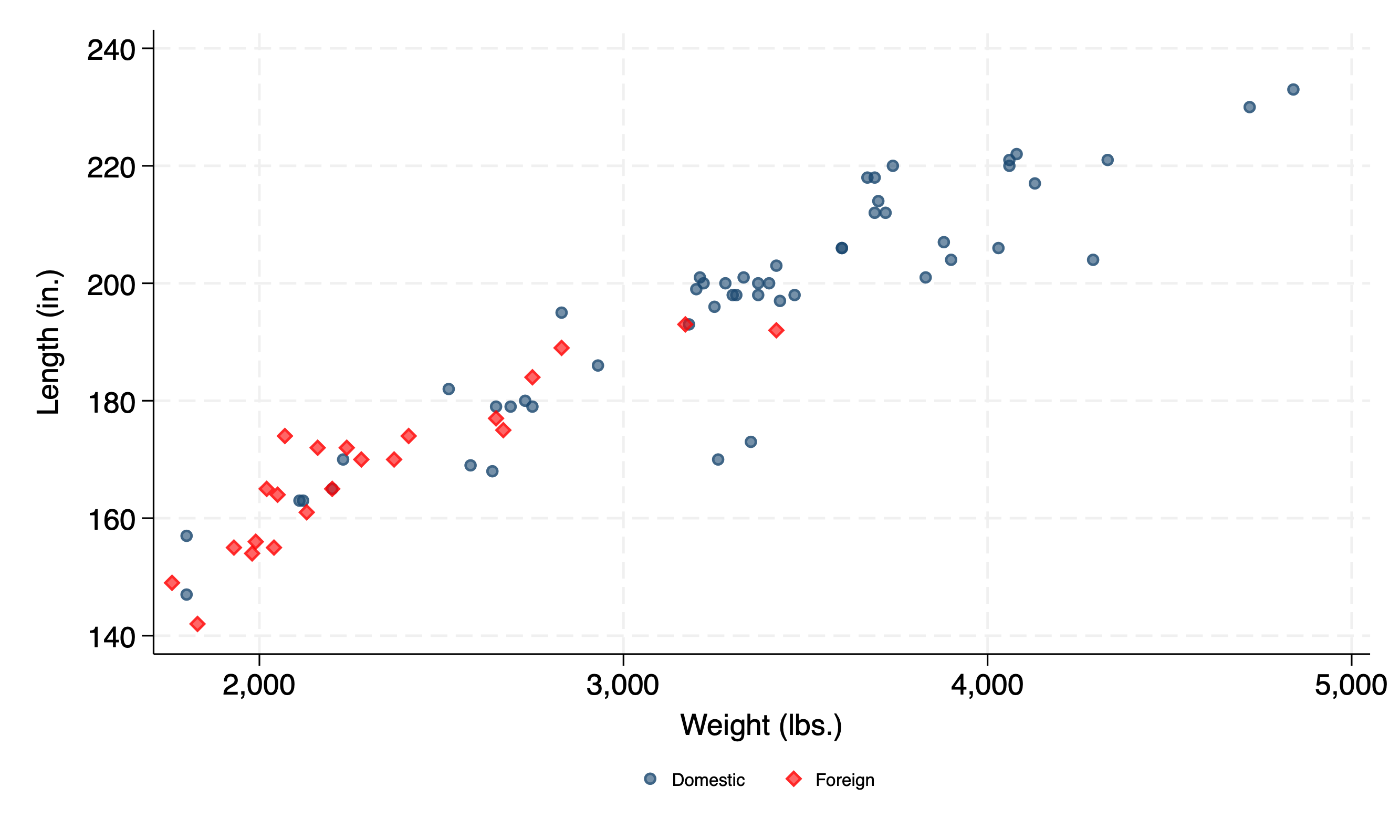}&
		\includegraphics[width=5cm]{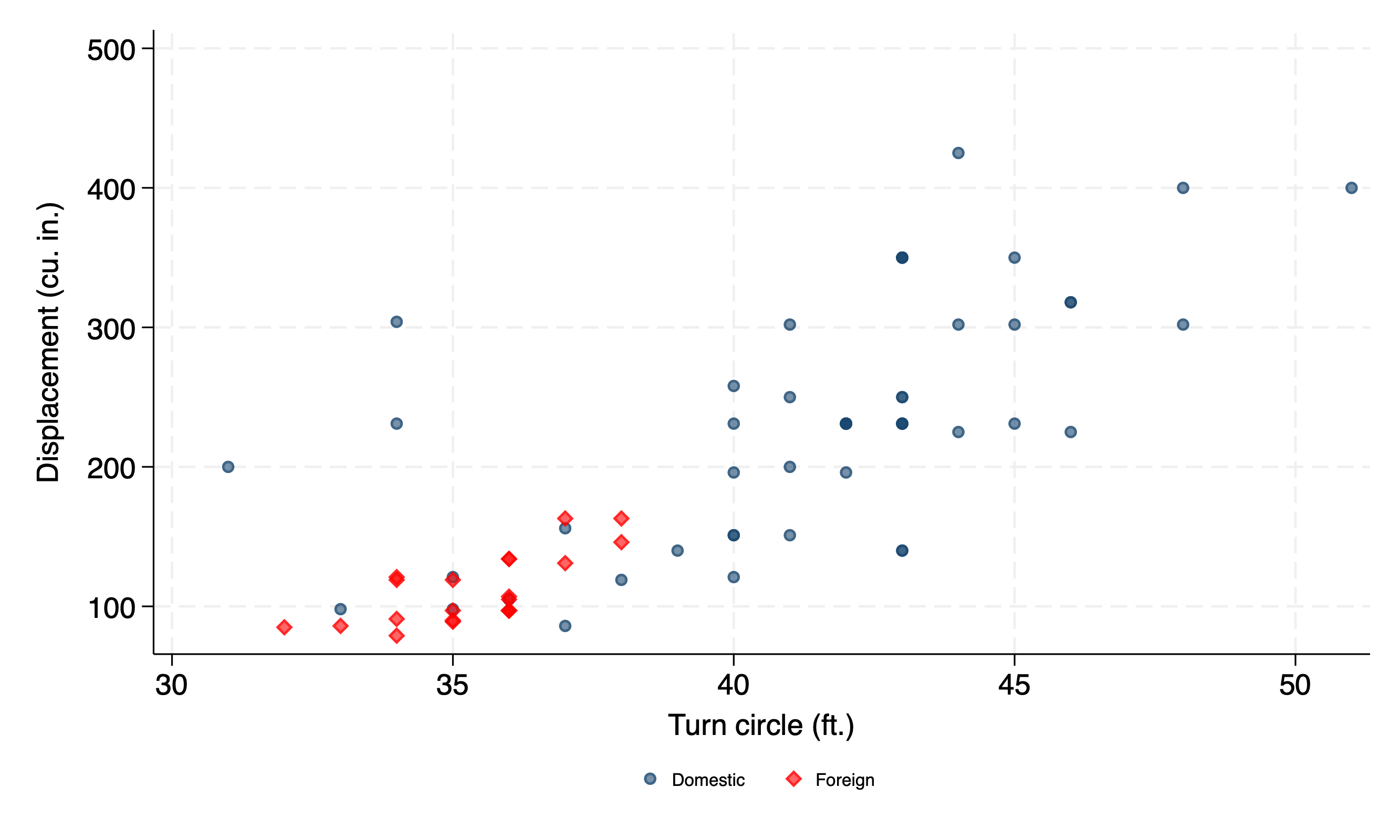}\\
		(a) MPG and Trunk & (b) Weight and Length & (c) Turn and Displacement \\
		\includegraphics[width=5cm]{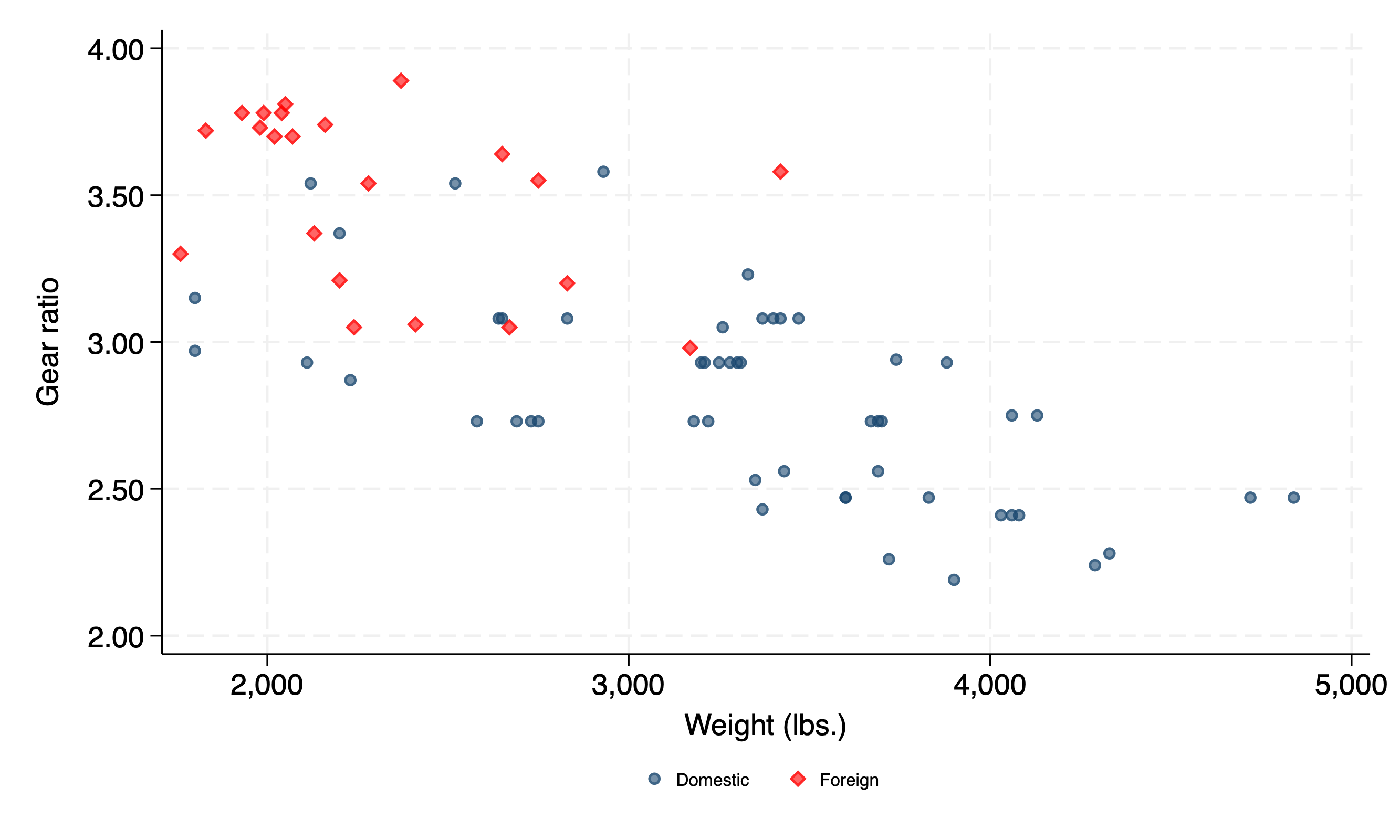}&
		\includegraphics[width=5cm]{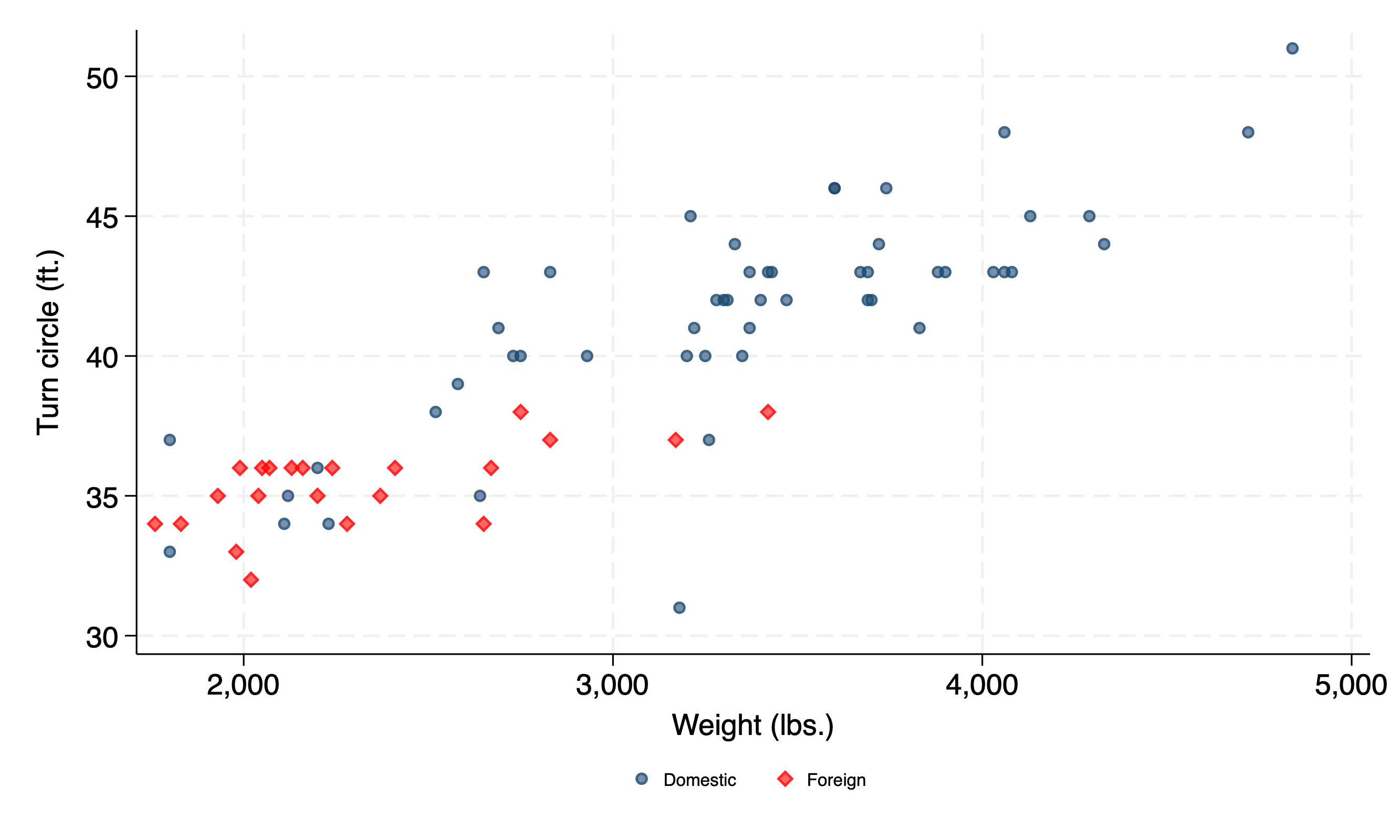}&
		\includegraphics[width=5cm]{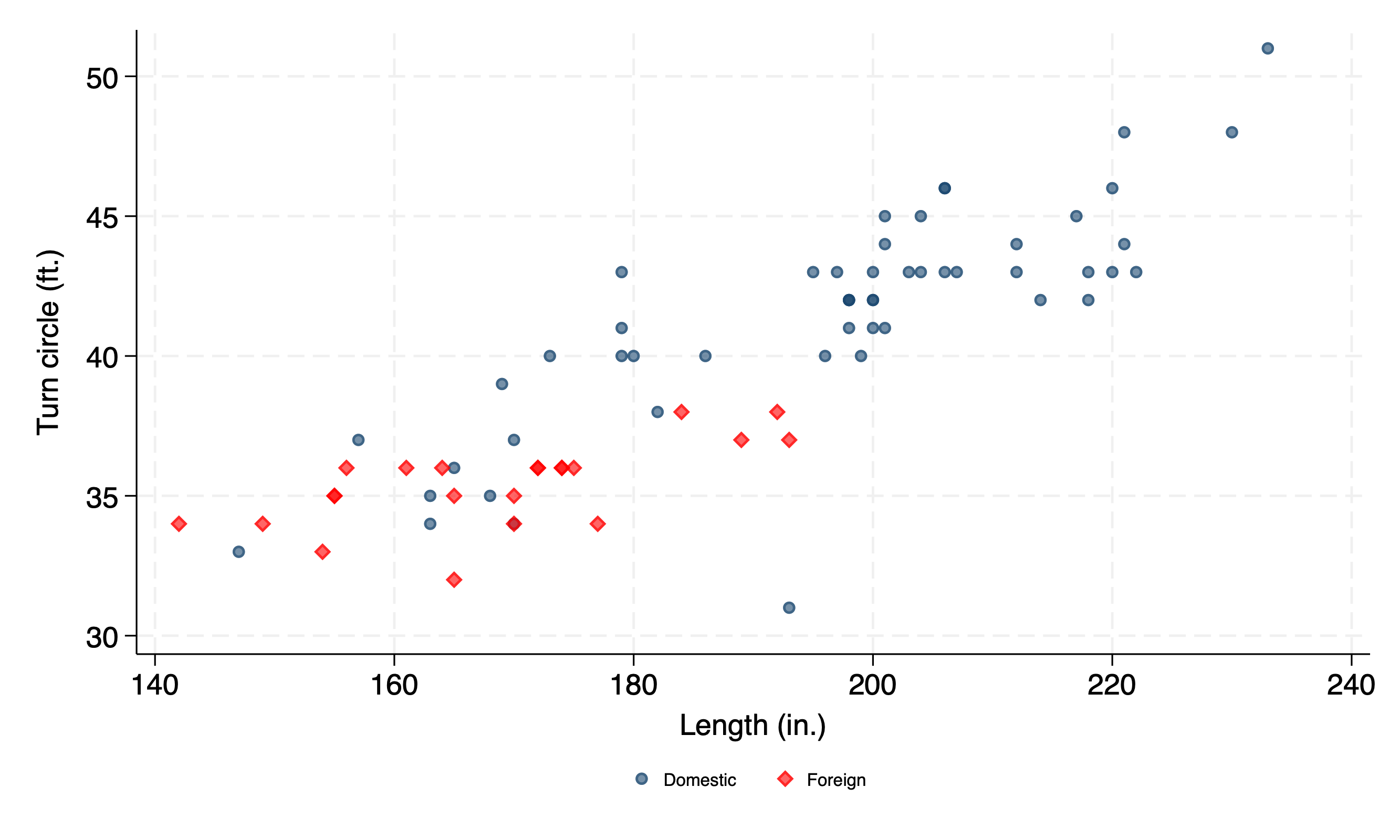}\\
		(d)Gears and Weight & (e) Weight and Turn & (f) Length and Turn\\
	\end{tabular}
	\raggedright
	\footnotesize{Notes: Selected pairwise scatterplots of the variables used in our analysis of the built in auto dataset of 1978 cars. For further descriptions of the variables see the Stata documentation and Table \ref{tab:auto1}. Coloration of the points is according to whether the car is from a foreign manufacturer, with red being foreign and blue domestic. $N=74$.}
\end{figure}

Evidence, particularly our X example and \cite{matejka2017same}, emphasises the importance of looking at the joint distribution of $X$. \cite{anscombe1973graphs} reminds on the importance of looking at relationships between $X$ and $Y$ graphically. As a first step to exploring the relationships within the data we generate pairwise scatter plots for all of the variables in the dataset. Selected plots are shown in Figure \ref{fig:auto1}. The code to make the remainder of the pairwise plots is included within the .do file on the GitHub site. 

Figure \ref{fig:auto1} shows that there is a negative association between mpg and the trunk size, as well as between gears and weight. However, the weakness of the correlations is also evident. There are stronger correlations in the other panels of Figure \ref{fig:auto1}. The segregation of foreign and domestic cars appears too. The foreign cars are seen to occupy the part of the space with lighter weight, lower displacement and smaller turning circles. This section asks whether combining the variables produces a fuller segregation between foreign and domestic, and whether the joint distribution of characteristics is informative on price. 

\begin{mybox}[label=box:auto1]{Production of TDABM with \texttt{ballmapper()}}
	To prepare the data, a foreign dummy is prepared and the axis variables are standardized:
	\begin{lstlisting}[language=Stata]
		gen double for_num = (foreign == 1)
		foreach v in mpg trunk weight length turn displacement gear_ratio {
			egen std_`v' = std(`v')
		}
	\end{lstlisting}
	The TDABM graph is then produced with $\varepsilon=1.50$. Here the foreign variable is the coloration:
	\begin{lstlisting}[language=Stata]
	ballmapper std_\ast, epsilon(1.5) color(for_num) layout repulsion(0.05) attraction(0.01) filename("foreign_std15") labels
	\end{lstlisting}
\end{mybox}

\begin{figure}
	\begin{center}
		\caption{TDABM Graphs of Auto Data: Price}
		\label{fig:auto2}
		\begin{tabular}{c c}
			\multicolumn{2}{c}{\includegraphics[width=10cm]{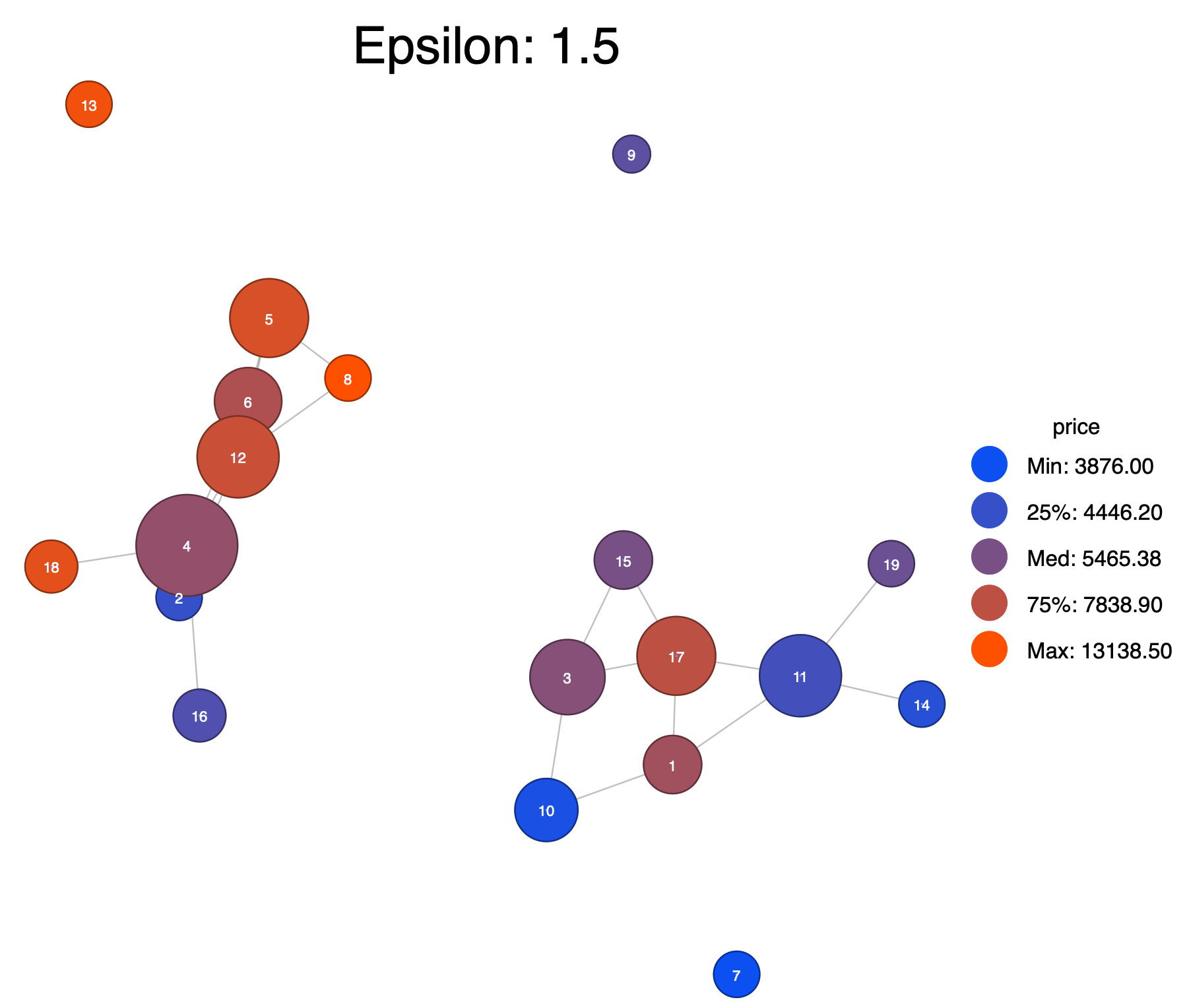}}\\
			\multicolumn{2}{c}{(a) $\varepsilon=1.50$}\\
			\includegraphics[width=6cm]{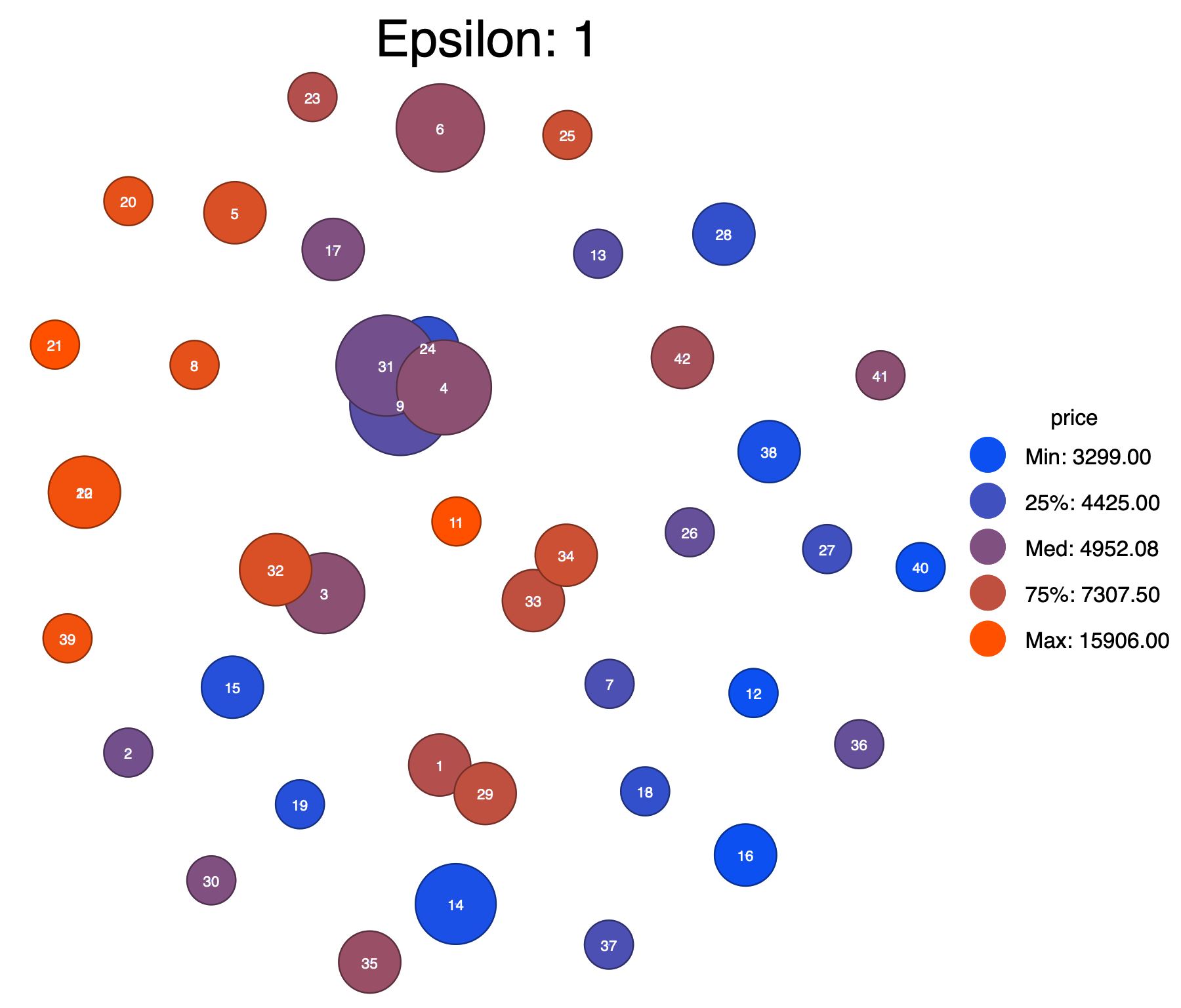} &
			\includegraphics[width=6cm]{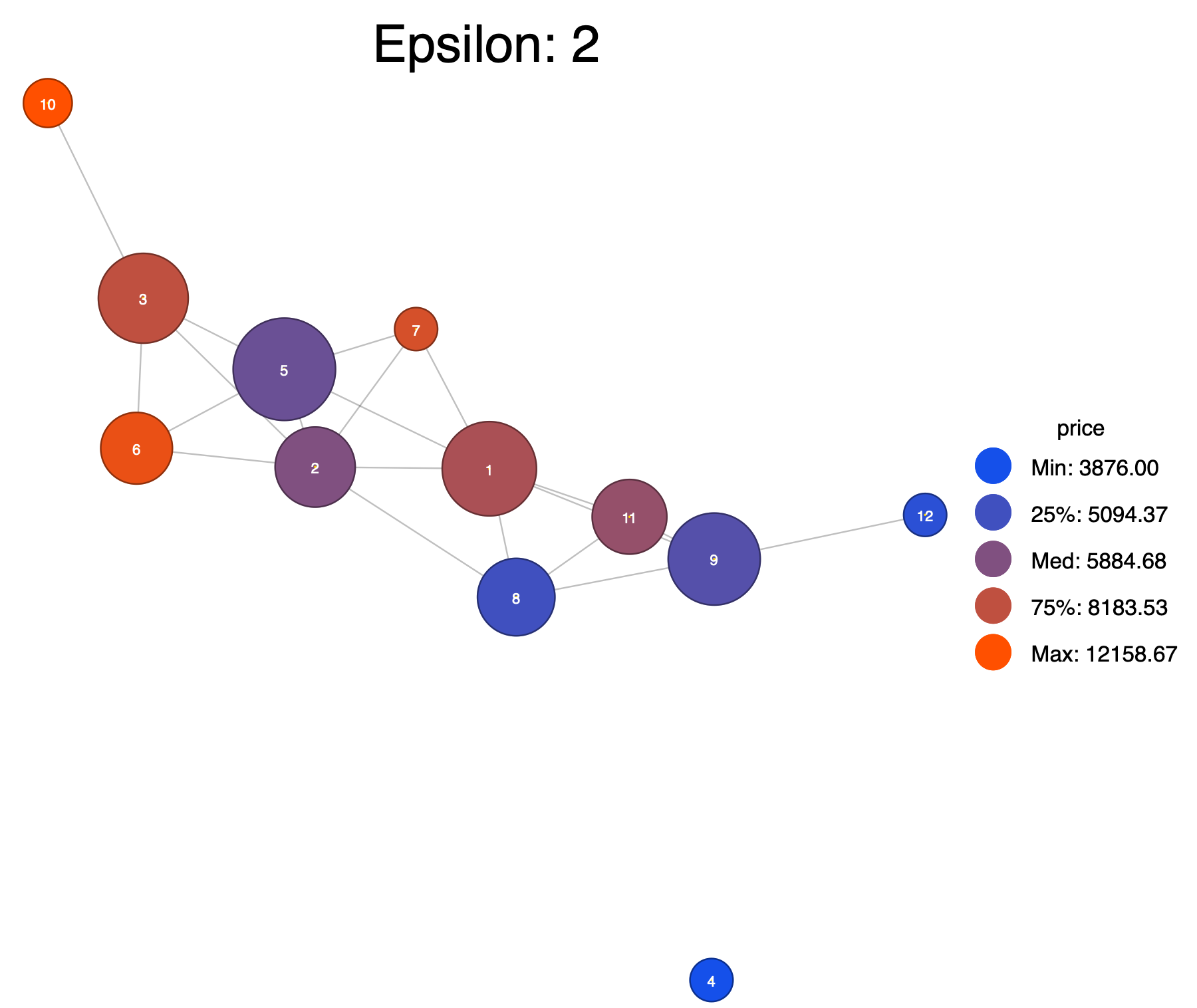} \\
			(b) $\varepsilon = 1.00$ & (c) $\varepsilon=2.00$ \\
			\includegraphics[width=6cm]{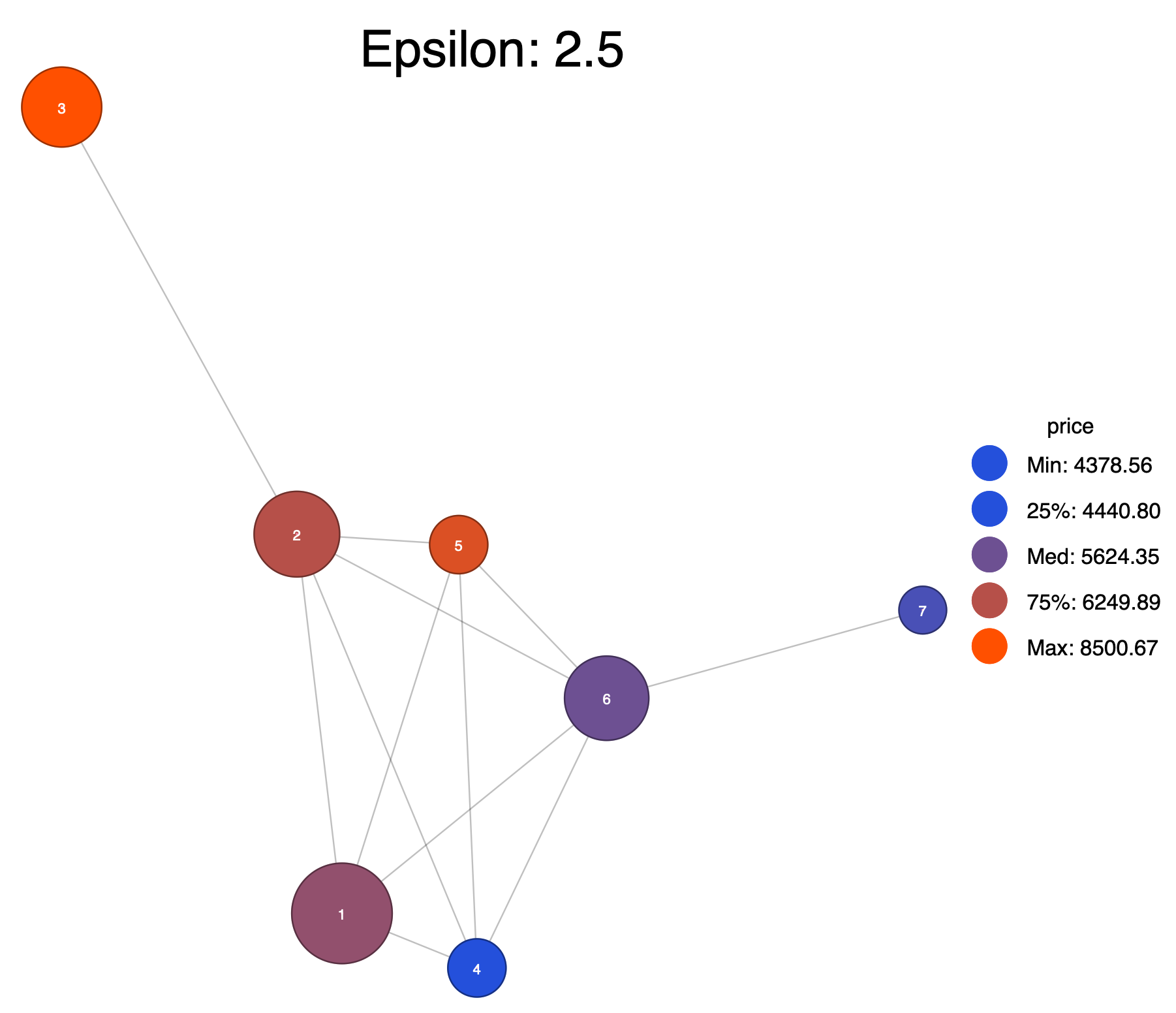} &
			\includegraphics[width=6cm]{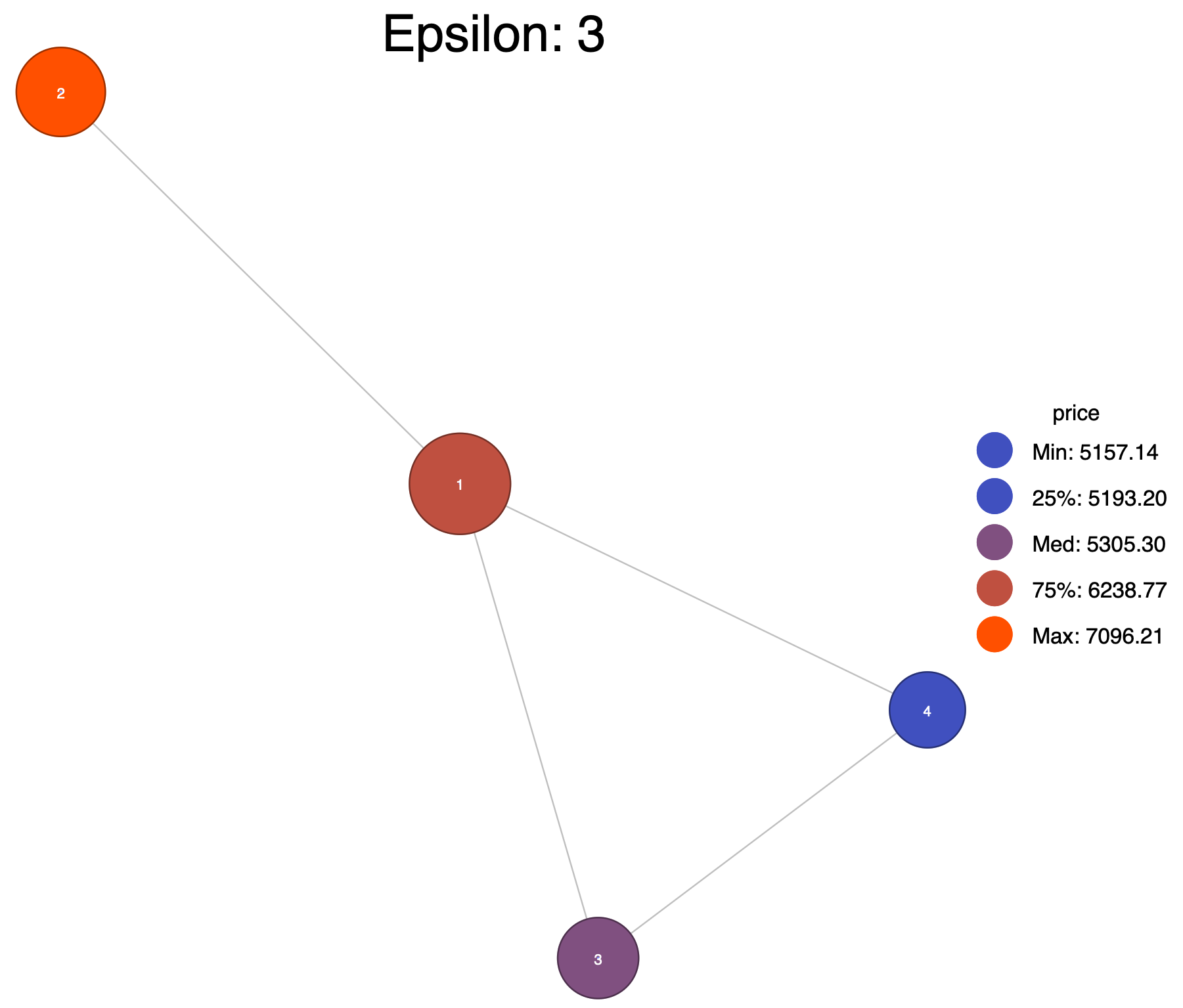} \\
			(d) $\varepsilon = 2.50$ & (e) $\varepsilon=3.00$
		\end{tabular}
	\end{center}
	\raggedright
	\footnotesize{Notes: TDABM plots of the auto data with the stated ball radii. The axis variables used are mpg, trunk, weight, length, turn, displacement and gear\_ratio. All axis variables are standardized prior to applying the \texttt{ballmapper()} function. Coloration is according to the average price in \textdollar 's. Analysis uses Stata's built in auto dataset of 1978 cars. For further descriptions of the variables see the Stata documentation and Table \ref{tab:auto1}. $N=79$ }
\end{figure}

\begin{figure}
	\begin{center}
		\caption{TDABM Graphs of Auto Data: Foreign Brands}
		\label{fig:auto3}
		\begin{tabular}{c c}
			\multicolumn{2}{c}{\includegraphics[width=10cm]{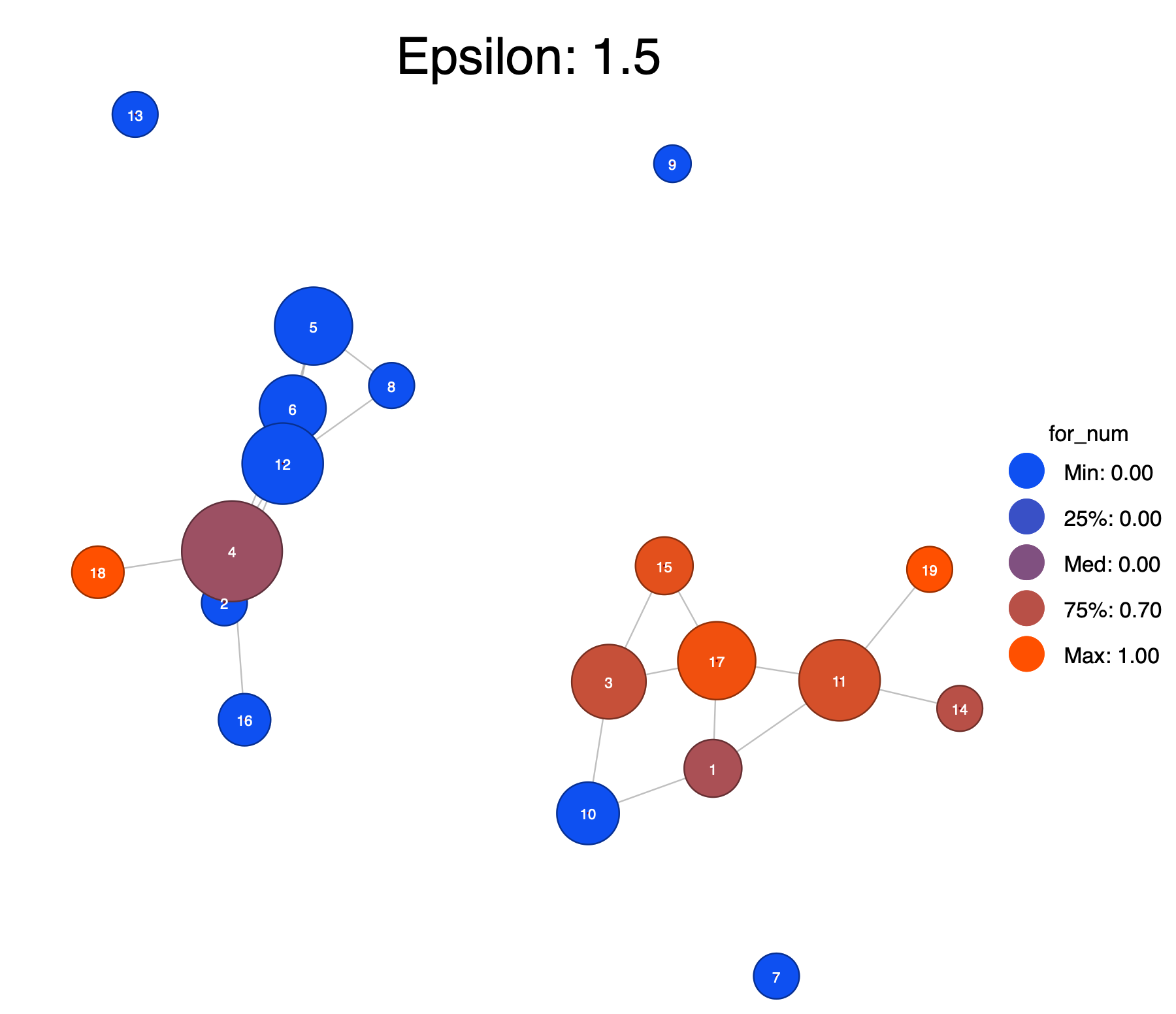}}\\
			\multicolumn{2}{c}{(a) $\varepsilon=1.50$}\\
			\includegraphics[width=6cm]{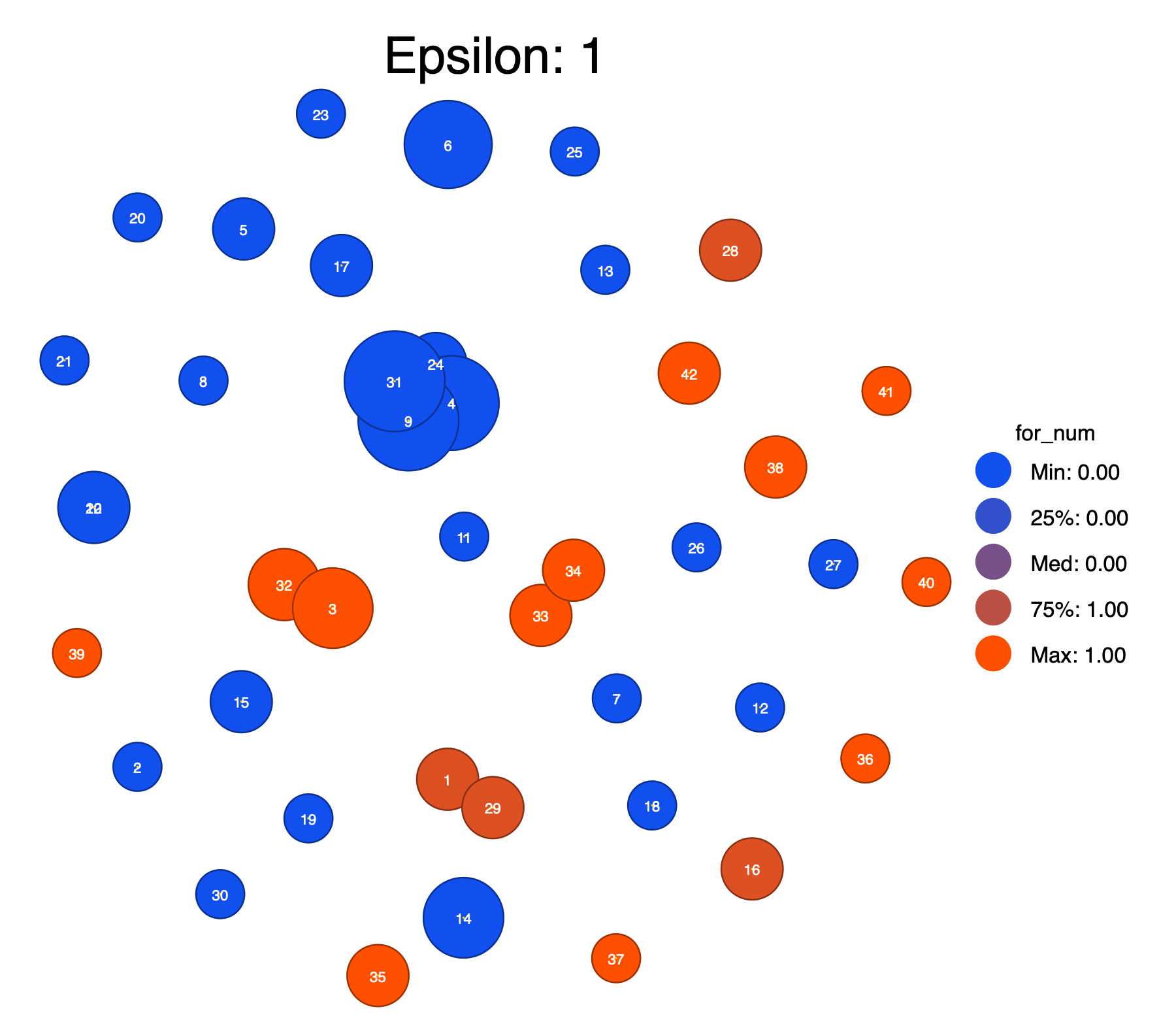} &
			\includegraphics[width=6cm]{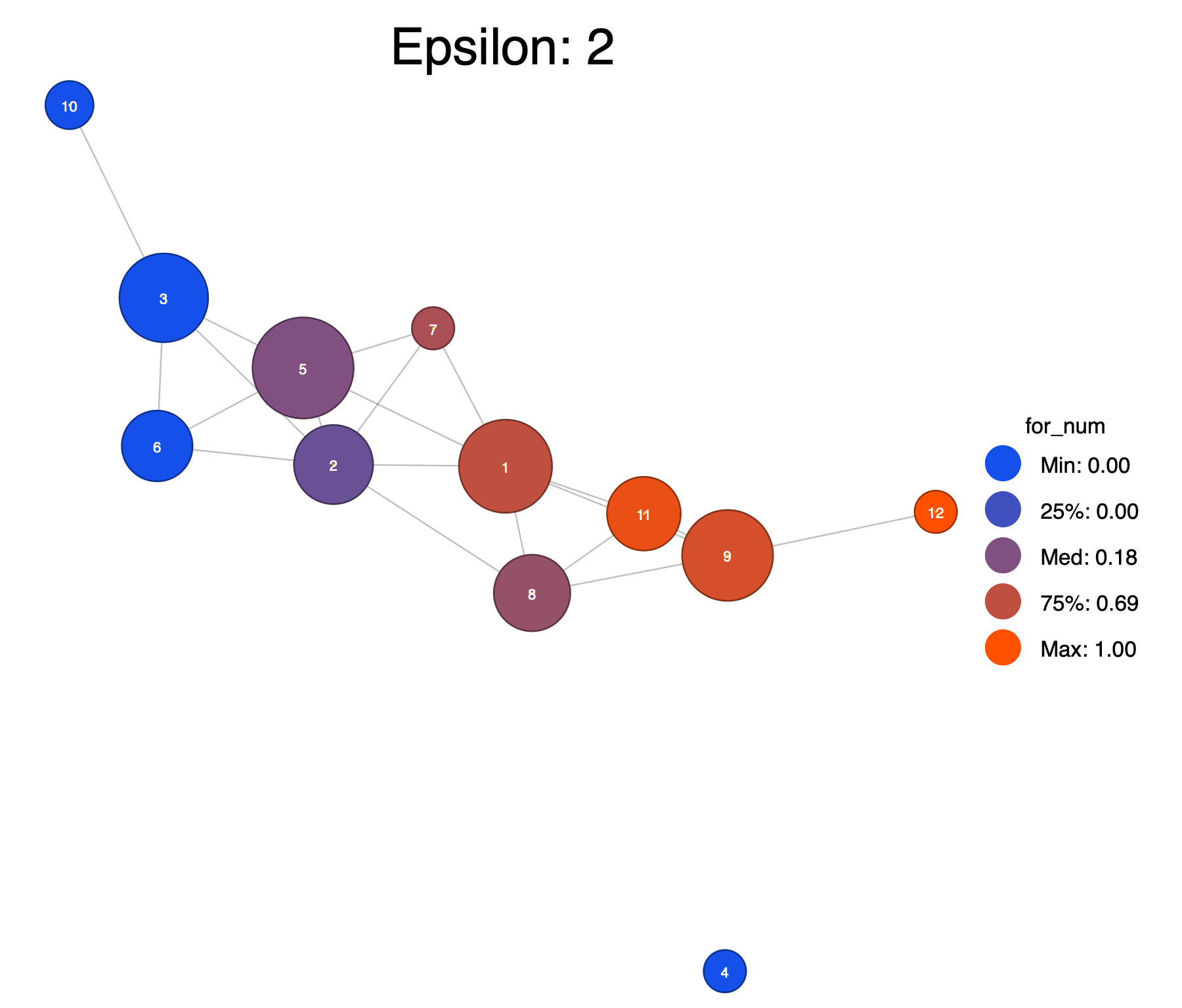} \\
			(b) $\varepsilon = 1.00$ & (c) $\varepsilon=2.00$ \\
			\includegraphics[width=6cm]{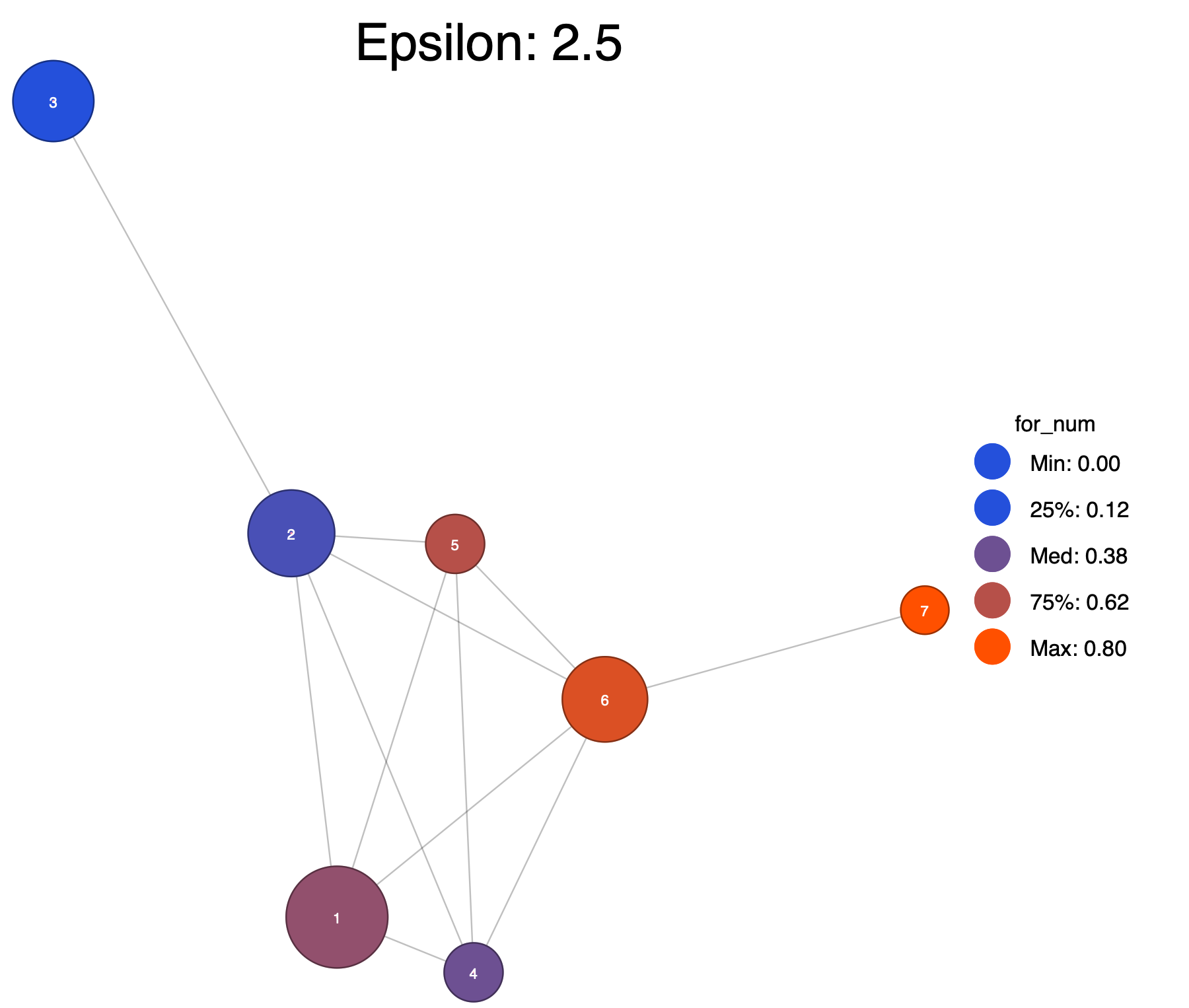} &
			\includegraphics[width=6cm]{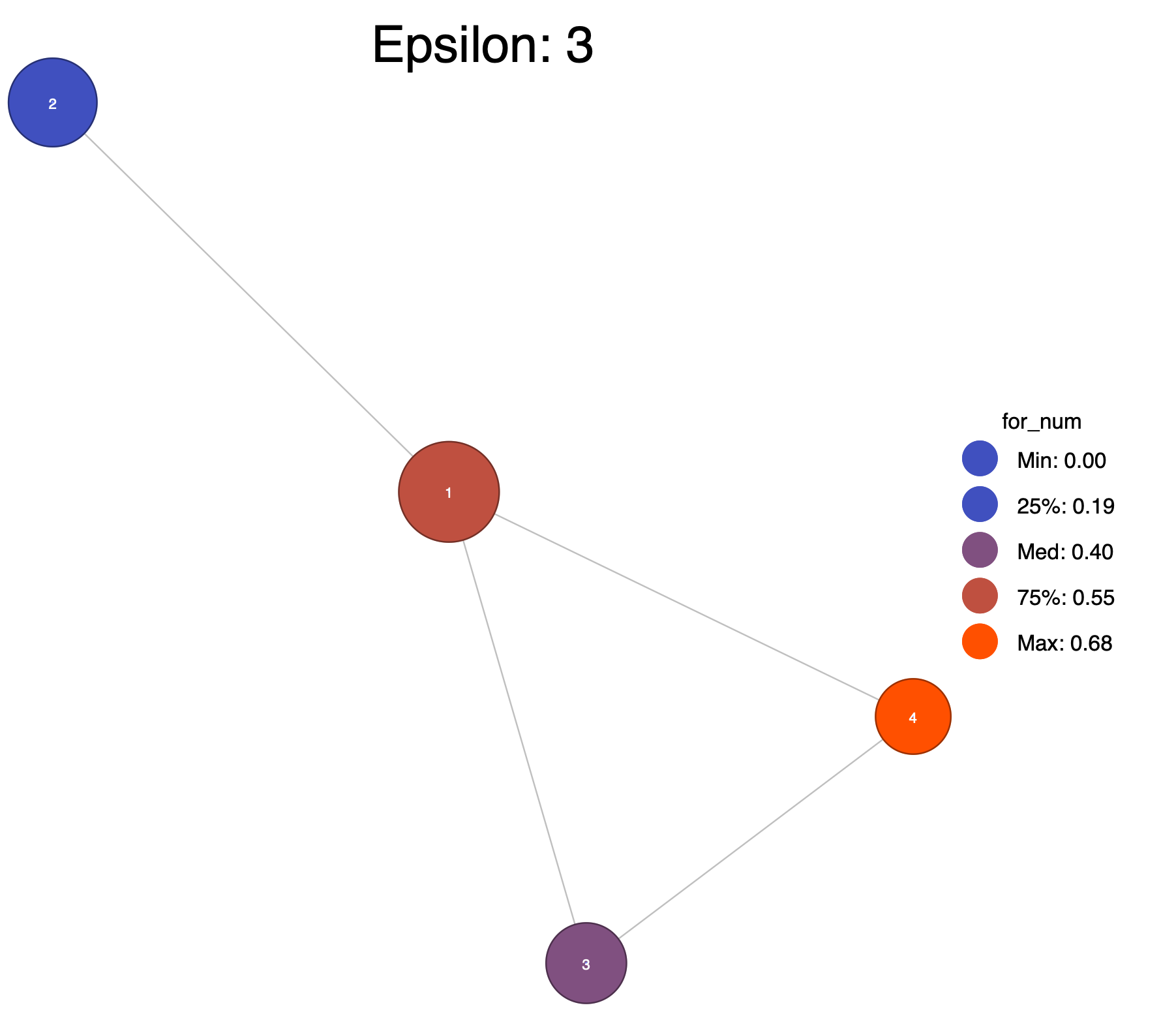} \\
			(d) $\varepsilon = 2.50$ & (e) $\varepsilon=3.00$
		\end{tabular}
	\end{center}
		\raggedright
	\footnotesize{Notes: TDABM plots of the auto data with the stated ball radii. The axis variables used are mpg, trunk, weight, length, turn, displacement and gear\_ratio. All axis variables are standardized prior to applying the \texttt{ballmapper()} function. Coloration is according to the proportion of observations within the ball that are produced by foreign manufacturers. Analysis uses Stata's built in auto dataset of 1978 cars. For further descriptions of the variables see the Stata documentation and Table \ref{tab:auto1}. $N=79$ }
\end{figure}

Using the code in Box \ref{box:auto1} we construct TDABM graphs for the auto data. We produce a total of 10 plots, using radius of $\varepsilon = 1, 1.5, 2, 2.5, 3$, plotting the results in Figures \ref{fig:auto2} and \ref{fig:auto3}. Figure \ref{fig:auto2} is colored according to the average price in the ball. Figure \ref{fig:auto3} is colored according to the proportion of models within the ball which are from foreign manufacturers. 
 
Figure \ref{fig:auto2} shows the data arranges as two connected components at $\varepsilon=1.50$. These 2 components form as the radius increases, but combine for $\varepsilon=2.00$. As the radius increase to $\varepsilon=2.50$ and $\varepsilon=3.00$, panels (d) and (e), the number of balls shrinks further. There is evidence of a dense core with smaller balls at each end, only at $\varepsilon=3.00$ does one of the two extremes merge in. Our focus is on the $\varepsilon=1.50$ case. Panel (a) shows that the higher prices are in the left of the two large groups. The right of the two groups has more blue balls with lower prices. Of the three disconnected balls, ball 13 is a high price ball and balls 7 and 8 are lower priced. We note that there is price variation amongst both of the connected components, balls 2 and 16 having lower prices but being connected with the higher priced balls, and ball 17 having near median prices in the lower price group. 

Looking at the brand nationality, the split between the two connected component groups is clear. The left group is almost entirely domestic, including the cheaper cars of balls 2 and 16. The right group is almost entirely foreign manufactured, apart from ball 10 and its overlaps into balls 1 and 3. All of the disconnected balls are domestic. Although the split is not complete, panel (a) of Figure \ref{fig:auto3} shows why a classifier may be able to perform well with this data. Increasing the radius to $\varepsilon=2.00$ shows that the foreign cars are to the right hand end of the space, whilst the domestics are to the left. It is therefore possible that a plane could be fit through the data by a linear discriminant classifier. Panels (d) and (e) show that the pattern holds to higher $\varepsilon$.

In order to see more information about these balls we can find the average values of each of the balls using the \texttt{ballsummary()} command. The code is provided in Box \ref{box:auto2}. The resultant summary is in Table \ref{tab:auto3}.

\begin{mybox}[label=box:auto2]{Production of Ball Summaries with \texttt{ballsummary()}}
	Having run the \texttt{ballmapper()} at $\varepsilon=1.50$ we may generate the summary. Recall that the summary will only work on the most recently implemented run of \texttt{ballmapper()}:
	\begin{lstlisting}[language=Stata]
		ballsummary mpg trunk weight length turn displacement gear_ratio price foreign, csvfile("auto_means12")
	\end{lstlisting}
	The summary is placed into file \texttt{auto\_means.csv}.
\end{mybox}

\begin{table}
	\begin{center}
		\caption{Ball Summary Statistics for Auto Data}
		\label{tab:auto3}
		\begin{tabular}{l c c c c c c c c c c}
			\hline
			Ball &	mpg	&trunk&	weight&	length&	turn	&displacement	&gear\_ratio&	price&	foreign&	Size \\
			\hline
			1&	22.50&	9.250&	2713&	182.8&39.75&	131.5&	3.438&	5725&	0.25&	4\\
			2&	18.00&	12.00&	3390&	185.0&	41.50&254.0&	2.545&	4352&	0	&2\\
			3	&22.38&	11.88&	2573&	170.8&	36.13&	127.3&	3.129&	5465&	0.625	&8\\
			4&	18.82&	15.29&	3283&	198.5&	41.65&	211.7&	2.939&	5619&	0.118&	17\\
			5&	15.67&	18.78&	3949&	213.3&	43.78&	333.0&	2.416&	8551&	0	&9\\
			6&	17.17&19.83&	3728&	214.8&	43.17&	233.2&	2.770&	6483&	0	&6\\
			7&	27.50&	9.500&	2170&	166.5&	34.00&	267.5&	2.900&	3876	&0&	2\\
			8&	18.50&	15.00&	4160&	205.0&	44.00&	350.0&	2.325&	13139&	0&	2\\
			9&	22.00&	17.00&	3180&	193.0&	31.00&	200.0&	2.730&	4504&	0&	1\\
			10&	23.00&	8.600&	2680&	177.2&	40.60&	146.6&	2.800&	4010&	0&	5\\
			11&	29.30&	8.700&	2056&	160.1&	34.40&	96.90&	3.524&	4446&	0.700&	10\\
			12&	16.20&	16.70&	3833&	209.0&	44.10&	309.2&	2.504&	7839&	0&	10\\
			13&	12.00&	20.00&	4780&	231.5&	49.50&	400.0&	2.470&	12546&	0&	2\\
			14&	32.50&	10.00&	2000&	161.0&	36.00&	91.50&	3.090&	4087&	0.5&	2\\
			15&	24.00&	16.00&	2063&	159.3&	35.75&	99.00&	3.575&	5081&	0.75&	4\\
			16&	18.67&	11.00&	3430&	200.3&	42.33&	231.0&	3.080&	4480&	0&	3\\
			17&	23.89&	10.67&	2302&	171.1&35.78&	111.6&	3.684&	6738&	0.889&	9\\
			18&	16.00&	14.33&	3140&	191.3&	37.33&	152.3&	3.253&	11558&	1&	3\\
			19&	38.00&	13.00&	2045&	159.5&	35.50&	93.50&	3.795&	4598&	1&	2\\
			\hline
		\end{tabular}
	\end{center}
	\raggedright
	\footnotesize{Notes: Summary statistics for the 19 balls obtained in a TDABM analysis of the auto data with $\varepsilon=1.50$.  For further descriptions of the variables see the Stata documentation and Table \ref{tab:auto1}. $N=74$.}
\end{table}

Table \ref{tab:auto3} shows that there is notable variation in all of the variables across the balls. This is as would be expected since the balls cover different parts of the space. However, Table \ref{tab:auto3} represents confirmation that the variation applies in a multi-dimensional application of TDABM. Balls 2 and 16 were highlighted in the earlier discussion as lower priced balls connected into the higher priced component. We see ball 2 contains just 2 cars, both domestically produced. Ball 16 has 3 cars. By looking at the merged dataframe in Stata we see that ball 2 contains the AMC Pacer and Chevrolet Nova. The 3 cars in Ball 16 are the Buick Skylark, the Pontiac Firebird and the Pontiac Pheonix. For those familiar with American cars of 1978, more exploration can then be done. For our purpose, this illustrates how information about the balls is readily obtained from the data.

Balls 1, 14 and 15 provide interest since they contain a maximum of 4 cars and are mixed between domestic and foreign. Ball 1 contains the AMC Concord, Ford Mustang, Plymouth Sapporo, and Datsun 810. The Datsun 810 is the foreign car in this set. Ball 14 contains the Plymouth Champion and the Toyota Corolla. Ball 15 contains Plymouth Horizon, the Fiat Strada, Volkswagen Rabbit and the Volkswagen Sirocco. Here it is interesting to see which imported cars have similar properties to domestic cars as they would be competitors for customers who seek cars in that part of the characteristics space. Our use of the auto dataset allows us to say something of competition between cars, but does not get at causality as data is limited. 

The second summary is produced for a single variable but gives more detailed summary statistics. We wish to understand more about the outcome variables, price and proportion which are foreign. The code in Box \ref{box:auto3} produces Table \ref{tab:auto4}. The option to generate box plots is also activated.

\begin{mybox}[label=box:auto3]{Production of Ball Summaries with \texttt{variablesummary()}}
	Having run the \texttt{ballmapper()} at $\varepsilon=1.50$ we may generate the detailed summary of single variables. 
	\begin{lstlisting}[language=Stata]
		variablesummary foreign, boxplot boxfile("foreign_15_box") csvfile("foreign_15_stats")
		variablesummary price, boxplot boxfile("price_15_box") csvfile("price_15_stats")
	\end{lstlisting}
	The summary is placed into the stated .csv file and the boxplots are saved as the given .png file.
\end{mybox}

\begin{table}
	\begin{center}
		\caption{Outcome Summary Table for Auto Data}
		\label{tab:auto4}
		\begin{tabular}{l l c c c c c c cc}
		$Y$	& Ball & Mean & Std Dev & Min & q25	& q50 & q75	& Max & $N_b$\\
		\hline
		Price (\textdollar)&	1&	5725&	1946.6&	4099&	4143&	5336.5&	7307.5&	8129&	4\\
		&	2&	4352&	561.4	&3955	&3955&	4352&	4749&	4749&	2\\
		&	3&	5465&	1930.4&	3799&	4062.5	&5183&	5849&	9735&	8\\
		&	4&	5619&	2506.9&	3291&	4181&	4733&	5189&	11995&	17\\
		&	5&	8551&	3026.5&	5705&	6165&	7827&	10371&	14500&	9\\
		&	6&	6483&	1961.6&	4890&	5705&	5793&	6342&	10372&	6\\
		&	7&	3876&	816.0&	3299&	3299&	3876&	4453&	4453&	2\\
		&	8&	13139&	3913.8&	10371&	10371&	13139&	15906&	15906&	2\\
		&	9&	4504&	&	4504&	4504&	4504&	4504&	4504&	1\\
		&	10&	4010&	246.1&	3667&	3829&	4172&	4187&	4195&	5\\
		&	11&	4446&	831.1	&3748&	3895&	4192&	4589&	6486&	10\\
		&	12	&7839&	3799.4&	3955&	5379&	6094.5&	10371&	14500&	10\\
		&	13&	12546&	1482.8&	11497&	11497&	12546&	13594&	13594&	2\\
		&	14&	4087&	478.7&	3748&	3748&	4086.5&	4425&	4425&	2\\
		&	15&	5081&	1190.5&	4296	&4389&	4589.5&	5773.5&	6850&	4\\
		&	16&	4480&	428.8&	4082&	4082&	4424&	4934&	4934&	3\\
		&	17&	6738&	1521.8	&4697	&6229	&6486	&7140&	9735&	9\\
		&	18&	11558&	1692.8&	9690&	9690&	11995&	12990&	12990&	3\\
		&	19&	4598	&1130.7&	3798&	3798&	4597.5&	5397&	5397&	2\\
		Foreign & 1	&0.25&	0.5	&0&	0&	0&	0.5&	1&	4\\
		&2	&0&	0&	0&	0&	0&	0&	0&	2\\
		&3	&0.625&	0.518&	0&	0&	1&	1&	1&	8\\
		&4	&0.118&	0.333	&0&	0&	0&	0&	1&	17\\
		&5	&0&	0&	0&	0&	0&	0&	0&	9\\
		&6	&0	&0	&0&	0&	0&	0&	0&	6\\
		&7	&0&	0&	0	&0&	0&	0&	0&	2\\
		&8	&0	&0&	0&	0&	0&	0&	0&	2\\
		&9	&0	&&	0&	0&	0&	0&	0&	1\\
		&10	&0	&0	&0&	0&	0&	0&	0&	5\\
		&11	&0.700&	0.484&	0&	0&	1&	1&	1&	10\\
		&12	&0&	0&	0&	0&	0	&0&	0&	10\\
		&13	&0	&0	&0&	0	&0&	0	&0&	2\\
		&14	&0.500&	0.708&	0&	0&	0.5	&1&	1&	2\\
		&15	&0.750&	0.500&	0&	0.5	&1&	1&	1&	4\\
		&16	&0&	0	&0&	0&	0&	0&	0&	3\\
		&17	&0.889&	0.333&	0&	1&	1&	1&	1&	9\\
		&18	&1&	0&	1	&1	&1	&1	&1	&3\\
		&19	&1&	0&	1&	1&	1&	1&	1&	2\\
		\hline
		\end{tabular}
	\end{center}
	\raggedright
	\footnotesize{Notes: Summary statistics for the 19 balls obtained in a TDABM analysis of the auto data with $\varepsilon=1.50$.  Figures report the mean, standard deviation, minimum, 25th percentile, median, 75th percentile and maximum for each ball. Two variables are considered, being the outcome variables from the TDABM plots, price and foreign ownership percentage. For further descriptions of the variables see the Stata documentation and Table \ref{tab:auto1}. $N=74$.}
\end{table}

\begin{figure}
	\begin{center}
		\caption{Boxplot of Price Variation within Balls for Auto Data}
		\label{fig:auto4}
		\includegraphics[width=10cm]{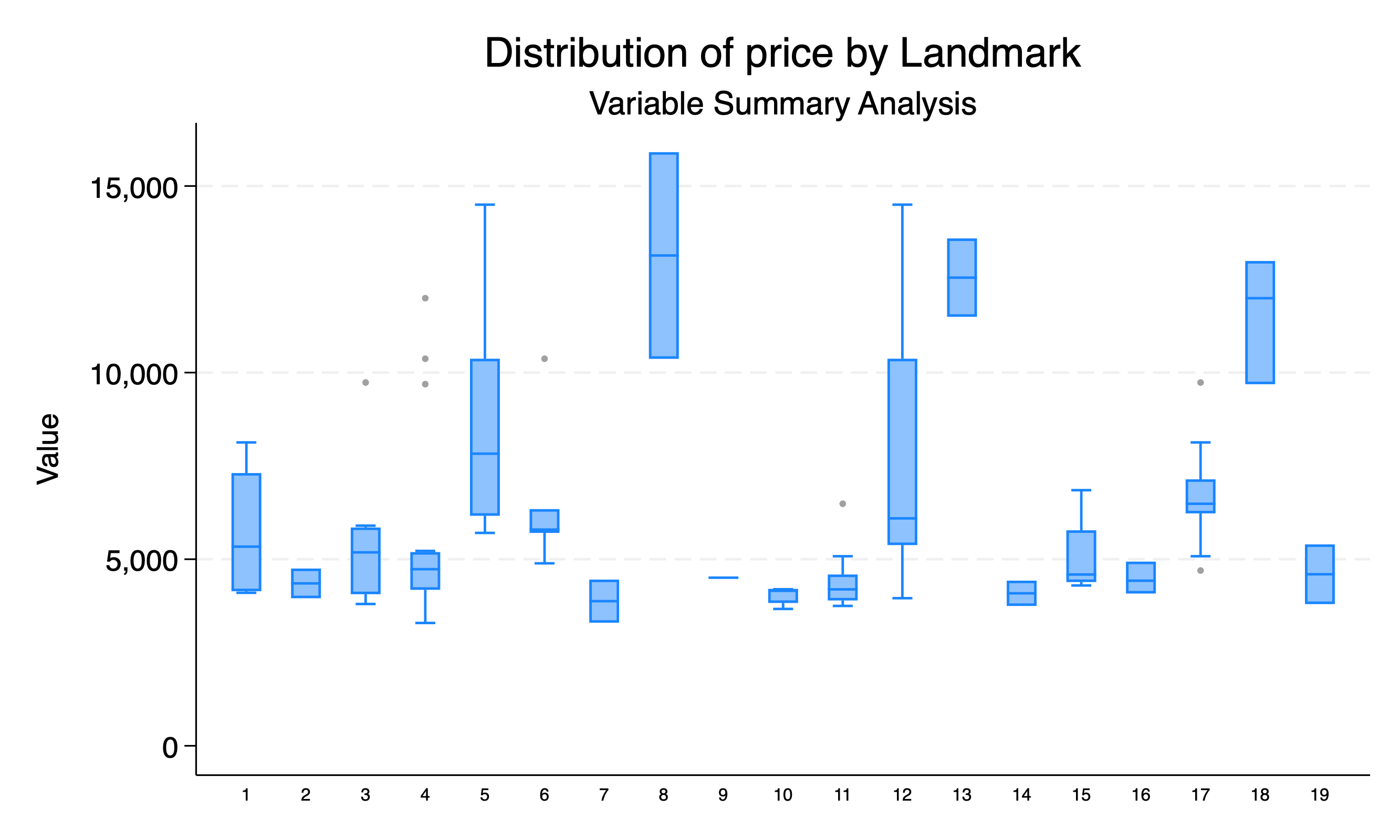}
	\end{center}
		\raggedright
	\footnotesize{Notes: Boxplots of the price variation within the 19 balls obtained in a TDABM analysis of the auto data with $\varepsilon=1.50$.  Plots are based on the minimum, 25th percentile, median, 75th percentile and maximum for each ball. For further descriptions of the variables see the Stata documentation and Table \ref{tab:auto1}. $N=74$.}
\end{figure}

Table \ref{tab:auto4} is split into two panels. The data shows that there is variation within the balls, the standard deviation ranges from less than 10\% of the average price up to 33\% of the price. This suggests that the characteristics used to produce the TDABM plot do have some relationship with the price of the cars. However, it would also be understood that there are other factors which influence price that are not captured. Variation of prices within balls is evidence of further explanatory variables being needed to model prices. For the foreign ownership, we see that most balls are either entirely domestic, or entirely foreign. We have already looked at some of the mixed balls. The variation of prices within balls is also seen in the corresponding boxplot of Figure \ref{fig:auto4}.

Figure \ref{fig:auto4} demonstrates that there are some balls with high price and some with low. This is consistent with the messaging from the TDABM plot in Figure \ref{fig:auto2}. Balls 5 and 12 have the largest range of prices contained within. Ball 5 has 9 cars and ball 12 has 10. Both balls are entirely domestic. Hence it may be understood that these balls represent cars from brands with differing levels of prestige, hence different abilities to charge a higher price for the same vehicles. There are several cars which appear in both balls, including the Dodge Magnum which has a price at the low end of the range, and the Cadillac Eldorado which has a price at the higher end of the range. TDABM is agnostic to the branding, but the placement of cars in the space can inspire the user to tell the story of why these cars find themselves in the same part of the characteristic space. 

The analysis of this section shows how a TDABM graph can be produced by the \texttt{ballmapper()} function and then interrogated using the other features of the \texttt{ballmapper} package. The data may then be further interrogated and explored by the collective expertise of the users. Our analysis shows the potential of the modelled features in classifiers. We also show the need to consider other variables in modelling price. 

\section{Summary}
\label{sec:summary}

This guide to the \texttt{ballmapper} package has shown how Topological Data Analysis Ball Mapper (TDABM) \citep{dlotko2019ball} may be applied in Stata. TDABM is a model free approach to multivariate data visualization and understanding. \cite{anscombe1973graphs} and \cite{matejka2017same} make clear arguments on the importance of visualizing data. TDABM offers a means to overcome the limitations of being able to include only 2 dimensions on the page. The abstract maps of the dataspace produced can be used to drive discursive analysis, inform modelling and evaluate models. These abilities have been exploited in the literature \citep{qiu2020refining, rudkin2023economic, rudkin2024return,benites2025topology, tubadji2025cultural}, and many others. The \texttt{ballmapper} package allows Stata users to realize these benefits.

The Stata code is a basic implementation which provides for the most common workflows of TDABM application. For those who wish to examine the impact of landmark selection, or undertake extensive analysis across radii, it is necessary to either write further functions or employ the Python or R libraries\footnote{See \cite{rudkin2025introductionp} for Python and \cite{rudkin2025introduction} for R.}. In this guide we have not considered the evaluation of statistical models. However, to do so is straightforward since you will have a column of predicted values and residuals that can be used as coloring variables in the TDABM. The boxplots and summaries become means of visually inspecting whether the residuals are indeed independent of the joint distribution of characteristics. Future work will consider extensions of the \texttt{ballmapper} package to cover methodological innovations. 

\bibliography{tdabmpres}
\bibliographystyle{apalike}

\newpage
\appendix
\setcounter{table}{0}
\setcounter{figure}{0}
\renewcommand{\thetable}{A\arabic{table}}
\renewcommand{\thefigure}{A\arabic{figure}}
\numberwithin{equation}{section}

\section{Additional Tables}

This appendix features the full ball summary tables that are abridged within the main paper. Because there are 79 data points in the full dataset, we only provide the full list of 79 balls here. Tables \ref{tab:x2a} and \ref{tab:x2a2} provide the summary of the 5 $Y$ variables, whilst Tables \ref{tab:x3a} and \ref{tab:x3a2} provides a detailed summary of the value of $Y_1$ across the dataset.

\begin{table}
	\begin{center}
		\caption{Summary of Y Variables X Dataset (Balls 1 to 40)}
		\label{tab:x2a}
		\begin{tabular}{l c c c c c c}
			\hline
			Ball & $Y_1$ & $Y_2$ & $Y_3$ & $Y_4$ & $Y_5$ & Size \\
			\hline
			1	&0.656	&1&	65.43&	-0.016 &	0&	44\\
			2	&0.136	&1&	82.75&	-0.039 &	0&	40\\
			3	&-2.102&1&	79.77&	-0.060&	0&	11\\
			4	&-2.329	&1&	101.5&	0.384&	0&	3\\
			5	&1.110	&1&	90.63&	-0.195&	0&	15\\
			6&	2.524&	1&	78.00&	0.379&	0	&9\\
			7&	-0.614&	1.138&	64.06&	0.389	&0&	29\\
			8&	-1.702&	3.222&	42.15&	-0.594&	0&	9\\
			9&	-1.027&	1	&128.7&	1.366	&0&	2\\
			10&	0.644&	3.609&	38.40&	-0.292&	0&	23\\
			11&	11.28&	2.167&	65.44&	-0.059&	0&	24\\
			12&	9.193&	3.455&	43.60&	0.352&	0&	22\\
			13&	12.92&	2&	83.73&	-0.052&	0	&38\\
			14&	14.51&	2	&104.8&	0.271&	0	&5\\
			15&	13.66&	2&	94.98&	-0.063&	0&	11\\
			16&	11.75&	2&	69.47&	-0.089&	0&	38\\
			17&	11.81&	2&	70.03&	0.016&	0&	50\\
			18&	10.05&	3.143&	51.11&	0.075&	0&	28\\
			19&	-12.17&	3&	75.39&	-0.385&	0&	39\\
			20&	-13.01&	3&	85.15&	-0.128&	0	&30\\
			21&	-10.86&	3.154&	59.37&	0.083&	0&	26\\
			22&	-14.03&3&	96.89&	0.001&	0&	12\\
			23&	-10.83&3&	61.05&	0.046&	0&	5\\
			24&	-10.48&	3&	56.37&	-0.156&	0&	12\\
			25&	-15.65&	3&	124.5&	0.064&	0&	2\\
			26&	-12.82&	3&	82.40&	-0.320&	0&	46\\
			27&	-14.01&3&	97.54&	-0.850&	0&	2\\
			28&	-7.179&	6.680&	26.52&	0.165&	0&	25\\
			29&	-0.535&	4	&71.43&	0.122&	0	&43\\
			30&	1.349&	4&	77.05&	0.091&	0&	29\\
			31&	-1.474&	4&	85.80&	0.020&	0&	25\\
			32&	-2.753&	4&	115.5&	-0.293&	0&	2\\
			33&	-1.862	&4&	63.13&	0.566&	0&	12\\
			34&	0.008&	7.286&	31.42&	-0.371&	0&	28\\
			35&	0.006&	4	&87.18&-0.087&	0&	27\\
			36&	1.082&	4	&93.88&	-0.131&	0&	13\\
			37&	1.761&	6.667&	34.48&	-0.568&	0&	9\\
			38&	0.130&	4.258&	55.13&	0.261&	0&	31\\
			39&	1.726&	4&	61.58&	-0.068&	0&	8\\
			40&	-0.151&	4&	101.60&	-0.393&	0&	5\\
			\hline
		\end{tabular}
	\end{center}
	\raggedright
	\footnotesize{Notes: Table provides the mean values of each of the 5 $Y$ variables used in coloration of the X dataset. The underlying dataset has 900 observations on 2 variables, $X_1$ and $X_2$. The data is translated to have 9 groups of 100 points centred on (-6,6) (-3,3), (3,3), (6,6), (0,0), (-3,-3), (3,-3), (-6,-6), and (6,-6). The colorations are given as $Y_1 = X_1 + X_2 + \theta$ where $\theta \sim N(0,0.2)$, $Y_2$ is the group number, $Y_3 = X_1^2 + X_2^2 + \theta$ where again $\theta \sim N(0,0.2)$,  $Y_4 = \phi$ where $\phi \sim N(0,1)$, and $Y_5$ takes the value 1 when $0<X_1<3$ and $0<X_2<3$ are both satisfied.}
\end{table}

\begin{table}
	\begin{center}
		\caption{Summary of Y Variables X Dataset (Balls 41-79)}
		\label{tab:x2a2}
		\begin{tabular}{l c c c c c c}
			\hline
			Ball & $Y_1$ & $Y_2$ & $Y_3$ & $Y_4$ & $Y_5$ & Size \\
			\hline
			41&	-0.498&	5&	18.49&	-0.091&	0&	50\\
			42&	-1.092&	6.412&	6.601&	-0.243&	0&	17\\
			43&	0.774&	5&	22.43&	-0.007&	0	&31\\
			44&	2.930&	7&	6.70&	-0.285&	0.625&	8\\
			45&	0.415&	5&	13.93&	-0.169&	0	&33\\
			46&	-0.968&	4.692&	29.62&	-0.278&	0&	26\\
			47&	-2.397&	5&	18.74&	1.010	&0&	9\\
			48&	-2.626&	5.400&	10.99&	1.101	&0&	5\\
			49&	2.132&	5 &26.19&	-0.853&	0&	5\\
			50&	5.161&	6.188&	14.60&	0.130	&0.313&	16\\
			51&	5.026&	6	&13.67&	0.115&	0.391&	23\\
			52&	5.781&	6&	19.25&	-0.370&	0&	12\\
			53&	6.473&	5.900&	21.65&	-0.116	&0.100&	40\\
			54&	2.953&	7.5	&5.351&	-0.103&	1&	16\\
			55&	7.166&	5.733&	26.39&	0.229&	0&	30\\
			56&	8.798&	5.5	&40.69&1.018&	0&	8\\
			57&	6.209&	6	&20.56&	0.212&	0&	12\\
			58&	-5.449&	7	&15.43&	0.143&	0&	49\\
			59&	-6.777&	7	&27.64&	-0.078&	0&	10\\
			60&	-4.242&	7.095&	10.40&	-0.061&	0&	21\\
			61&	-6.663&	7&	23.13&	0.172	&0&	33\\
			62&	-4.240&	7.308&	9.815&	-0.089&	0&	26\\
			63&	-6.165&	7&	22.75&	0.270&	0&	6\\
			64&	-5.886&	7&	18.24&	0.029&	0&	27\\
			65&	-7.053&	7&	27.87&	-0.045&	0&	8\\
			66&	1.281	&7.886&	21.80&	-0.161&	0&	35\\
			67&	-0.199&	8&	18.93&	-0.154&	0&	51\\
			68&	-2.085&	8&	18.74&	0.035&	0&	12\\
			69&	2.581&	8	&16.87&	0.600&	0&	5\\
			70&	-1.851&	8&	29.46&	-0.195&	0&	11\\
			71&	-0.641&	8.240&	9.457&	-0.056&	0&	25\\
			72&	3.267&	7.200&	29.81&	-0.552&	0&	5\\
			73&	0.435&	8.034 & 14.77&	-0.156&	0	&33\\
			74&	0.349&	9&	0.489	&0.245	&0.327&	49\\
			75&	-1.718&	8.813&	3.073&	0.159&	0	&16\\
			76&	-1.210&	8.806&	1.469&	0.138&	0&	31\\
			77&	1.737&	9&	4.252	&-0.506	&0.250&	4\\
			78&	1.471&	8.739&	2.217&	-0.432&	0.609&	23\\
			79&	0.339&	8.692&	1.546&	-0.009&	0.115&	26\\
			\hline
		\end{tabular}
	\end{center}
	\raggedright
	\footnotesize{Notes: Table provides the mean values of each of the 5 $Y$ variables used in coloration of the X dataset. The underlying dataset has 900 observations on 2 variables, $X_1$ and $X_2$. The data is translated to have 9 groups of 100 points centred on (-6,6) (-3,3), (3,3), (6,6), (0,0), (-3,-3), (3,-3), (-6,-6), and (6,-6). The colorations are given as $Y_1 = X_1 + X_2 + \theta$ where $\theta \sim N(0,0.2)$, $Y_2$ is the group number, $Y_3 = X_1^2 + X_2^2 + \theta$ where again $\theta \sim N(0,0.2)$,  $Y_4 = \phi$ where $\phi \sim N(0,1)$, and $Y_5$ takes the value 1 when $0<X_1<3$ and $0<X_2<3$ are both satisfied.}
\end{table}

Tables \ref{tab:x2a} and \ref{tab:x2a2} show the expected structures manifest in the balls. Mean values of $Y_1$ range from -3 to 3, corresponding to cases where the ball is located on either edge of the sub-cloud. Values of $Y_2$ are often whole numbers, implying that the ball only covers one of the sub-clouds. However, there are decimals where a ball covers multiple sub-clouds. Because of the way in which the sub-clouds are numbered there is limited immediate merit in studying the values of the decimals. $Y_3$ is quadratic meaning that all of the means are positive. For the sub-clouds at the ends of the X shape the sum of the squares is much larger than those sub-clouds closer to the center. Hence we see balls like 14 which has a mean value of $Y_3$ of 104.8. $Y_4$ is noise and so there should be no pattern observed. We see that there is indeed no pattern. For $Y_5$, the value of 1 is observed for any point which has $0<X_1<3$ and $0<X_2<3$. Many of these points with $Y_5=1$ are in the sub-cloud centered on (0,0) and have low values of $Y_3$. Others are in the subcloud centered on (3,3) and have higher values of $Y_3$. Hence we see the expected patterns.
 
\begin{table}
	\begin{center}
		\caption{Example Table Lines From $Y_1$ Summary X Dataset (Balls 1-40)}
		\label{tab:x3a}
		\begin{tabular}{l c c c c c c c c}
			\hline
			Ball & Mean & Std. Dev. & Min & q25 & q50 & q75 & Max & Size\\
			\hline
			1	&0.656	&0.770	&-0.696 &	0.030 &	0.542&	1.228 &	2.699&	44\\
			2	&0.136	&0.775 &-1.652 &	-0.578 &	0.158 &	0.785&	1.732 &	40\\
			3	&-2.101&	0.752&	-3.379&	-2.608 &	-2.279 &	-1.431 &	-0.924&	11\\
			4	&-2.329&	0.593&	-2.756&	-2.756&	-2.579 &	-1.652 &	-1.652 &	3\\
			5	&1.110	&0.520&	0.344&	0.918&	1.053 &	1.376&	2.456 &	15\\
			6&	2.524&	0.742&	1.400&	2.144&	2.533&	2.704&	3.912&	9 \\
			7&	-0.614&	0.843&	-2.643&	-0.701&	-0.526&	0.001&	0.721&	29\\
			8&	-1.702&	0.601&	-2.466&	-2.019&	-1.591&	-1.541&	-0.411&	9\\
			9&	-1.027&	0.131&	-1.120&	-1.120&	-1.027&	-0.935&	-0.935&	2\\
			10&	0.644&	0.843&	-0.664&	0.003&	0.530&	1.203&	2.324&	23\\
			11&	11.28&	0.725&	9.513&	10.88&	11.36&	11.71&	12.74&	24\\
			12&	9.190&	1.026&	7.364&	8.253&	9.271&	9.930&	10.55&	22\\
			13&	12.91&	0.746&	11.73&	12.29&	12.86&	13.33&	14.44&	38\\
			14&	14.51&	0.553&	13.96&	14.05&	14.44&	14.78&	15.32&	5\\
			15&	13.66&	0.542&	12.74&	13.27&	13.92&	14.13&	14.20&	11\\
			16&	11.75&	0.819&	10.36&	11.03&	11.66&	12.44&	13.25&	38\\
			17&	11.81&	0.779&	10.44&	11.27&	11.70&	12.44&	13.37&	50\\
			18&	10.05&	1.101&	8.038&	9.210&	9.969&	10.94&	11.63&	28\\
			19&	-12.17&	0.685&	-13.74&	-12.83&	-12.19&	-11.77&	-10.67&	39\\
			20&	-13.01&	0.711&	-14.49&	-13.63&	-12.98&	-12.45&	-11.63&	30\\
			21&	-10.86&	0.894&	-12.19&	-11.66&	-10.84&	-10.35&	-9.06&	26\\
			22&	-14.03&	0.568&	-15.03&	-14.36&	-13.80&	-13.64&	-13.29&	12\\
			23&	-10.83&	0.442&	-11.40&	-11.15&	-10.77& -10.44&-10.39&	5\\
			24&	-10.48&	0.637&	-11.32&	-10.91&	-10.54&	-10.13&	-9.06&	12\\
			25&	-15.65&	1.194&	-16.50& -16.50&	-15.65&	-14.81& -14.81&	2\\
			26&	-12.82&	0.818&	-15.02&-13.42&	-12.78&	-12.17&	-11.62&	46\\
			27&	-14.02&	0.389&	-14.29&	-14.29&	-14.02&	-13.74&	-13.74&	2\\
			28&	-7.179&	0.716&	-9.288&	-7.667&	-7.054&	-6.565&	-6.245&	25\\
			29&	-0.535&	0.765&	-2.123&	-1.054&	-0.526&	0.184&	0.982&	43\\
			30&	1.349&	0.799&	0.100&	0.555&	1.618&	1.786&	3.180&	29\\
			31&	-1.474	&0.889&	-3.111&	-2.180&	-1.558&	-0.648&	-0.074&	25\\
			32&	-2.753	&0.796&	-3.315&	-3.315&	-2.753&	-2.190&	-2.190&	2\\
			33&	-1.862&	0.653&	-2.881&	-2.495&	-1.793&	-1.257&	-1.054&	12\\
			34&	0.008&	0.938&	-1.517&	-0.902&	0.194&	0.867&	1.426&	28\\
			35&	0.006&	0.837&	-1.757&	-0.604&	-0.074&	0.543&	1.717&	27\\
			36&1.082&	0.724&	-0.148&	0.510&	1.275&	1.640&	2.163&	13\\
			37&	1.761&	0.751&	0.896&	1.108&	1.483&	2.261&	3.102&	9\\
			38&	0.130&	0.788&	-1.168&	-0.463&	0.192&	0.871&	1.774&	31\\
			39&	1.726&	0.755&	0.705&	1.135&	1.590&	2.487&	2.678&	8\\
			40&	-0.151&	0.414&	-0.604&	-0.358&	-0.156&	-0.148&	0.510&	5\\
			\hline
		\end{tabular}
	\end{center}
	\raggedright
	\footnotesize{Notes: Values provide the mean, standard deviation, the minimum, maximum, quartiles and median of $Y_1$ for each ball in the TDABM plot. $Y_1 = X_1 + X_2 + \theta$, where $\theta \sim N(0,0.2)$ is a random noise component. The underlying dataset has 900 observations on 2 variables, $X_1$ and $X_2$. The data is translated to have 9 groups of 100 points centred on (-6,6) (-3,3), (3,3), (6,6), (0,0), (-3,-3), (3,-3), (-6,-6), and (6,-6). }
\end{table}

\begin{table}
	\begin{center}
		\caption{Example Table Lines From $Y_1$ Summary X Dataset (Balls 41-79)}
		\label{tab:x3a2}
		\begin{tabular}{l c c c c c c c c}
			\hline
			Ball & Mean & Std. Dev. & Min & q25 & q50 & q75 & Max & Size\\
			\hline
			41&	-0.498&	0.796&	-2.173&	-0.987&	-0.511&	0.051&	0.794&	50\\
			42&	-1.092&	0.701&	-1.998&	-1.664&	-1.299&	-0.712&	0.231&	17\\
			43&	0.774&	0.726&	-0.539&	0.354&	0.743&	1.203&	2.324&	31\\
			44&	2.930&	0.848&	1.959&	2.288&	2.693&	3.665&	4.191&	8\\
			45&	0.415&	0.784&	-0.967&	-0.140&	0.408&	1.034&	2.153&	33\\
			46&	-0.968&	0.772&	-2.466&	-1.539&	-0.968&	-0.411&	0.779&	26\\
			47&	-2.397&	0.702&	-3.371&	-3.077&	-2.173	&-1.851&	-1.503&	9\\
			48&	-2.626&	0.754&	-3.479&	-3.333&	-2.492&	-1.998&	-1.829&	5\\
			49&	2.132&	0.558&	1.556&	1.799&	1.988&	2.324&	2.996&	5\\
			50&	5.161&	0.904&	3.053&	4.759&	5.026&	5.877&	6.669&	16\\
			51&	5.026&	0.666&	3.580&	4.500&	5.096&	5.525&	6.212&	23\\
			52&	5.781&	0.728&	4.831&	5.357&	5.586&	6.236&	7.178&	12\\
			53&	6.473&	0.816&	5.096&	5.754&	6.661&	6.990&	8.295&	40\\
			54&	2.953&	1.097&	1.378&	2.218&	2.601&	3.746&	4.964&	16\\
			55&	7.166&	0.716&	5.689&	6.853&	7.028&	7.522&	8.861&	30\\
			56&	8.798&	0.702&	7.826&	8.367&	8.579&	9.358&	9.950&	8\\
			57&	6.209&	0.763&	4.865&	5.850&	6.139&	6.796&	7.507&	12\\
			58&	-5.449&	0.872&	-7.001&	-6.229&	-5.347&	-4.836&	-3.770&	49\\
			59&	-6.777&	0.785&	-7.774&	-7.337	&-6.783&	-6.460&	-5.292&	10\\
			60&	-4.242&	0.778&	-5.317&	-4.866&	-4.148&	-3.598&	-2.635&	21\\
			61&	-6.663&	0.639&	-7.899&	-7.027&	-6.638&	-6.343&	-5.203&	33\\
			62&	-4.240&	0.859&	-5.375&	-4.866&	-4.418&	-3.740&	-2.325&	26\\
			63&	-6.165&	0.491&	-6.687&	-6.487&	-6.268&	-5.971&	-5.310&	6\\
			64&	-5.886&	0.806&	-7.615&	-6.522&	-5.844&	-5.299&	-4.523&	27\\
			65&	-7.053&	0.569&	-7.899&	-7.526&	-6.945&	-6.587&	-6.409&	8\\
			66&	1.281&	0.807&	-0.246&	0.649	&1.184&	1.770&	3.158&	35\\
			67&	-0.199&	0.802&	-1.784&	-0.676&	-0.245&	0.416&	1.770&	51\\
			68&	-2.085&	0.518&	-2.708&	-2.504&	-2.171&	-1.729&	-1.058&	12\\
			69&	2.581&	0.415&	2.159&	2.301&	2.449&	2.812&	3.182&	5\\
			70&	-1.851&	0.561&	-2.708&	-2.324&	-1.784&	-1.246&	-1.128&	11\\
			71&	-0.641&	0.759&	-2.108&	-0.963&	-0.418&	-0.144&	1.002&	25\\
			72&	3.267&	0.684&	2.261&	3.102&	3.158&	3.904&	3.910&	5\\
			73&	0.435&	0.855&	-1.027&	-0.245&	0.379&	1.091&	2.449&	33\\
			74&	0.349&	0.680& -0.945&	-0.060&	0.346&	0.815&	1.861&	49\\
			75&	-1.718&	1.066&	-3.738&	-2.409&	-1.761&	-0.697&	-0.251&	16\\
			76&	-1.210&	0.886&	-3.132&	-1.866&	-1.135&	-0.528&	0.228&	31\\
			77&	1.737&	0.965&	0.778&	1.069&	1.558&	2.405&	3.053&	4\\
			78&	1.471&	0.736&	0.230&	0.867&	1.266&	2.109&	2.745&	23\\
			79&	0.339&	0.803&	-1.349&	-0.060&	0.495 &	1.081&	1.620&	26\\
			\hline
		\end{tabular}
	\end{center}
	\raggedright
	\footnotesize{Notes: Values provide the mean, standard deviation, the minimum, maximum, quartiles and median of $Y_1$ for each ball in the TDABM plot. $Y_1 = X_1 + X_2 + \theta$, where $\theta \sim N(0,0.2)$ is a random noise component. The underlying dataset has 900 observations on 2 variables, $X_1$ and $X_2$. The data is translated to have 9 groups of 100 points centred on (-6,6) (-3,3), (3,3), (6,6), (0,0), (-3,-3), (3,-3), (-6,-6), and (6,-6). }
\end{table}

Tables \ref{tab:x3a} and \ref{tab:x3a2} also show the expected pattern. Means for $Y_1$ are close to 0 when the ball is near the centre of a cloud on the top left to bottom right diagonal of the X. To the lower left balls have strong negative values of $Y_1$, as seen for balls like 19 to 27. Meanwhile, balls 10 to 17 can be seen to have high values of $Y_1$. Balls 10 to 17 are in the sub-cloud centered on (6,6) at the top right of the X. The range of values within a $Y_1$ ball corres

\section{Full Intuition Exposition}

\begin{figure}
	\begin{center}
		\caption{Step-by-Step Construction of Ball Mapper Plot (Balls 1 to 16)}
		\label{fig:astep}
		\begin{tabular}{c c c c}
			\includegraphics[width=4cm]{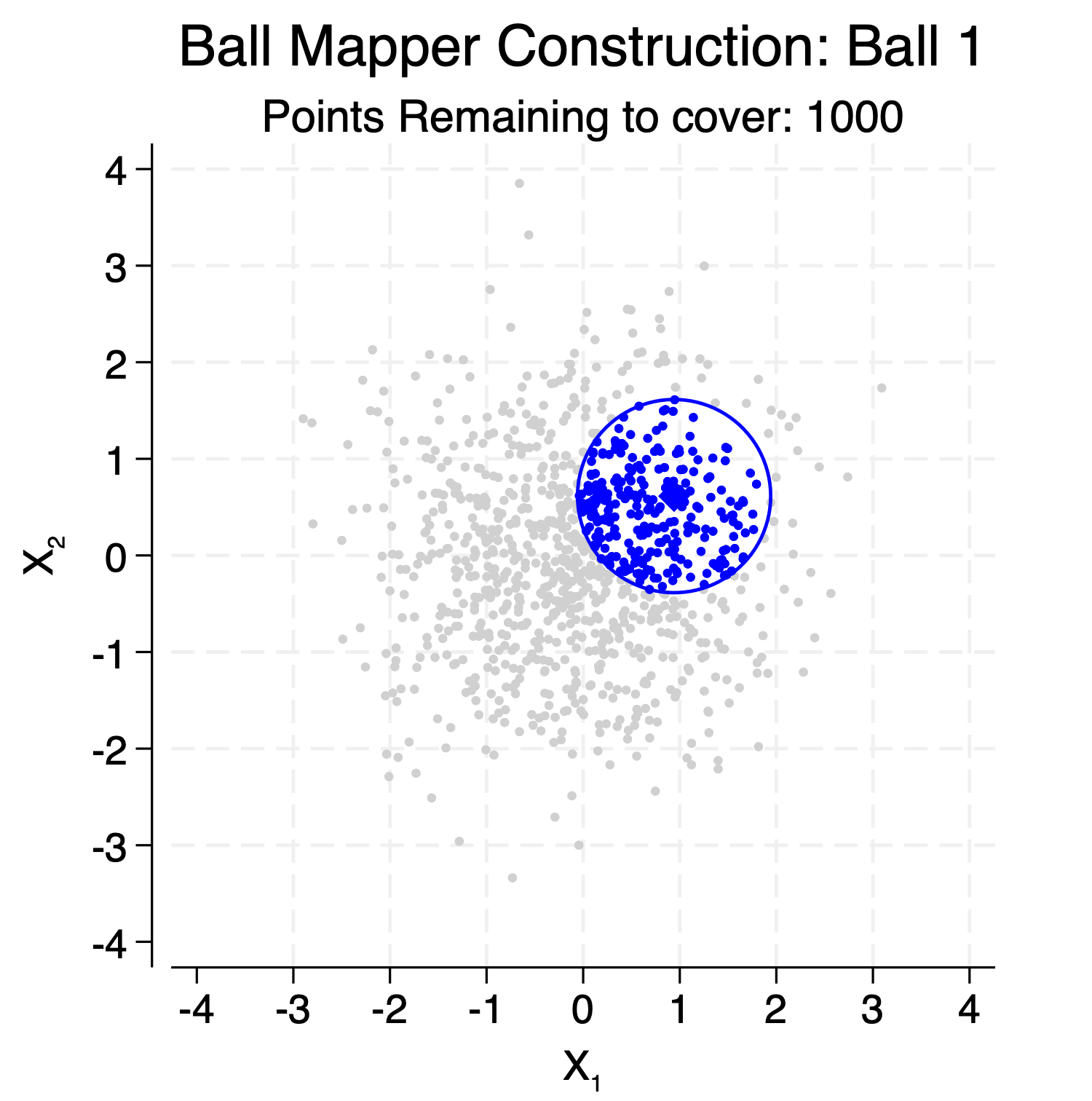}&
			\includegraphics[width=4cm]{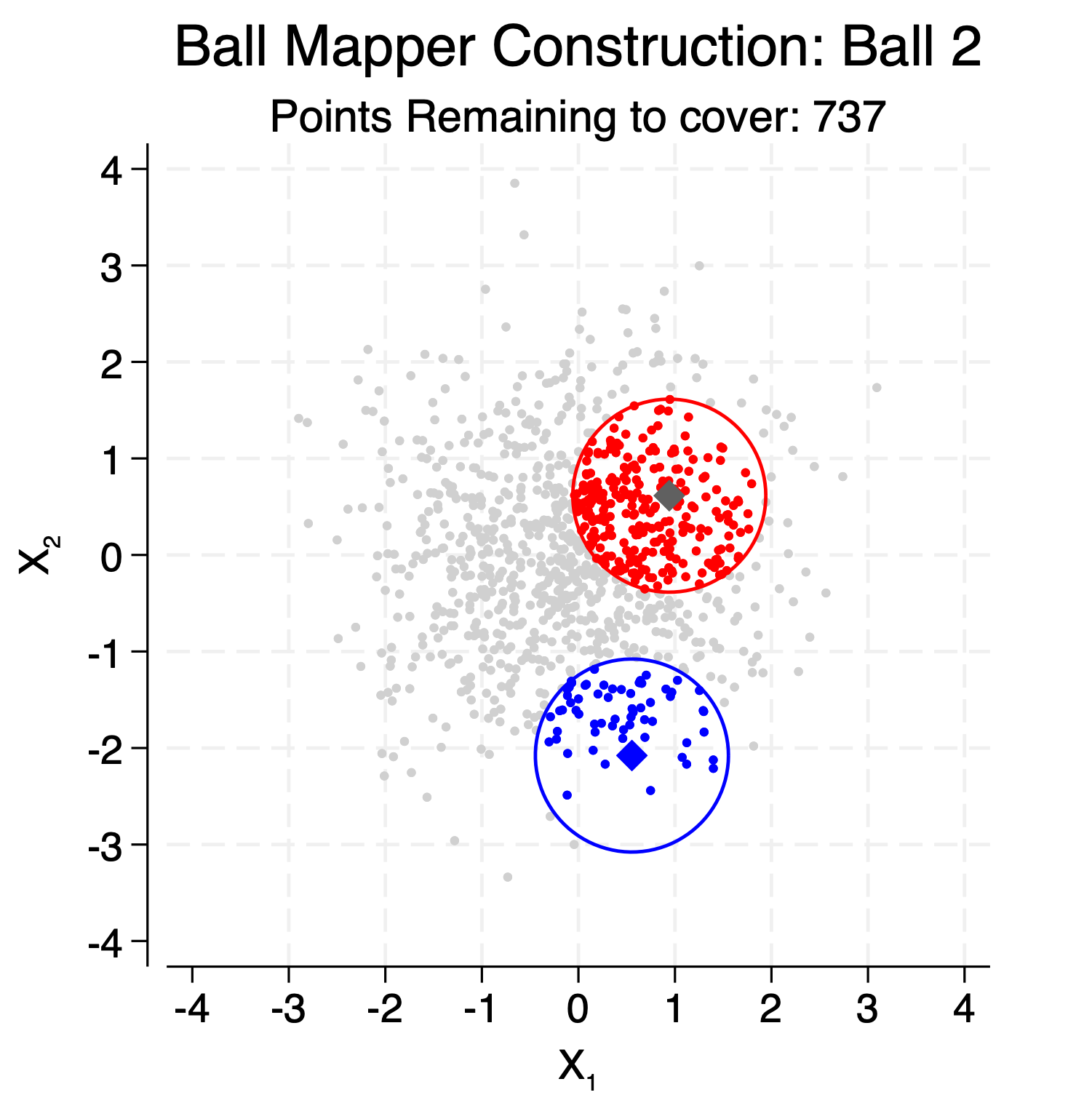}&
			\includegraphics[width=4cm]{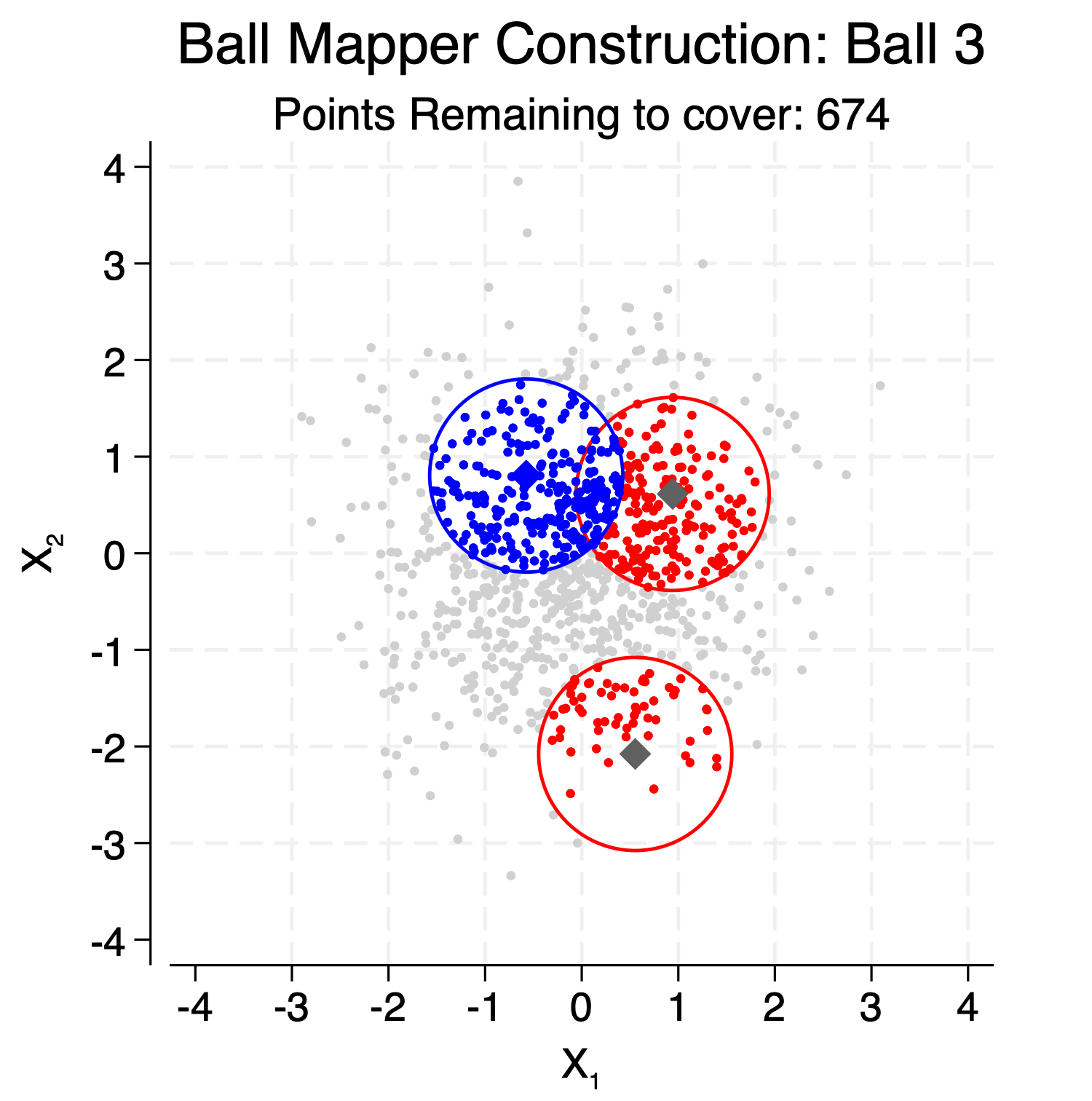}&
			\includegraphics[width=4cm]{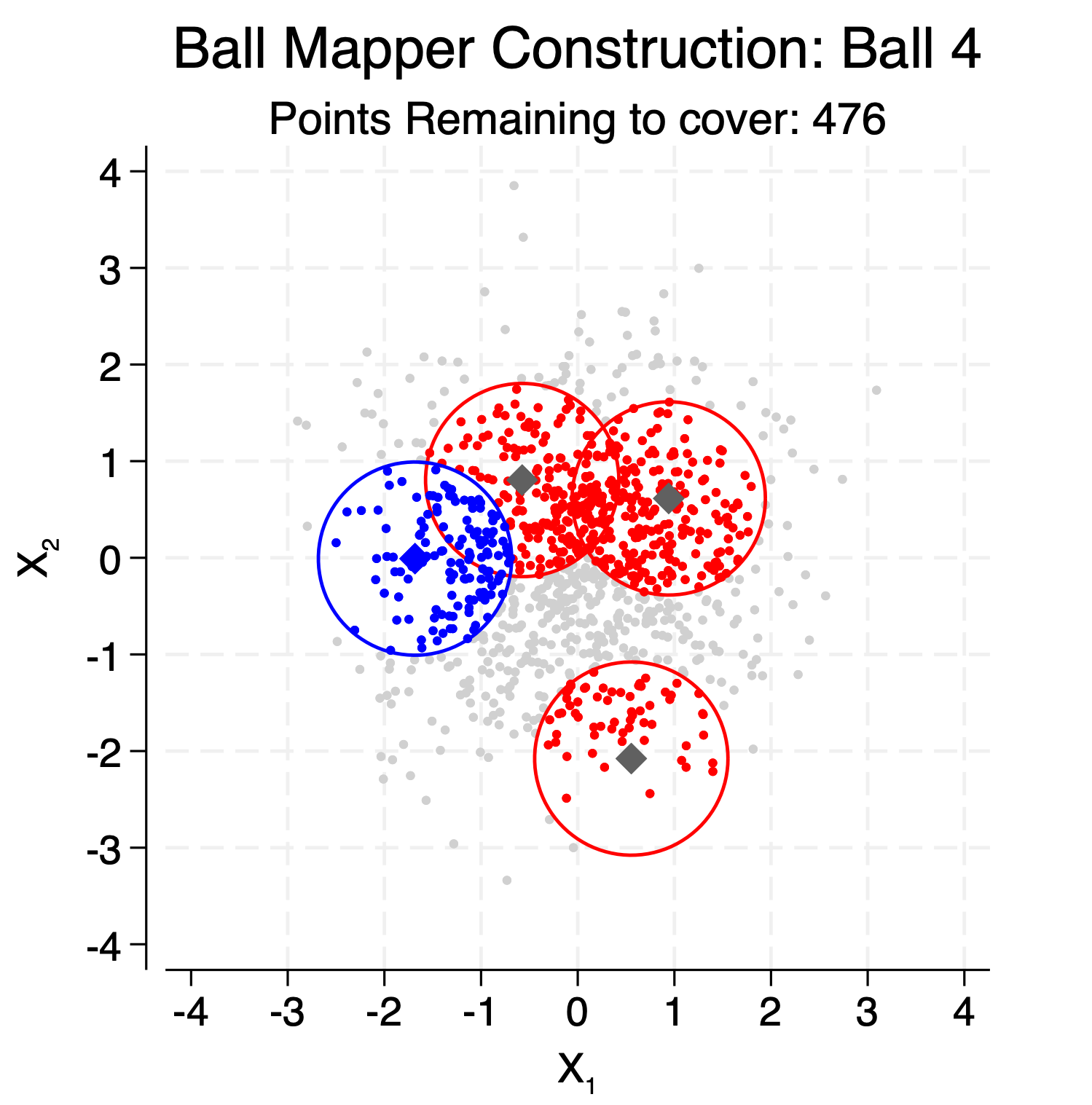}\\
			\includegraphics[width=4cm]{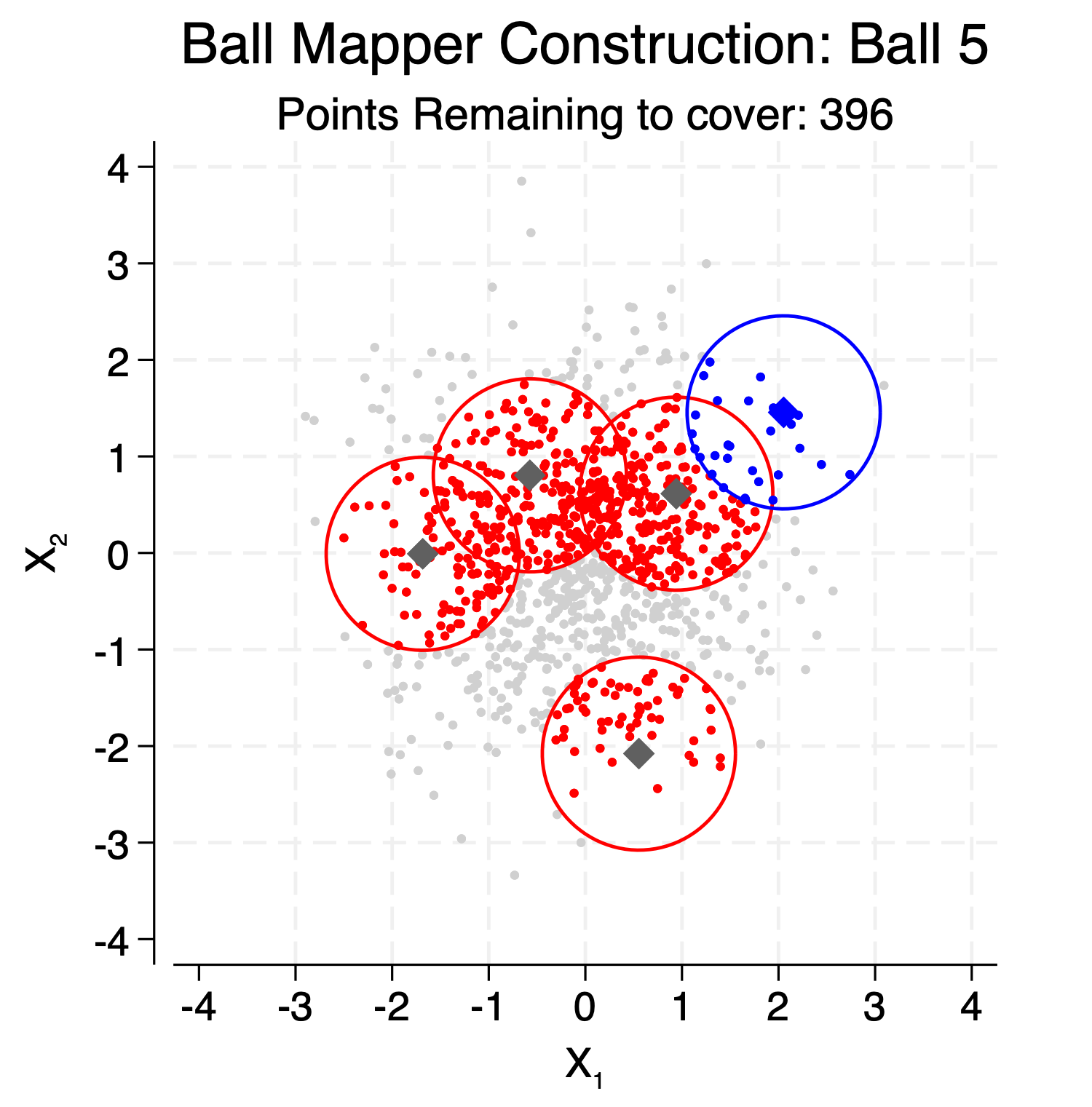}&
			\includegraphics[width=4cm]{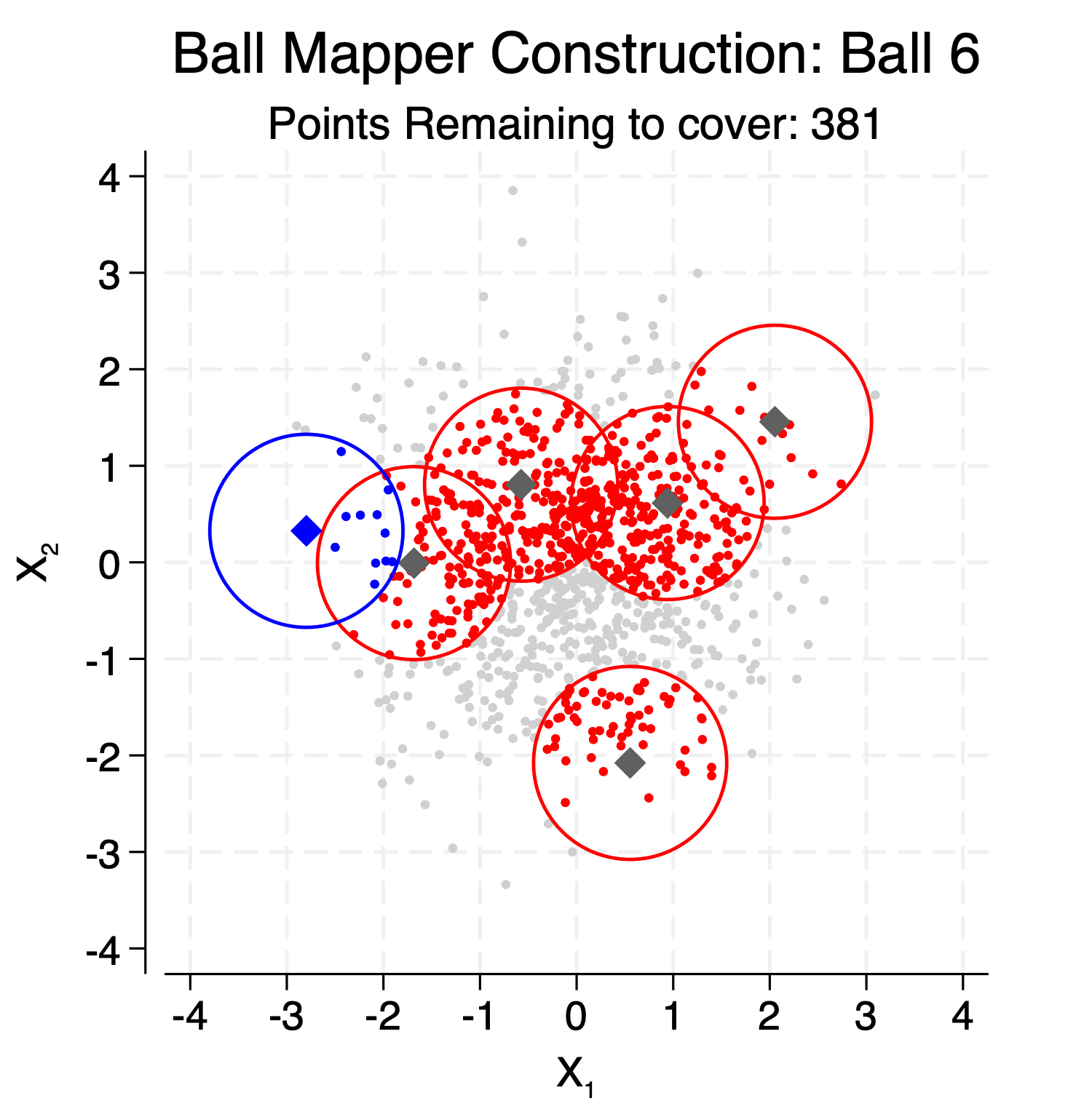}&
			\includegraphics[width=4cm]{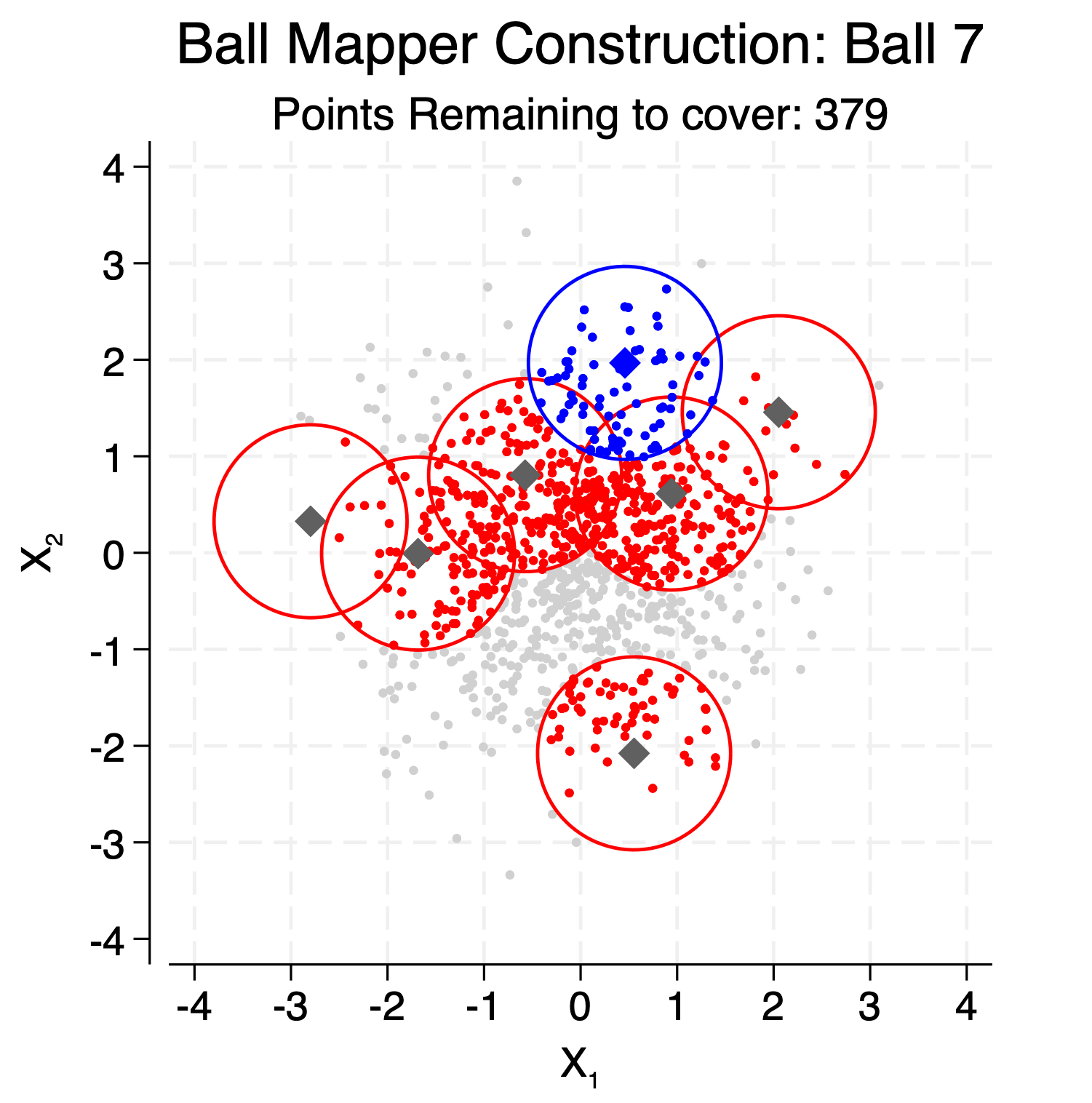}&
			\includegraphics[width=4cm]{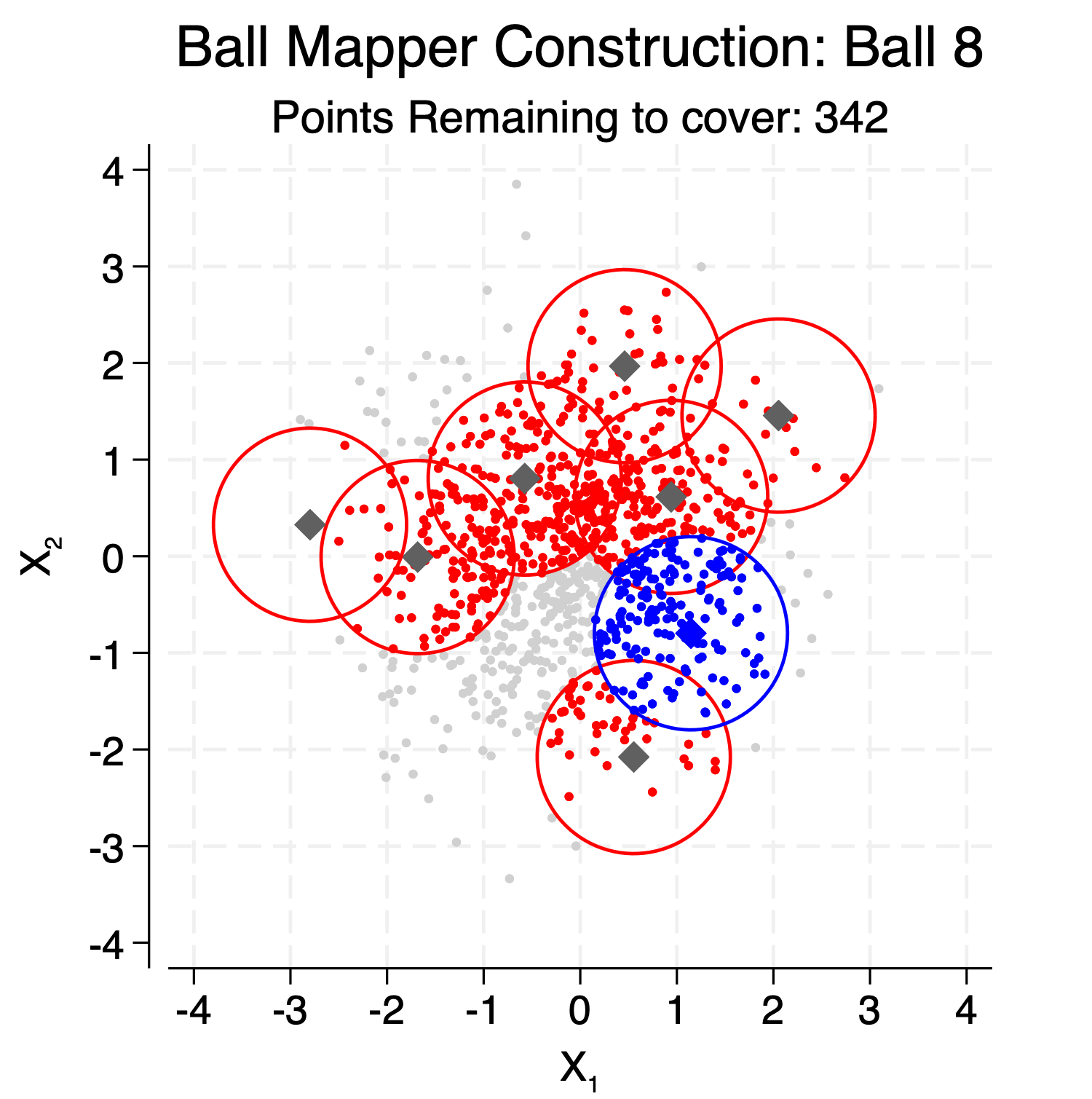}\\
			\includegraphics[width=4cm]{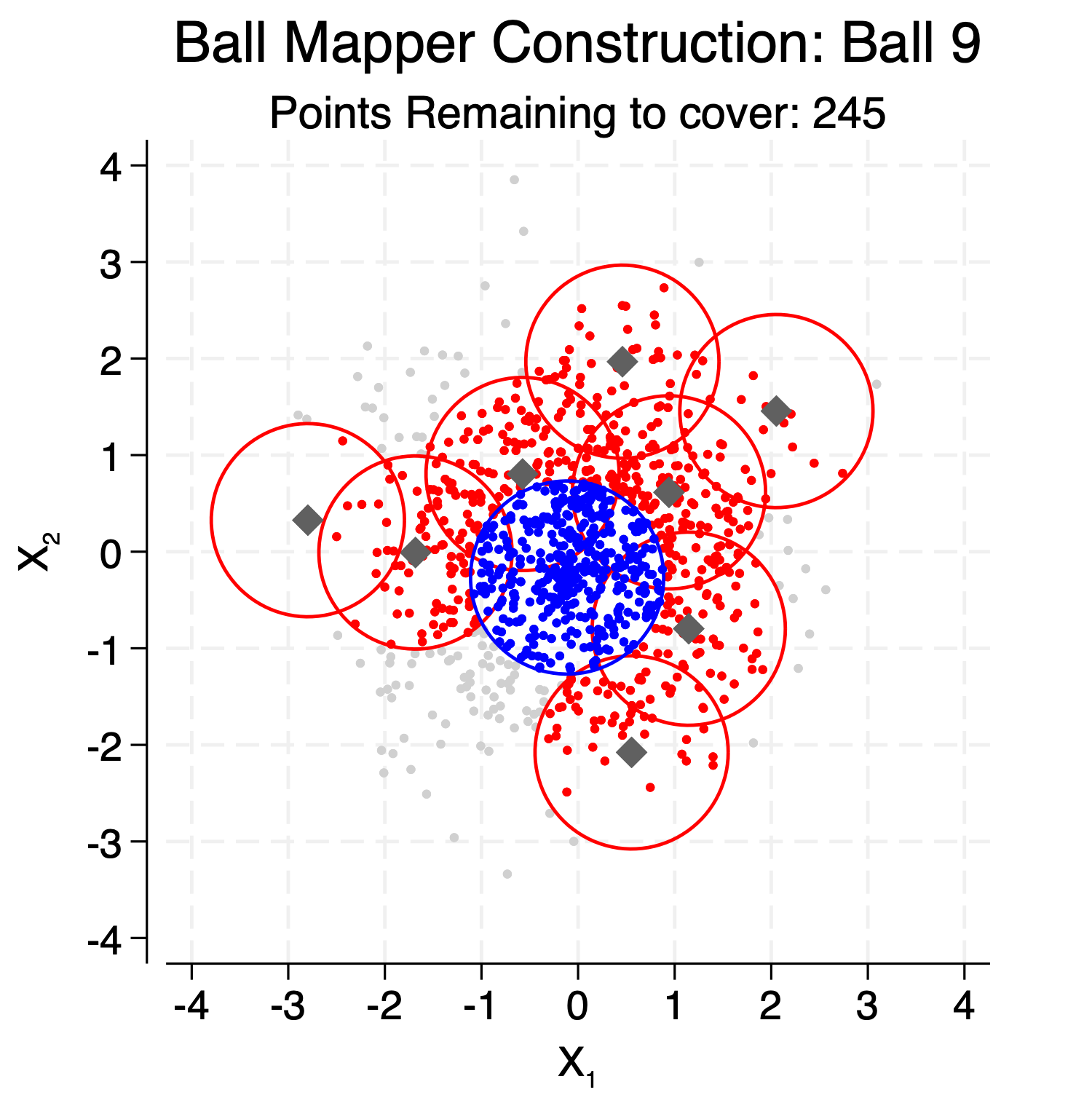}&
			\includegraphics[width=4cm]{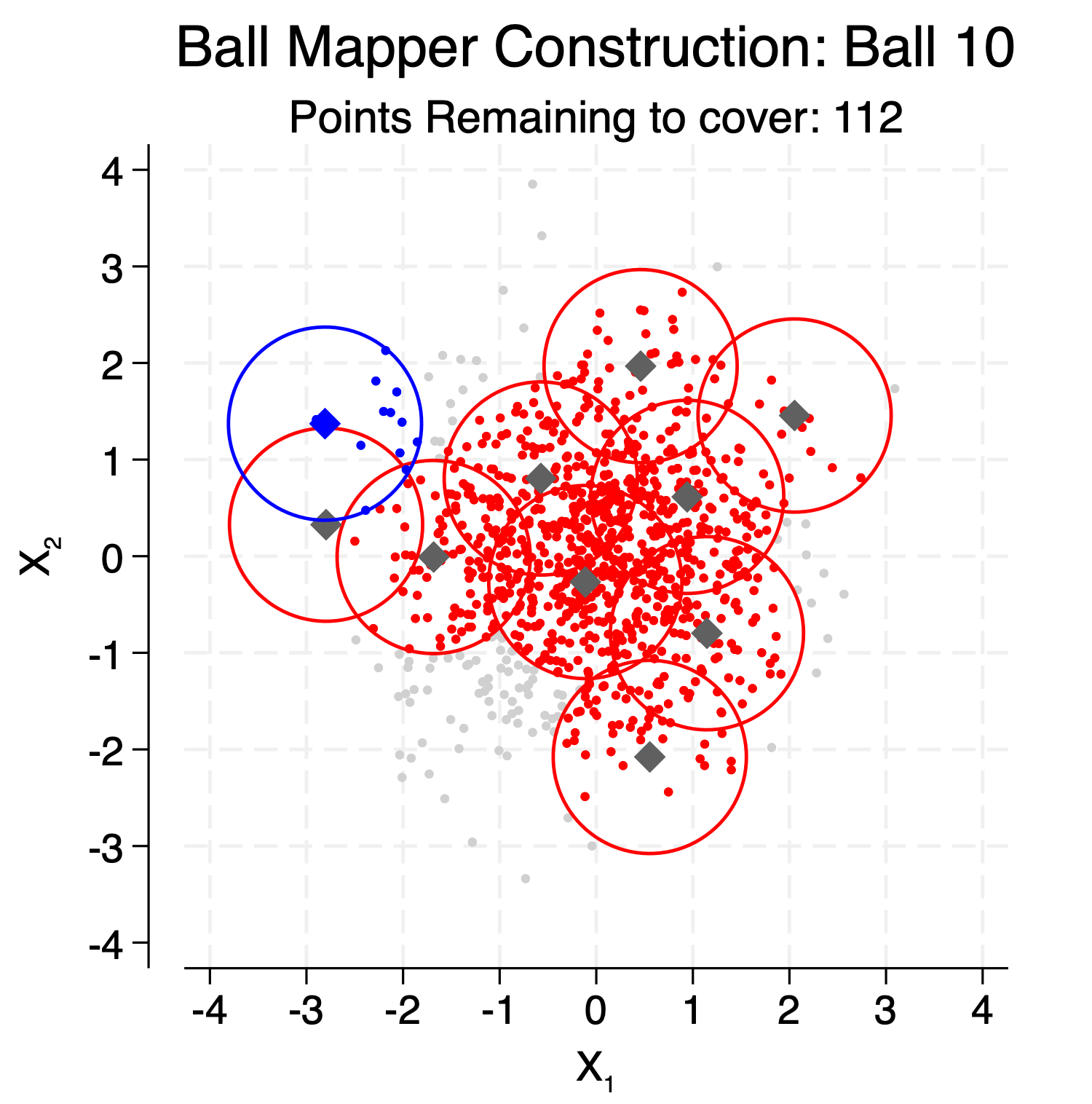}&
			\includegraphics[width=4cm]{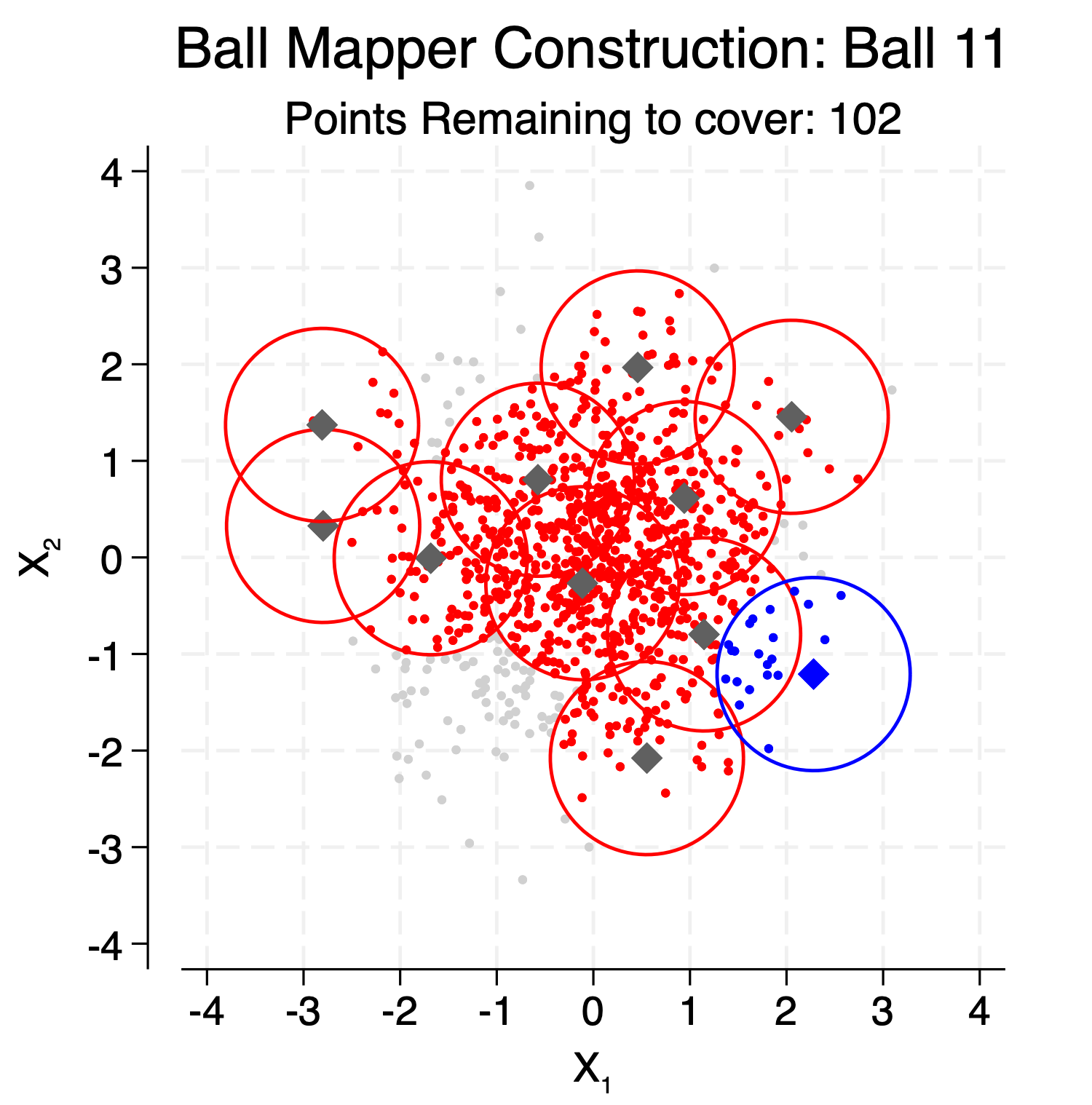}&
			\includegraphics[width=4cm]{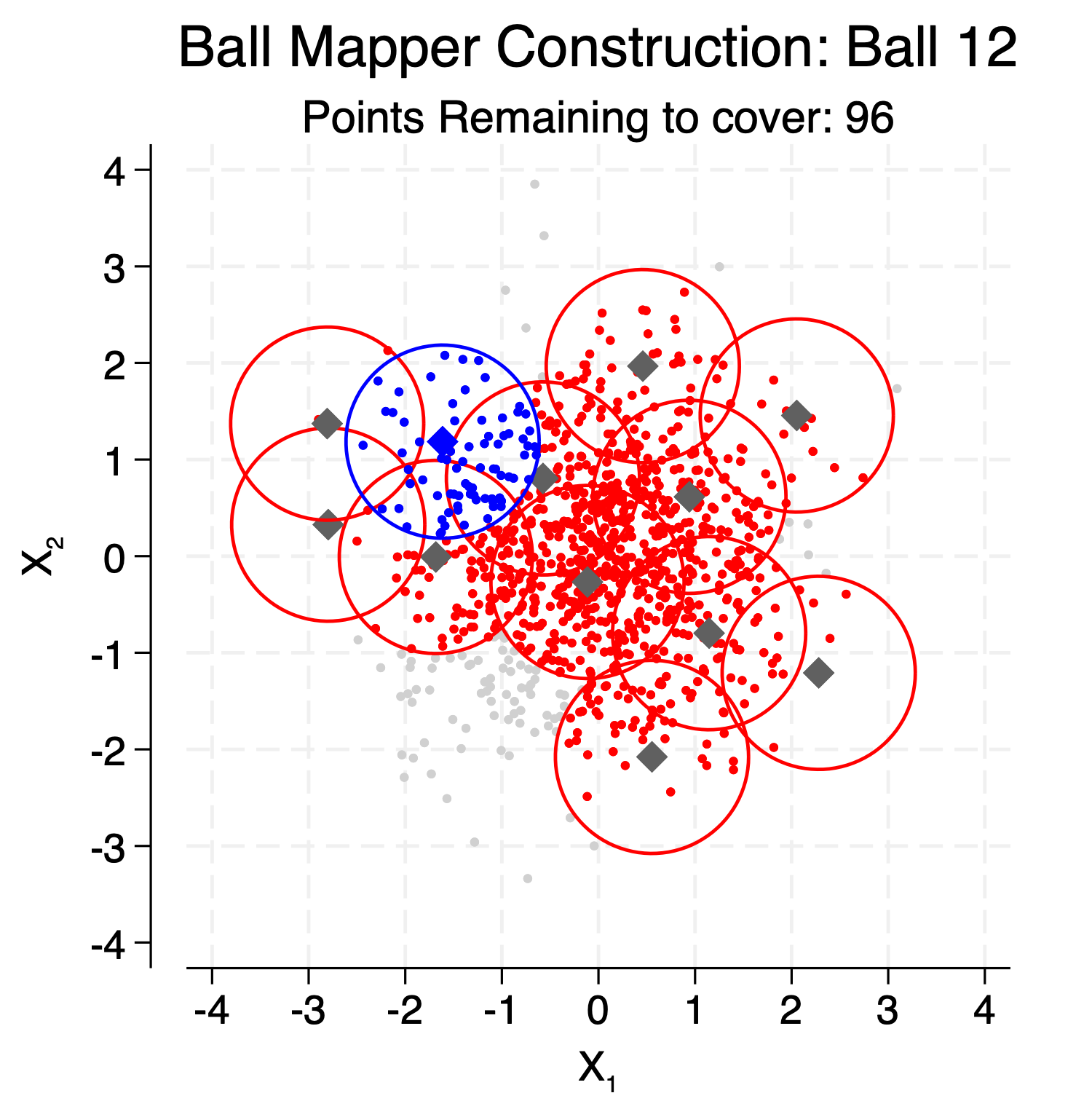}\\
			\includegraphics[width=4cm]{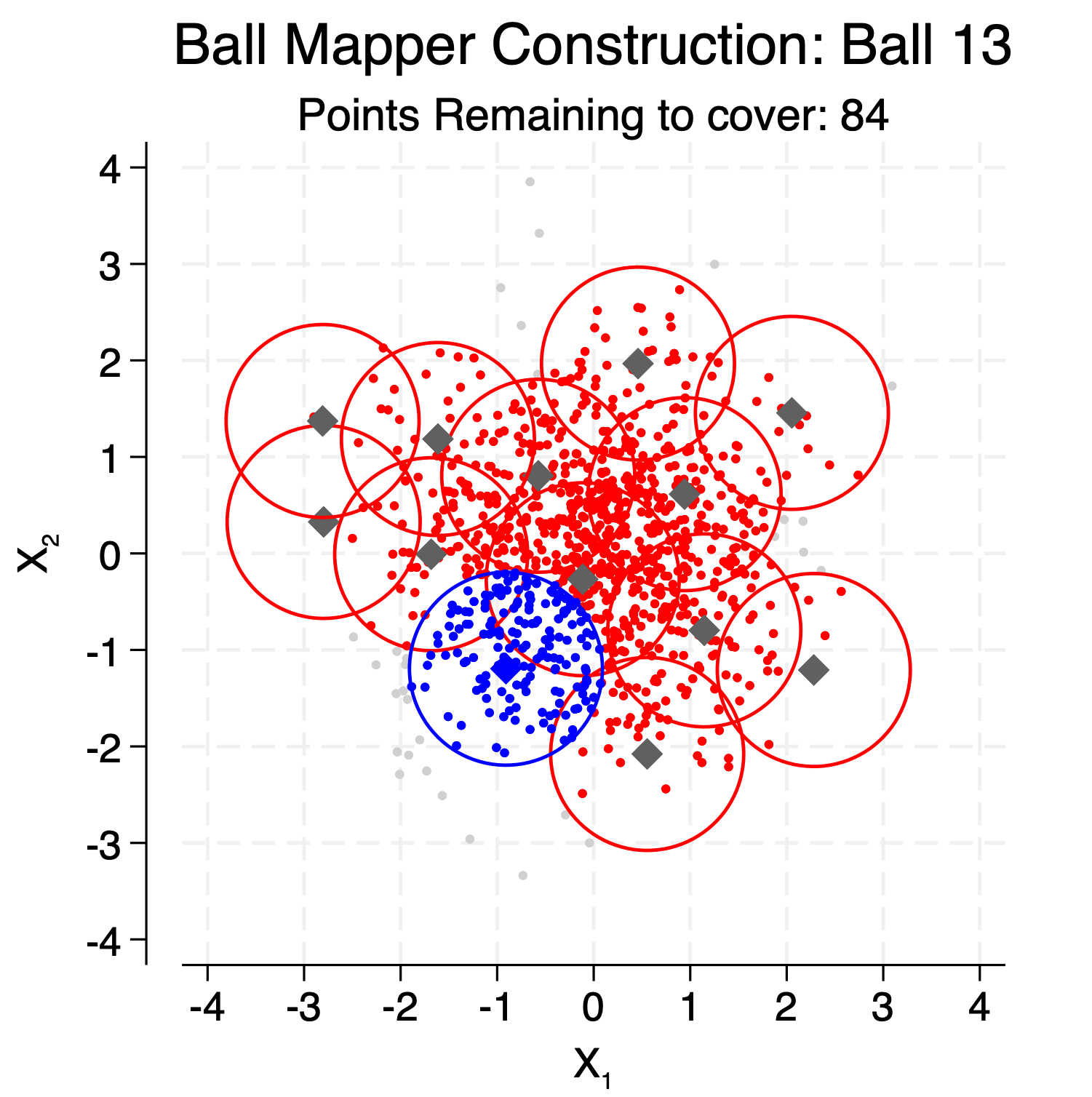}&
			\includegraphics[width=4cm]{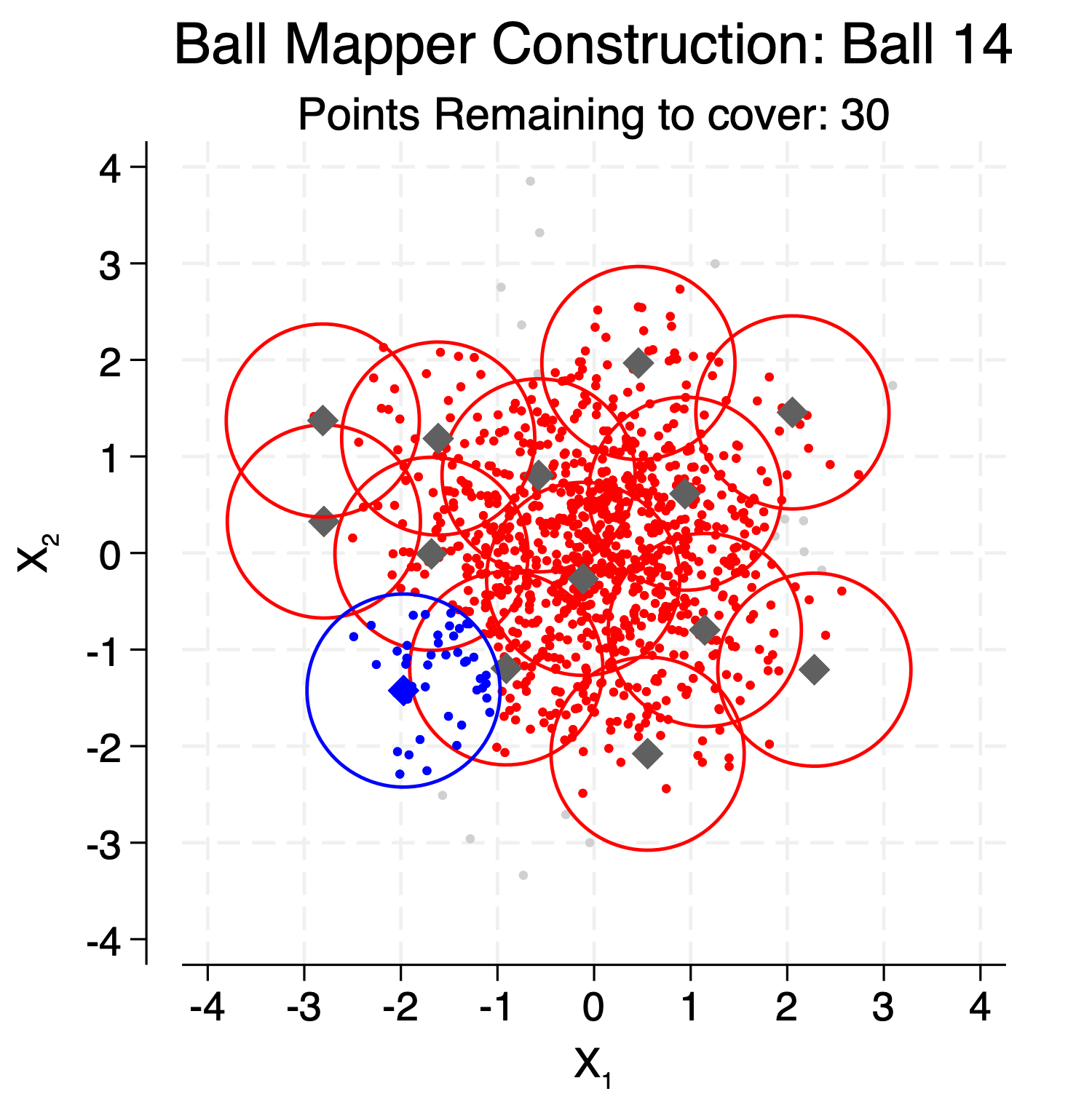}&
			\includegraphics[width=4cm]{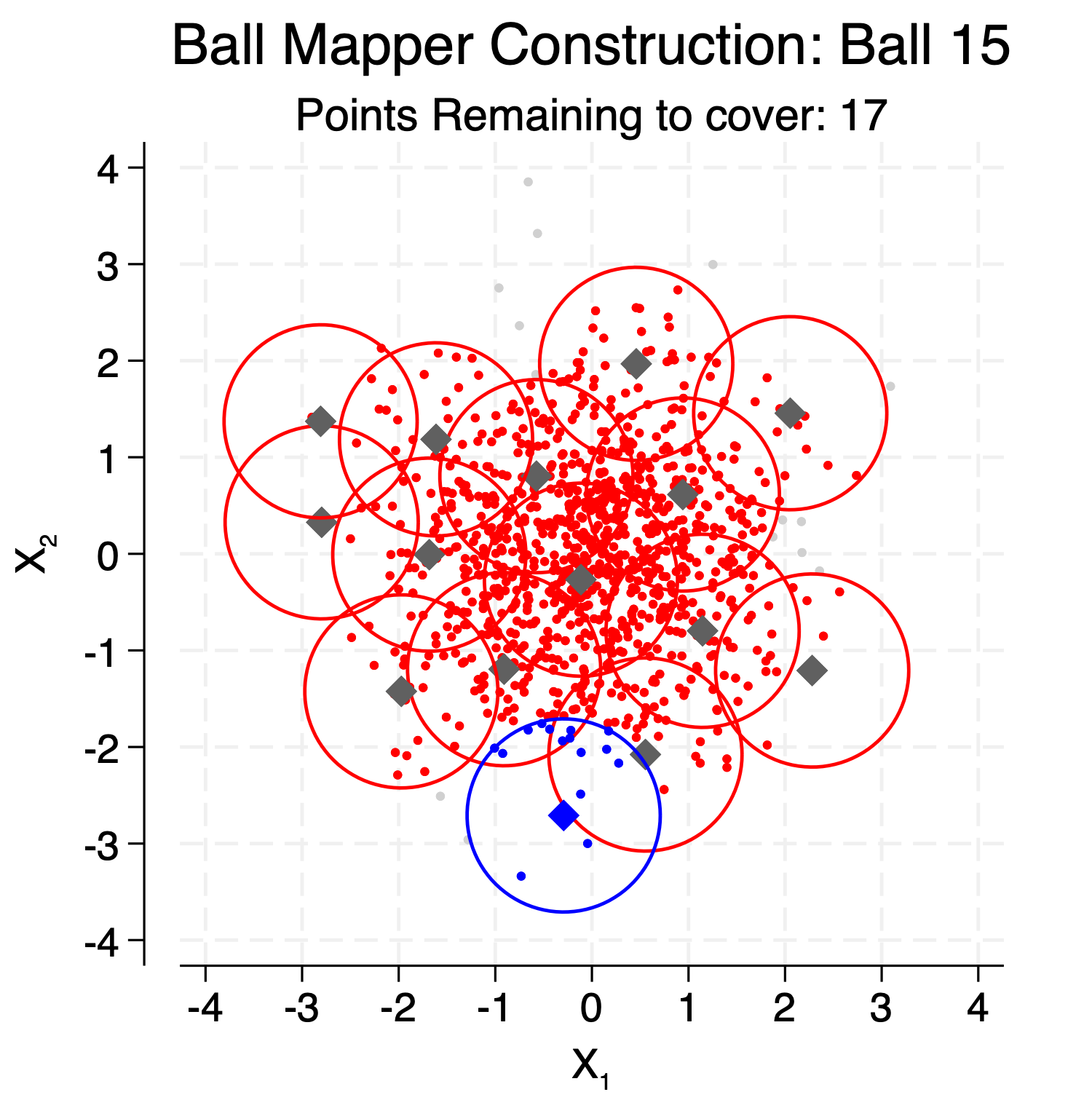}&
			\includegraphics[width=4cm]{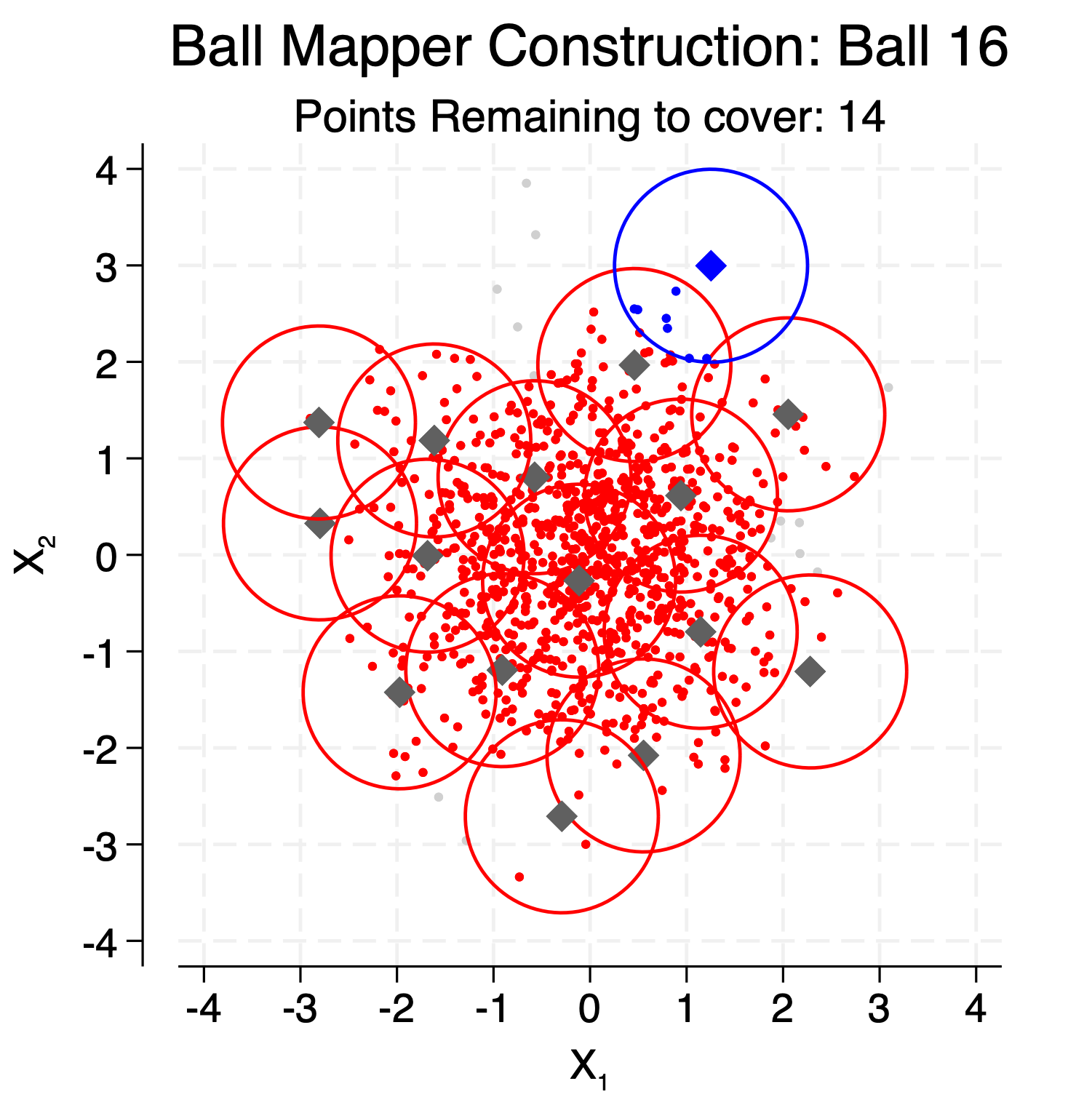}\\
		\end{tabular}
	\end{center}
	\raggedright
	\footnotesize{Notes: Construction of the TDABM coverage of a bivariate Gaussian cloud. Points already covered are colored red, the specific ball being created in the panel is colored blue. Here $X_1 \sim N(0,1)$ and $X_2 \sim N(0,1)$ and the ball radius is $\varepsilon = 1.00$.}
\end{figure}

\begin{figure}
	\begin{center}
		\caption{Step-by-Step Construction of Ball Mapper Plot (Balls 17 to 21)}
		\label{fig:astep5}
		\begin{tabular}{c c c c}

			\includegraphics[width=4cm]{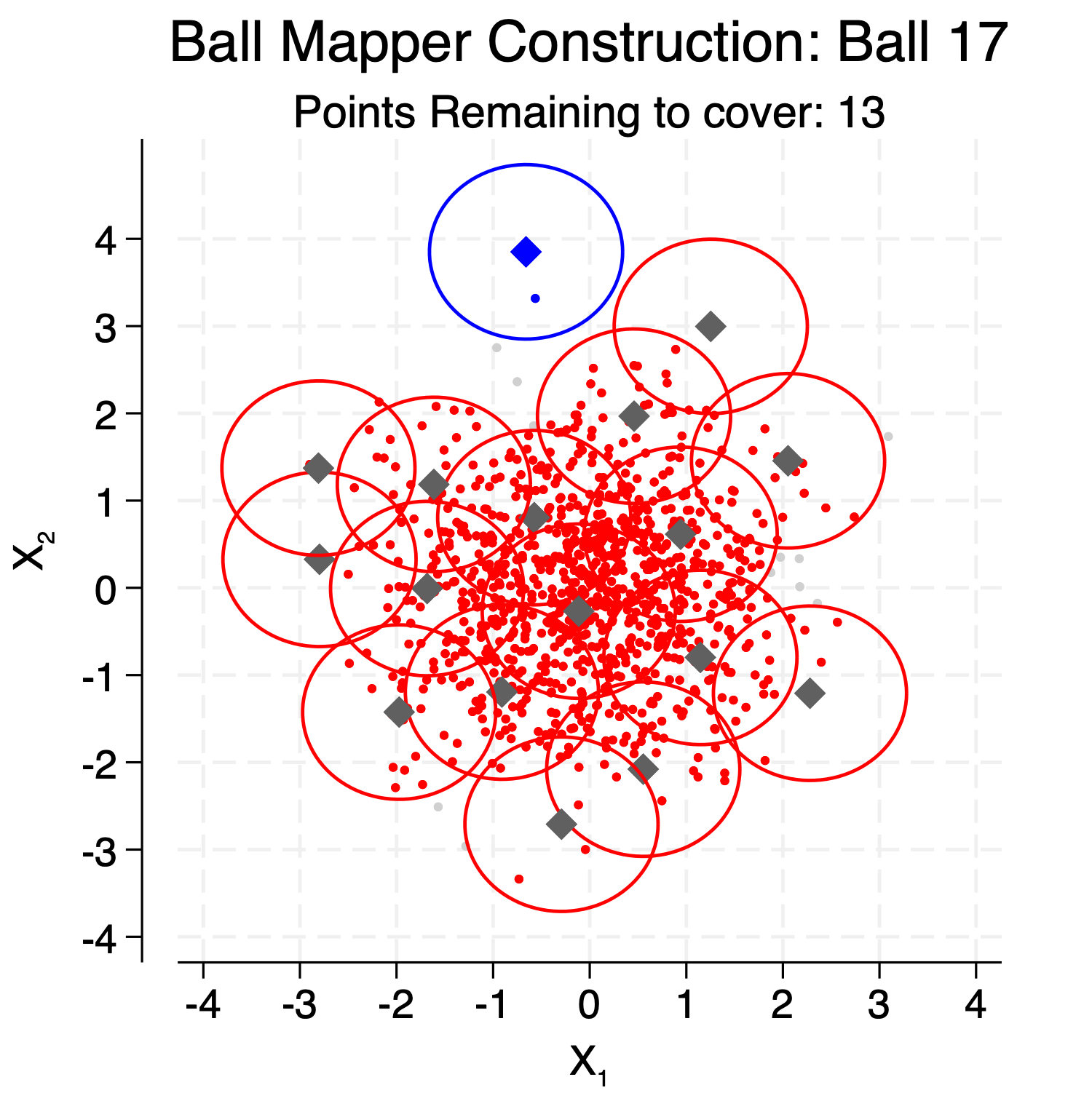}&
			\includegraphics[width=4cm]{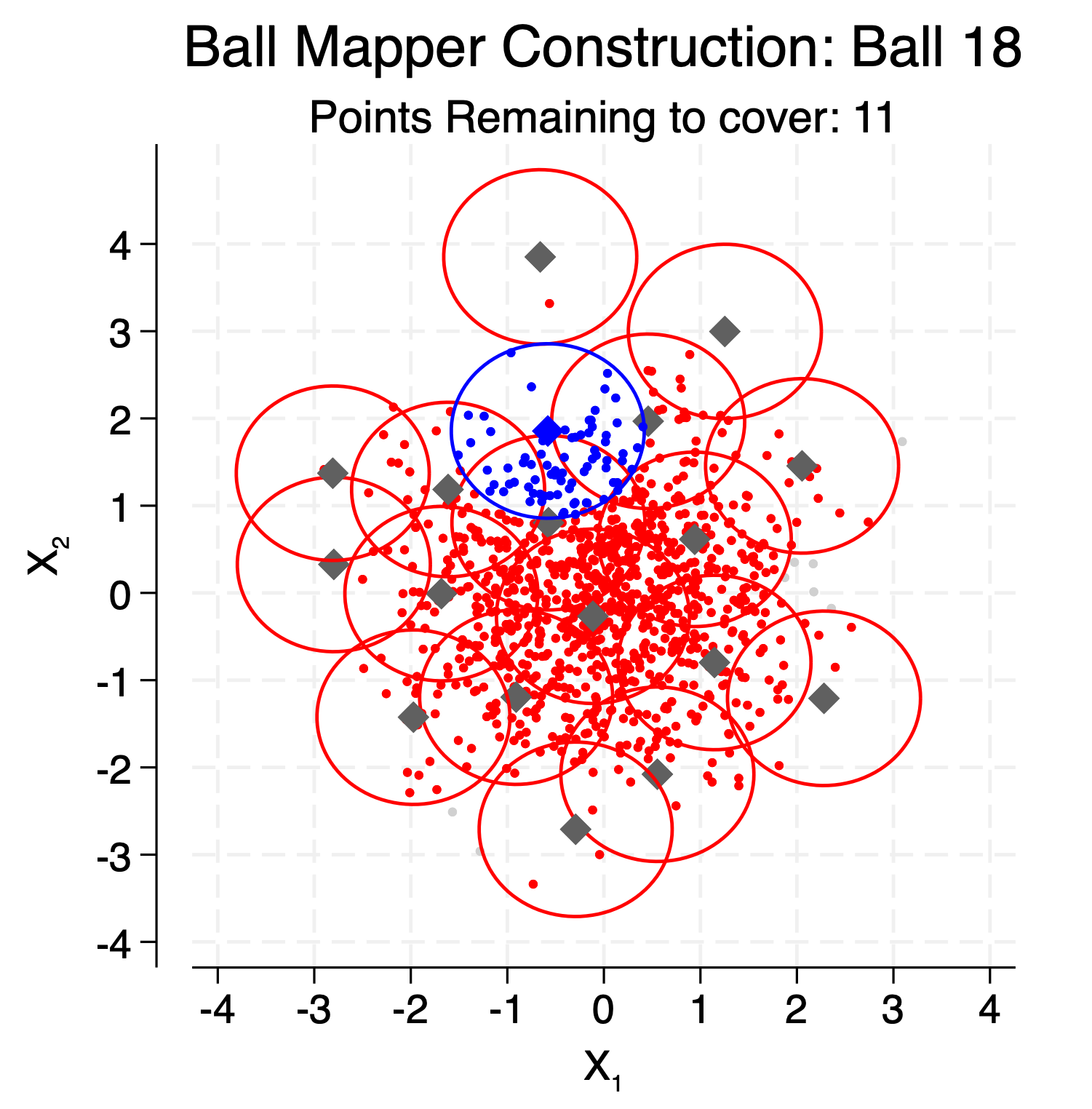}&
			\includegraphics[width=4cm]{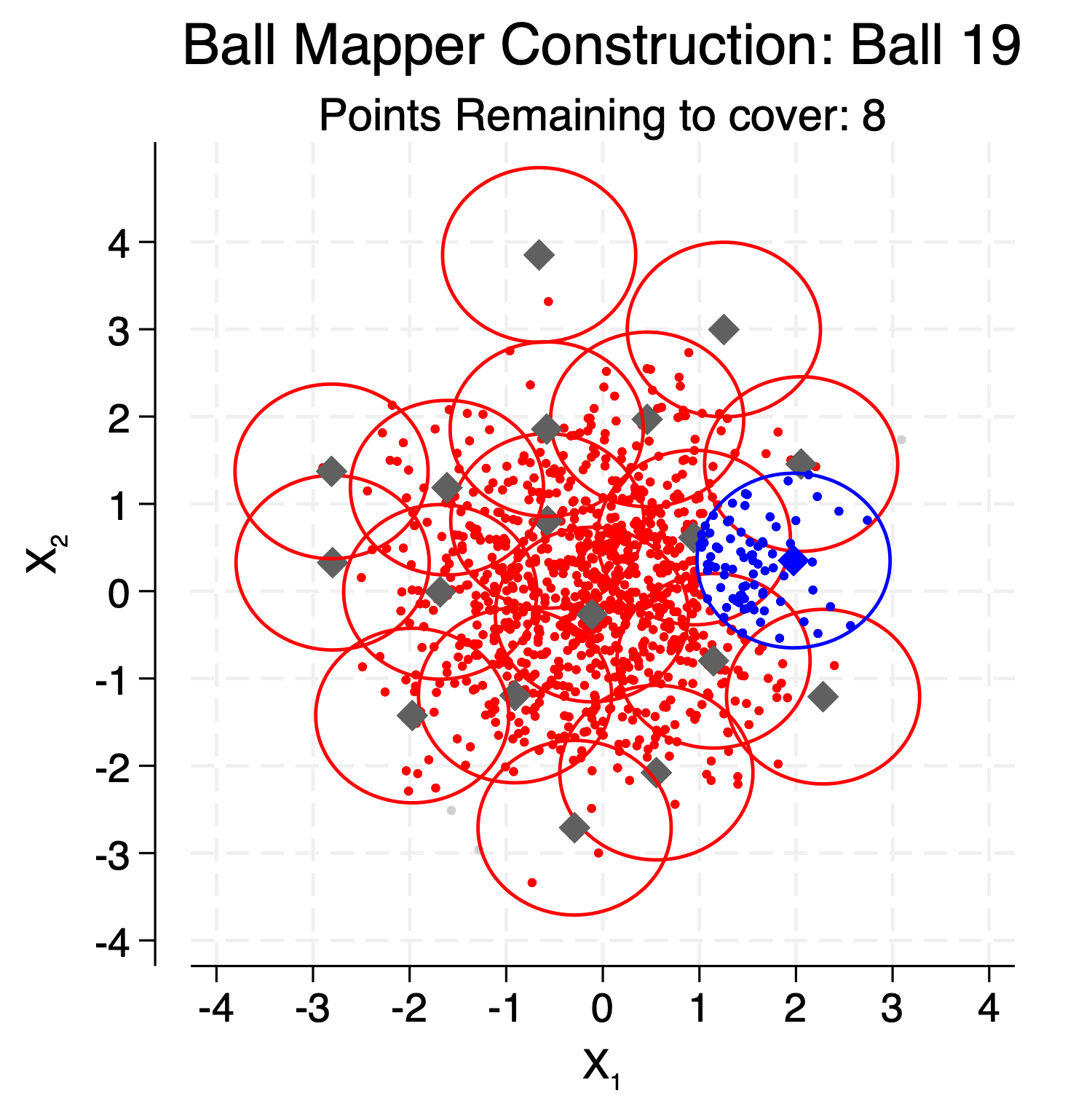}&
			\includegraphics[width=4cm]{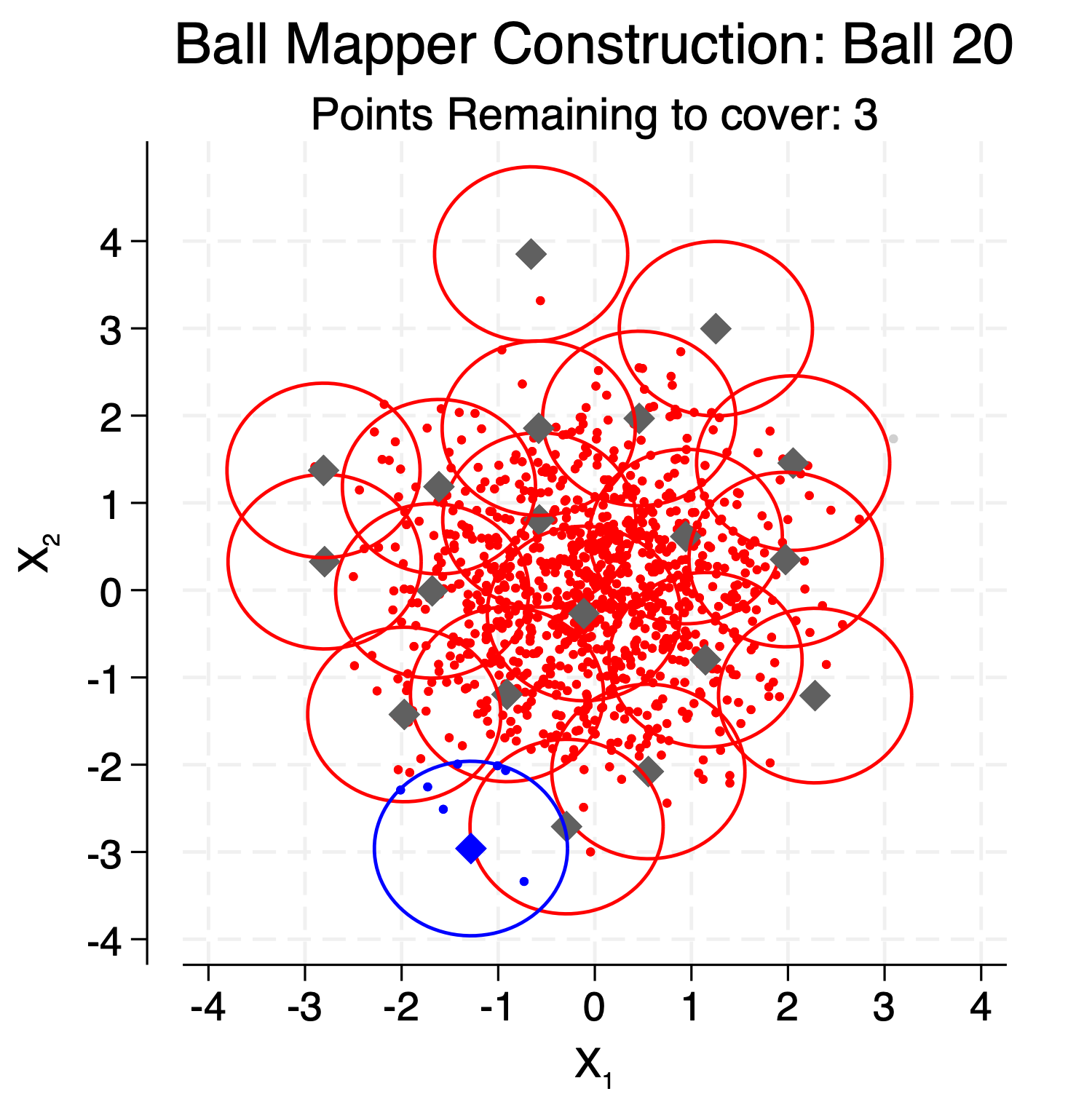}\\
			\includegraphics[width=4cm]{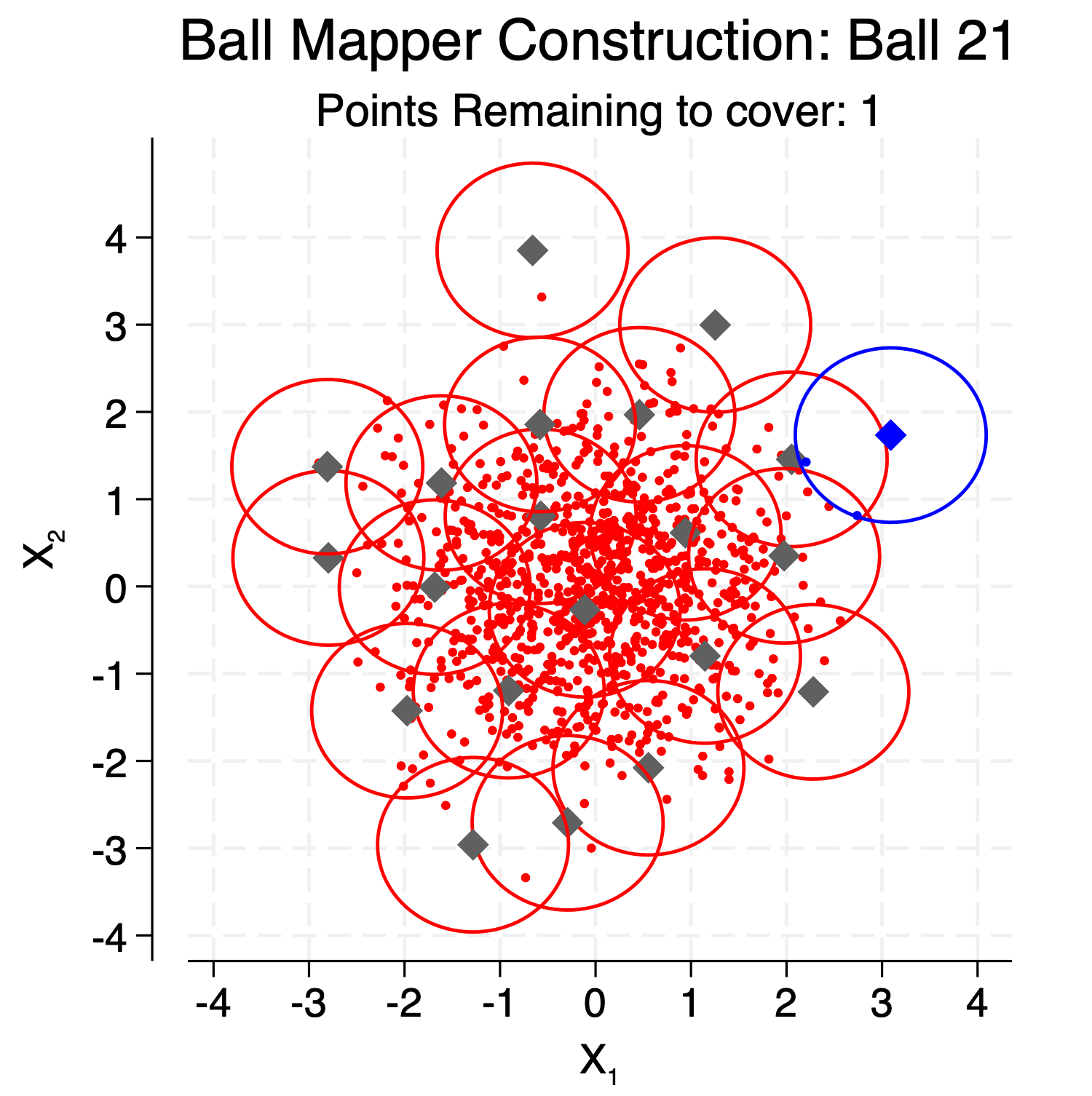}&&&\\
		\end{tabular}
	\end{center}
	\raggedright
	\footnotesize{Notes: Construction of the TDABM coverage of a bivariate Gaussian cloud. Points already covered are colored red, the specific ball being created in the panel is colored blue. Here $X_1 \sim N(0,1)$ and $X_2 \sim N(0,1)$ and the ball radius is $\varepsilon = 1.00$.}
\end{figure}

In order to illustrate the construction of the Topological Data Analysis Ball Mapper (TDABM) graph in detail, this appendix provides a walk through for a 2-dimensional dataset. The data used is that which is used in the intuition section of the main paper. Formally, there are 1000 observations ($N=1000$) on 2 variables, $K=2$, in the $N \times K$ dataset $X$. Each variable is drawn independently from a standard normal distribution, such that $X_1 \sim N(0,1)$ and $X_2 \sim N(0,1)$. The resulting point cloud is a Gaussian cloud of dimension 2. Figures \ref{fig:astep} and \ref{fig:astep5} work step-by-step through the addition of the balls.

The first ball is created around the first data point in $X$. In panel (a), the ball is shown in blue with a large diamond showing landmark 1. All of the blue points are covered by Ball 1. In panel (b), a second ball is added. The second ball is drawn around point number 2 in the dataset. The new ball covers the blue points, whilst the points covered by Ball 1 are shown as red. Here there is no overlap between the first two balls. Ball 3 is drawn close to ball 1, with an overlap in evidence. Ball 4 then sits to the left of ball 3, whilst ball 5 sits above ball 1 towards the top right of the space. Ball 6 picks up a few points to the left of the plot. Ball 7 sits towards the top right of the plot, having overlap with balls 1 and 3. The 8th ball is added to link balls 1 and 2. Ball 9 sits in the dense part of the cloud. Ball 10 is then just above ball 6 on the left of the plot, again only including a few further points. Ball 11 is to the lower right. Balls 12, 13 and 14 are further examples of balls which fill gaps between existing balls. Ball 15 is then to the bottom center of the plot. Ball 15 towards the bottom of the plot and picks up a small set of points therein. Balls 16 and 17 are to the upper edge of the plot, whilst ball 18 works to fill in another gap in the denser part of the space. Ball 19 is to the center right, again picking up a more dense part of the space. Balls 20 and 21 are to the edge of the space, completing the cover.

\begin{figure}
	\begin{center}
		\caption{Towards A TDABM Plot}
		\label{fig:astep2}
		\begin{tabular}{c c}
			\includegraphics[height=7cm]{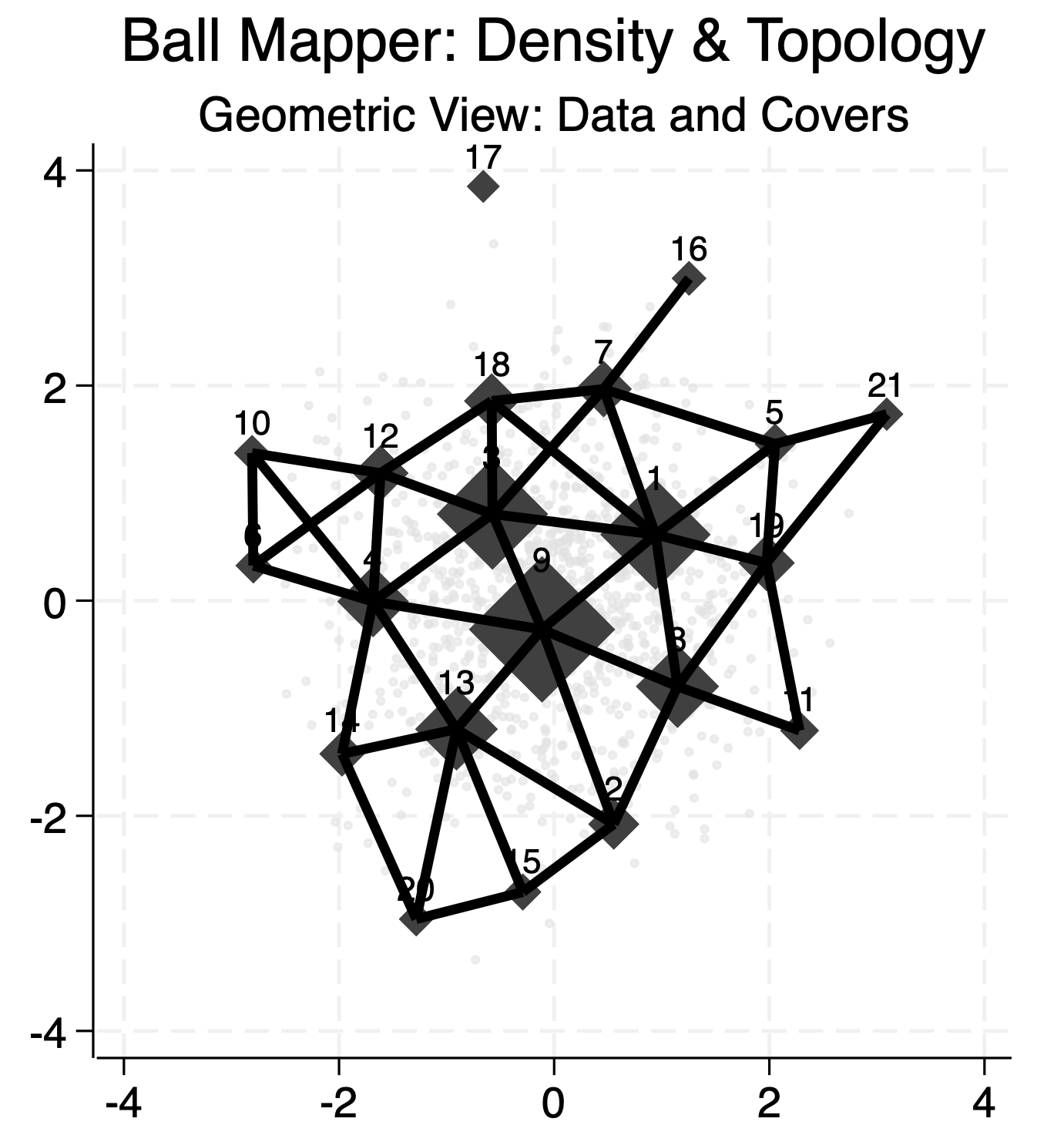}&
			\includegraphics[height=7cm]{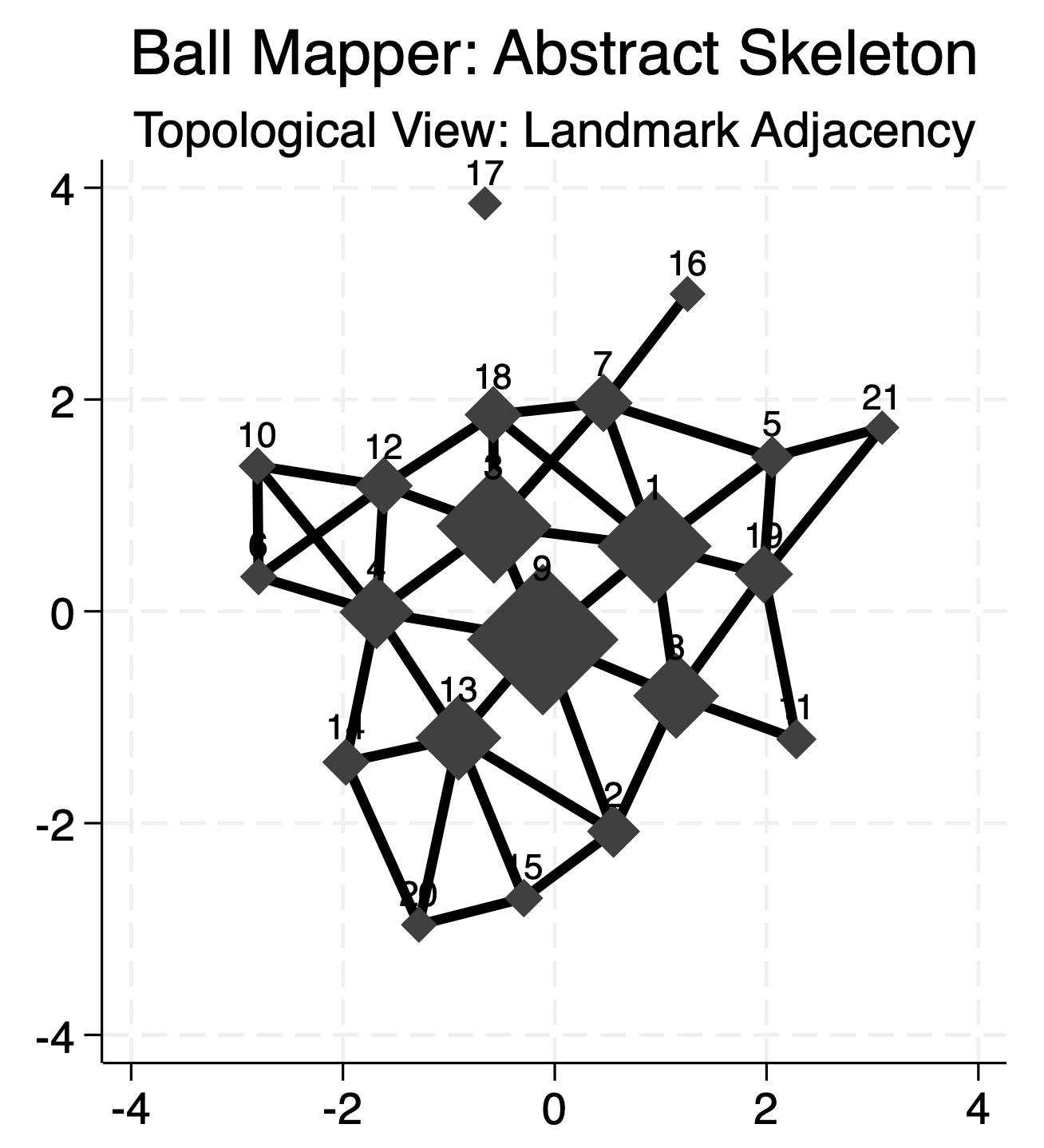}\\
			(a) Sizing, Edges and Labels & (b) Removing other points \\
		\end{tabular}
	\end{center}
	\raggedright
	\footnotesize{Notes: Construction of the TDABM graph on a 2-variable Gaussian cloud. The data here has 1000 observations with $X_1 \sim N(0,1)$ and $X_2 \sim N(0,1)$. In the building of a TDABM graph, once all points are covered the landmarks are sized according to the number of points in their respective balls and edges are drawn between any pair of landmarks for whom the balls overlapped. Because the information from the data is then contained within the graph, the full set of datapoints are redundant. Panel (b) shows the same graph with the other data points removed. The final step would be to remove the axes and make the landmarks abstract.}
\end{figure}

The next stages of the graph construction are readily understood through Figure \ref{fig:astep2}. First the density of the data is captured through the size of the landmark points. Proximity of landmarks to one another is captured through the drawing of edges across any non-empty overlap. To aid understanding of the graph, the landmark points are numbered. Because we are using a scatter plot to demonstrate the functionality, the numbers are placed near the landmarks to which they refer. The balls remember the points that are within them as the information is stored by the algorithm. Individual data points are then redundant and may be removed. Panel (b) shows the graph without the full data points. Information on the structure and density remains. The final step here would be to remove the axes and make the plot abstract. The edges would be unchanged by the abstraction of the network. The resulting plot resembles a network, but is a topologically faithful representation of the underlying data.

In this demonstration, the underlying dataset had two dimensions. A scatterplot of two dimensional data is easy enough to understand and interpret. Consequently the process of abstraction, and the removal of data points from the plot, serve to make the plot less informative. However, in the case where there are more variables, and we cannot see the data in a single plot, the TDABM process permits the construction of an interpretable visualisation. 

\end{document}